\documentclass[twocolumn]{aastex62}
\usepackage{amsmath}

\newcommand{\del}{\partial}

\received{}
\revised{\today}
\accepted{}

\shorttitle{Mass Ejection of Binary Neutron Stars}
\shortauthors{Fujibayashi et al.}

\begin{document}

\title{Post-merger Mass Ejection of Low-mass Binary Neutron Stars
}

\correspondingauthor{Sho Fujibayashi}
\email{sho.fujibayashi@aei.mpg.de}

\author[0000-0001-6467-4969]{Sho Fujibayashi}
\affiliation{Max-Planck-Institut f\"ur Gravitationsphysik (Albert-Einstein-Institut), Am M\"uhlenberg 1, D-14476 Potsdam-Golm, Germany}

\author[0000-0002-4759-7794]{Shinya Wanajo}
\affiliation{Max-Planck-Institut f\"ur Gravitationsphysik (Albert-Einstein-Institut), Am M\"uhlenberg 1, D-14476 Potsdam-Golm, Germany}
\affiliation{Department of Engineering and Applied Sciences, Faculty of Science and Technology, Sophia University, 7-1 Kioicho, Chiyoda-ku, Tokyo, 102-8554, Japan}
\affiliation{Interdisciplinary Theoretical and Mathematical Science (iTHEMS) Research Group, RIKEN, Wako, Saitama, 351-0198, Japan}

\author[0000-0003-4988-1438]{Kenta Kiuchi}
\affiliation{Max-Planck-Institut f\"ur Gravitationsphysik (Albert-Einstein-Institut), Am M\"uhlenberg 1, D-14476 Potsdam-Golm, Germany}
\affiliation{Center for Gravitational Physics, Yukawa Institute for Theoretical Physics, Kyoto University, Kyoto, 606-8502, Japan}

\author[0000-0003-3179-5216]{Koutaro Kyutoku}
\affiliation{Interdisciplinary Theoretical and Mathematical Science (iTHEMS) Research Group, RIKEN, Wako, Saitama, 351-0198, Japan}
\affiliation{Center for Gravitational Physics, Yukawa Institute for Theoretical Physics, Kyoto University, Kyoto, 606-8502, Japan}
\affiliation{Department of Physics, Kyoto University, Kyoto, 606-8502, Japan}

\author[0000-0002-2648-3835]{Yuichiro Sekiguchi}
\affiliation{Center for Gravitational Physics, Yukawa Institute for Theoretical Physics, Kyoto University, Kyoto, 606-8502, Japan}
\affiliation{Department of Physics, Toho University, Funabashi, Chiba 274-8510, Japan}

\author[0000-0002-4979-5671]{Masaru Shibata}
\affiliation{Max-Planck-Institut f\"ur Gravitationsphysik (Albert-Einstein-Institut), Am M\"uhlenberg 1, D-14476 Potsdam-Golm, Germany}
\affiliation{Center for Gravitational Physics, Yukawa Institute for Theoretical Physics, Kyoto University, Kyoto, 606-8502, Japan}

\begin{abstract}

We study the post-merger mass ejection of low-mass binary neutron stars (NSs) with the system mass of $2.5\, M_\odot$, and subsequent nucleosynthesis by performing general-relativistic, neutrino-radiation viscous-hydrodynamics simulations in axial symmetry.
We find that the merger remnants are long-lived massive NSs surviving more than several seconds, irrespective of the nuclear equations of state (EOSs) adopted. The ejecta masses of our fiducial models are $\sim 0.06$--$0.1\, M_\odot$ (depending on the EOS), being $\sim 30\%$ of the initial disk masses ($\sim 0.15$--$0.3\, M_\odot$). Post-processing nucleosynthesis calculations indicate that the ejecta is composed mainly of light $r$-process nuclei with small amounts of lanthanides (mass fraction $\sim 0.002$--0.004) and heavier species due to the modest average electron fraction ($\sim 0.32$--0.34) for a reasonable value of the viscous coefficient. Such abundance distributions are incompatible with the solar $r$-process-like abundance patterns found in all measured $r$-process-enhanced metal-poor stars. Therefore, low-mass binary NS mergers should be rare. If such low-mass NS mergers occur, their electromagnetic counterparts, kilonovae, will be characterized by an early bright blue emission because of the large ejecta mass as well as the small lanthanide fraction.
We also show, however, that if the effective turbulent viscosity is very high, or there is an efficient mass ejection working in the early post-merger phase, the electron fraction of the ejecta could be low enough that the solar $r$-process-like abundance pattern is reproduced and the lanthanide fraction becomes so high that the kilonova would be characterized by early bright blue and  late bright red emissions.

\end{abstract}


\section{Introduction} \label{sec:intro}
Binary neutron star (NS) mergers are characterized by the following astrophysical aspects.
First, they are the sources of gravitational waves that can be detected by ground-based gravitational wave detectors \citep{Abadie2010a,Accadia2011a,Akutsu2018a}.
Second, the remnant of an NS merger, that is, either the massive NS or black hole (BH) surrounded by an accretion disk, could be a central engine of short gamma-ray bursts \citep[e.g.,][]{Eichler1989a}, for which the afterglow follows later on across a wide range of electromagnetic wavelengths \citep[from X-rays to radio,][ and references therein]{Metzger2012a,Hotokezaka2015a}. Third, NS mergers are the promising sites of the nucleosynthesis for about half of the elements heavier than iron---the rapid neutron capture ($r$-process) nuclei (\citealt{Lattimer1974a,Symbalisty1982a,Eichler1989a,Meyer1989a,Freiburghaus1999a}; see also a recent comprehensive review, \citealt{Cowan2019a}).
The decaying radioactive nuclei freshly synthesized in the merger ejecta release energy and power an ultraviolet-optical-infrared transient called kilonova or macronova \citep{Li1998a,Kulkarni2005a,Metzger2010a}. All of these (gravitational- and electromagnetic-wave) signals have been observed in the first gravitational-wave event from the binary NS merger, GW170817 \citep{Abbot2017a,Villar2017a,Abbot2017b,Lyman2018a,Alexander2017a}, while in the second binary NS merger event, GW190425 \citep{Abbott2020a}, no electromagnetic counterpart has been detected.
This indicates that there is a wide variety of possibilities in binary NS mergers reflected by the variation of system masses (the inferred values are 2.73--$2.78\,M_\odot$ for GW170817 and $\sim 3.4\,M_\odot$ for GW190425).

The merger of binary NSs is accompanied by successive mass ejection episodes as found in recent general-relativistic hydrodynamical simulations including weak interactions as well as temperature-dependent nuclear equations of state \citep[EOSs, e.g.,][]{Shibata2019a}.
During the merger stage, the so-called dynamical mass ejection occurs and the material of $\sim 10^{-3}$--$10^{-2}\, M_\odot$ is ejected by tidal torque and shock heating \citep[e.g.,][]{Hotokezaka2013a,Bauswein2013a,Sekiguchi2015a,Palenzuela2015a,Sekiguchi2016a,Foucart2016a,Shibata2017a,Radice2018a,Bovard2017a,Vincent2020a}. The resultant wide range of electron fraction, $Y_\mathrm{e}$ $\sim 0.1$--0.4, in the ejecta generally leads to the robust production of $r$-process nuclei with an abundance pattern that agrees reasonably with the solar $r$-process abundance distribution as shown by nucleosynthesis studies \citep{Wanajo2014a,Goriely2015a,Bovard2017a,Radice2018a}.

The post-merger mass ejection phase then follows for seconds, during which about 10--50\% of the material  in the accretion disk ($\sim 0.01$--$0.1\, M_\odot$) around the central remnant becomes unbound by viscous heating \citep[here the viscosity is believed to originate from magneto-rotational instability (MRI), e.g.,][]{Fernandez2013a,Just2015a,fujibayashi2018a,Siegel2018a,Fernandez2019a}.\footnote{We note that the neutrino drag effect may suppress the growth of MRI \citep{Guilet2017a}.
}
The central remnant can be either a (hyper-) massive NS or a BH depending primarily on the total mass of binary NSs as well as the nuclear EOS. In the case that the total mass is larger than a certain critical value determined for a given EOS, the merger remnants collapse into a BH promptly or after the phase of a short-lived hyper-massive NS \citep{Shibata2006a,Shibata2019a}.

In the BH accretion disk, the viscous effect heats up the material and transports the angular momentum outward in the accretion disk, leading eventually to mass ejection from the disk. The viscous heating generates thermal energy that induces the electron-positron pair creation, enhancing the positron capture by neutrons and increasing the electron fraction, $Y_\mathrm{e}$. The viscous expansion reduces the density of the disk material, and as a result, the electron degeneracy is decreased, increasing $Y_\mathrm{e}$ as well (at this stage, the role of neutrino irradiation on varying $Y_\mathrm{e}$ is subdominant). However, the average values of $Y_\mathrm{e}$ for the ejecta, $\langle Y_\mathrm{e} \rangle$, show some diversity in the literature: $\langle Y_\mathrm{e} \rangle \sim 0.1$--0.2 \citep{Fernandez2013a,Siegel2018a,Fernandez2019a}, $\langle Y_\mathrm{e} \rangle \sim 0.2$--0.3 \citep{Just2015a}, and $\langle Y_\mathrm{e} \rangle \sim 0.3$ \citep{Fujibayashi2020a,Fernandez2020a}. Accordingly, the nucleosynthesis outcomes differ from those of each other.

In the case that the total mass of binary NSs is smaller than a critical value, which is the main focus of this paper, a long-lived massive NS is formed and it emits copious neutrinos that further increase the value of $Y_\mathrm{e}$ in the ejecta. The average values of $Y_\mathrm{e}$ for the ejecta derived in numerical simulations are in reasonable agreement among different groups, $\langle Y_\mathrm{e} \rangle \sim 0.3$--0.4 (for the lifetimes of massive NSs longer than a few 100~ms). However the number of such work is still limited \citep{Metzger2014a,Perego2014a,Lippuner2017a,fujibayashi2018a}. Moreover, the numerical methods are rather different among these work.
In \citet{Metzger2014a}, \citet{Perego2014a}, and \citet{Lippuner2017a}, the central massive NS is not solved self-consistently.
By contrast, \citet{fujibayashi2018a} self-consistently evolved the post-merger system (both the remnant NS and disk around it) by adopting the 3D result of the dynamical phase of an equal-mass ($1.35\, M_\odot$) NS merger \citep{Sekiguchi2015a,Sekiguchi2016a} as the initial condition.

The resultant nucleosynthetic abundance distributions in the post-merger ejecta from such systems exhibit a pattern that is different from the solar $r$-process pattern, with small amounts of heavy $r$-process nuclei. This is due to  the modest neutron richness of the ejected material. Such $r$-process abundance patterns predicted from the above simulations for long-lived massive NSs disagree with the implication from the spectroscopic analyses of Galactic halo stars. To date, all measured $r$-process-enhanced stars appear to exhibit solar $r$-process-like abundance patterns, in particular for the range of atomic number $Z = 56$--78 \citep[Ba--Pt, e.g.,][]{Cowan2019a}, although few measurements of the second $r$-process peak (Te, $Z = 52$, only available from space) have been reported \citep{Roederer2012a}. Deviation from the solar $r$-process pattern, which can be seen for $Z = 38$--47 (Sr--Ag), is generally not larger than a factor of a few. 

However, the discovery of the lowest-mass double NS system to date, PSR J1946+2052 \citep{Stovall2018a}, strongly suggests the presence of long-lived massive NSs as merger remnants.
The system mass of J1946+2052 is only $2.50 \pm 0.04\, M_\odot$, substantially smaller than the typical value of $\sim$ 2.6--2.7\,$M_\odot$ for the observed Galactic double NS systems \citep[e.g.,][]{Tauris2017a} and 
those inferred for the NS mergers GW170817 \citep[2.73--$2.78\, M_\odot$,][]{Abbot2017a} and GW190425 \citep[$\sim 3.4\, M_\odot$,][]{Abbott2020a}.
Thus, the long-lived (possibly supramassive) NSs surrounded by accretion disks could be the outcomes for this class of binary NS mergers.

The purpose of this work is to study the mass ejection and subsequent nucleosynthesis of low-mass NS binaries near the low-mass end.
The paper is organized as follows. In \S \ref{sec:method} we summarize our fully general relativistic, neutrino-radiation, and viscous-hydrodynamics code. The system mass is taken to be $2.5\, M_\odot$ (and $2.7\, M_\odot$ for comparison purposes), being similar to that of J1946+2052. We consider an equal-mass case; each NS mass is $1.25\, M_\odot$ (and $1.35\, M_\odot$) because the mass ratio of two NSs in this case is likely close to unity due to the lower bound of NS mass $\sim 1.2\, M_\odot$ predicted from theoretical studies \citep{Mueller2016a,Suwa2018a}. Two temperature-dependent nuclear EOSs, DD2 \citep{banik2014a} and SFHo \citep{steiner2013a}, are adopted in the simulations.
In \S~\ref{sec:hydro}, the hydrodynamical outcomes are discussed.
The properties of the central massive NS, disk, and ejecta that are important for the nucleosynthesis are shown.
In particular, we describe how the electron fraction of the ejecta is determined by considering the timescales of the weak-interaction processes.
The thermodynamic histories extracted from each simulation by several thousand tracer particles are then used for nucleosynthesis calculations (\S~\ref{sec:nucleosynthesis}). The time variations of radioactive energies, which power kilonova emission, are also obtained from the nucleosynthesis calculations. Finally, we present the summary and conclusions of our study in \S~\ref{sec:summary}. 
Throughout this paper, we employ the geometrical units in which the speed of light $c$ and gravitational constant $G$ are set to unity.

\section{Method}\label{sec:method}
\subsection{Einstein's Equation}
We use a fully general relativistic, neutrino-radiation viscous-hydrodynamics code developed in our previous work \citep{fujibayashi2018a}.
Einstein's equation is solved with a version of the puncture-Baumgarte-Shapiro-Shibata-Nakamura (BSSN) formalism \citep{shibata1995a,baumgarte1999,marronetti2008} incorporating Z4c prescription \citep{Hilditch2013a} for the constraint violation propagation.
The quantities solved in this formalism are listed in Table\,\ref{tab:var}.
From the three-metric $\gamma_{ij}$ and extrinsic curvature $K_{ij}$, we define $\tilde{\gamma}_{ij}=\gamma^{-1/3}\gamma_{ij}$, $W=\gamma^{-1/6}$, $K=\gamma^{ij}K_{ij}$, $\tilde{A}_{ij}=\gamma^{-1/3} (K_{ij}-\gamma_{ij} K/3)$, and $F_i = \delta^{jk}\del_j \tilde{\gamma}_{ik}$.
Here $\gamma$ denotes the determinant of $\gamma_{ij}$ and the determinant of $\tilde{\gamma}_{ij}$ is assumed to be unity.
For the gauge conditions, we employ dynamical lapse and shift gauge conditions described in Eqs.~(1) and (2) in \cite{fujibayashi2017a}.
We introduce a new variable $\Theta$, which indicates the degree of the constraint violation, for constraint propagation in the Z4c formalism \citep{Hilditch2013a}.
For this change, the equations in the original BSSN formalism are slightly modified.
For example, $\hat{K} \equiv K - 2\Theta$ obeys the same evolution equation as $K$ \citep[see, for example,][]{Shibata2016a}.
We write the evolution equation of $\Theta$ as
\begin{align}
(\del_t -\beta^k \del_k) \Theta &= \frac{1}{2}\alpha \biggl[ R - \tilde{A}_{ij}\tilde{A}^{ij} \notag\\
&\hspace{10mm} + \frac{2}{3} (\hat{K} + 2\Theta) - 16\pi \rho_\mathrm{h}\biggr ]e^{-r^2/r_\mathrm{o}^2}, \label{eq:z4c}
\end{align}
where $\rho_\mathrm{h} \equiv T^{\alpha\beta}n_\alpha n_\beta$, $r=\sqrt{x^2+y^2+z^2}$, and $r_\mathrm{o}\approx 1000$\,km. Here we set $\kappa_1=\kappa_2=0$ in the full description of Z4c in \cite{Hilditch2013a}.
As Eq.~\eqref{eq:z4c} shows, we incorporate the constraint propagation only for a near zone to avoid the accumulation of the constraint violation only the strong field zone.

We adopt sixth-order Kreiss-Oliger dissipation to all the geometrical variables except for the lapse function to avoid numerical instability that could appear in employing the non-uniform grid in the numerical evolution.
We solve the Einstein's evolution equation in the Cartesian coordinates and employ the so-called cartoon method to impose axisymmetric conditions for the geometrical quantities \citep{Alcubierre2001a,Shibata2000a}.

\begin{table*}[t]
\caption{ List of the quantities evolved in our formalism}
\begin{center}
\begin{tabular}{lll}
\hline \hline
Notation & Definition & Reference\\
\hline
$\tilde{\gamma}_{ij} = \gamma^{-1/3} \gamma_{ij}$ & Conformal three-metric & \cite{shibata1995a}\\
$W=\gamma^{-1/6}$ & Conformal factor & \cite{marronetti2008}\\
$K=\gamma^{ij}K_{ij}$ & Trace of the extrinsic curvature $K_{ij}$ & \cite{shibata1995a}\\
$\tilde{A}_{ij}=\gamma^{-1/3} (K_{ij}-\gamma_{ij} K/3)$ & Trace-free part of $K_{ij}$ & Eq.~(2.9) in \cite{shibata1995a} \\
$F_{i} = \delta^{jk} \del_j \tilde{\gamma}_{ki}$& Auxiliary variable & Eq.~(2.19) in \cite{shibata1995a} \\
$\Theta$ & Hamiltonian constraint violation & \cite{Hilditch2013a} \\
\hline
\end{tabular}
\end{center}
\label{tab:var}
\end{table*}

\subsection{Radiation Viscous-hydrodynamics Equations} \label{subsec:rad}
We briefly describe the basic equations and our method to solve viscous-hydrodynamics equations with neutrino radiation transfer.
We solve the following evolution equations in a two-dimensional (2D) axisymmetric manner.
We decompose neutrinos into ``streaming" and ``trapped" components.
Then we also decompose the total energy-momentum tensor of the matter, which includes fluid and neutrinos, as
\begin{eqnarray}
T^{\alpha\beta}_\mathrm{(tot)} = T^{\alpha\beta} + \sum_i T_{(\nu_i, \mathrm{S})}^{\alpha\beta},
\end{eqnarray}
where $T^{\alpha\beta}$ is the sum of the energy-momentum tensors for the fluid $T_\mathrm{ (fluid)}^{\alpha\beta}$ and trapped neutrinos $T_{(\nu_i, \mathrm{T})}^{\alpha\beta}$, and $T_{(\nu_i, \mathrm{S})}^{\alpha\beta}$ is that for streaming neutrinos.
The index $i$ specifies the neutrino species.
Following our previous work \citep{fujibayashi2017a,fujibayashi2018a}, we consider three species of neutrinos: electron neutrinos $\nu_\mathrm{e}$, electron antineutrinos $\bar{\nu}_\mathrm{e}$, and the other neutrino species $\nu_\mathrm{x}$ which represents muon and tau neutrinos and those antineutrinos altogether.

The evolution equations for these energy-momentum tensors are
\begin{eqnarray}
\nabla _\beta T^{\alpha\beta} &=& -Q_\mathrm{(leak)}^\alpha = - \sum_i Q^\alpha_{\mathrm{(leak)}\nu_i},\label{eq:fluid}\\
\nabla _\beta T_{(\nu_i, \mathrm{S})}^{\alpha\beta} &=& Q^\alpha_{\mathrm{(leak)}\nu_i}, \label{eq:rad}
\end{eqnarray}
where $Q^\alpha_{\mathrm{(leak)}\nu_i}$ denotes the ``leakage" rate of $i$-th species of neutrinos.
Note that this leakage rate includes the heating due to neutrino absorption and pair-annihilation, as well as the cooling due to neutrino emission, as described in \cite{fujibayashi2017a}.\footnote{
We updated the electron and positron capture rates from those originally described in \cite{sekiguchi2010a}.
Previously we used the difference of the chemical potential of protons and neutrons as the $Q$-value in calculating those rates, while in the current work the mass difference between neutron and proton is used for this.
After this update, we calibrated the neutrino luminosity derived in our simulation for the core collapse of a 15\,$M_\odot$ solar-metallicity progenitor \citep{Woosley2002a} by comparing with those in \cite{Liebendoerfer2003a} and \cite{Janka2012b}, for appropriately setting the parameters of our leakage-based neutrino radiation transport method \citep[see][for the parameters]{sekiguchi2010a}.
Then, we performed 3D merger simulations again using the calibrated parameters.
We found that our previous prescription gives higher values of the electron fraction in the equilibrium of weak-interaction reactions.
This leads to an overestimation of the electron fraction of the ejecta in the previous work \citep{Sekiguchi2015a,Sekiguchi2016a,fujibayashi2018a}, typically by $\sim$ 0.05.
Our previous prescription also overestimated the dynamical ejecta mass as $\approx 0.011 M_\odot$, which is by $\sim 30$\% larger than that in this study (for SFHo-135; see Table~\ref{tab:3dmodel}).
}
To solve the evolution equations for streaming neutrinos, we employ Thorne's truncated moment formalism \citep{Thorne1981a,shibata2011a} with a closure relation \citep{levermore1984a,gonzalez2007a}.
For trapped neutrinos, we employ a leakage-based scheme developed in \cite{sekiguchi2010a}.
The detailed description of these schemes is found in \cite{sekiguchi2010a} and \cite{fujibayashi2017a}.

The energy-momentum tensor of a viscous fluid with trapped-neutrinos is written as
\begin{eqnarray}
T_{\alpha\beta} = \rho h u_\alpha u_\beta + Pg_{\alpha\beta} - \rho h
\nu \tau^0{}_{\alpha\beta},
\end{eqnarray}
where $\rho$ is the baryon rest-mass density, $h=1+\varepsilon + P/\rho$ is the specific enthalpy with the specific internal energy $\varepsilon$ and the pressure $P$, $u^\alpha$ is the four-velocity of the fluid, $\nu$ is the viscous coefficient, and $\tau^0{}_{\alpha\beta}$ is the viscous stress tensor, which is a symmetric tensor satisfying the relation $\tau^0{}_{\alpha\beta}u^\alpha=0$.

Following \cite{Israel1979a}, we write the evolution equation for the viscous stress tensor as
\begin{eqnarray}
{\cal L}_u \tau^0{}_{\alpha\beta} = -\zeta(\tau^0{}_{\alpha\beta} - \sigma_{\alpha\beta}), \label{eq:tau0}
\end{eqnarray}
with the shear tensor $\sigma_{\alpha\beta}$ defined by
\begin{eqnarray}
\sigma_{\alpha\beta} = h_\alpha{}^\mu h_\beta{}^\nu (\nabla_\mu u_\nu + \nabla_\nu u_\mu) = {\cal L}_u h_{\alpha\beta}.
\end{eqnarray}
Here, $h_{\alpha\beta} = g_{\alpha\beta} + u_\alpha u_\beta$.
$\zeta$ is a non-zero constant of (time)$^{-1}$ dimension, which is set to $3\times 10^4$\,rad/s in this work following our previous work \citep{fujibayashi2018a}.
The value is approximately four times larger than the largest angular velocity of the system, $\approx 7000$\,rad/s.
With this prescription, the tensor $\tau^0{}_{\alpha\beta}$ approaches the shear tensor $\sigma_{\alpha\beta}$ in a sufficiently short timescale of $\sim\zeta^{-1}\approx 0.03$\,ms.

We can rewrite Eq.~(\ref{eq:tau0}) as
\begin{eqnarray}
{\cal L}_u \tau_{\alpha\beta} = -\zeta \tau^0{}_{\alpha\beta}, \label{eq:tau}
\end{eqnarray}
where $\tau_{\alpha\beta} \equiv \tau^0{}_{\alpha\beta} -\zeta h_{\alpha\beta}$.
Thus, in addition to the hydrodynamics equations of Eq.~(\ref{eq:fluid}), we solve Eq.~(\ref{eq:tau}) as a basic equation that describes the evolution of $\tau_{\alpha\beta}$.
The details of the formulation of our general relativistic viscous-hydrodynamics are found in~\cite{shibata2017b}.

For the viscous coefficient, we use the so-called Shakura-Sunyaev description as
\begin{eqnarray}
\nu = \alpha_\mathrm{vis} c_\mathrm{s} H_\mathrm{tur}, \label{eq:nudef}
\end{eqnarray}
where $c_\mathrm{s}$ is the sound speed and $H_\mathrm{tur}$ is a typical scale height of the system \citep{Shakura1973a}.
We adopt $H_\mathrm{tur}=10$\,km following \cite{fujibayashi2018a}.

\subsection{Condition for Ejecta}
\label{subsec:ubound}
To study the property of the ejecta in numerical simulations, we have to identify the unbound material in a finite computational domain.
This gives us a non-trivial question about how we should determine the condition for the unbound material.
There are several possibilities for this condition.  In our previous numerical simulations for binary NS mergers \citep{Sekiguchi2015a,Sekiguchi2016a}, we
employed the condition,
\begin{align}
u_t + 1 < 0   \label{eq:ut1}
\end{align}
for the entire fluid elements in the computational domain, assuming that the dynamical ejecta has small internal energy which is much smaller than kinetic energy.
In our previous axisymmetric simulation for the evolution of the merger remnants \citep{fujibayashi2018a}, we employed the condition,
\begin{align} 
hu_t + 1 < 0  \label{eq:hut1}
\end{align}
for the fluid elements that have outgoing velocity (i.e., we took into account the thermal contribution for identifying the ejecta) because the kinetic energy per mass of the post-merger ejecta is much smaller than that of the dynamical ejecta.

In the present work, alternatively, we employ the following condition:
\begin{align}
h u_t + h_\mathrm{min} < 0, \label{eq:huth}
\end{align}
where $h_\mathrm{min}$ is the minimum value of the specific enthalpy in the chosen tabulated EOS ($\approx 0.9987$).
The reason for this choice is that the atomic mass unit $m_u$ is used as the mass per baryon in the EOS we employed, and thus, the specific enthalpy can be smaller than unity for the case that the actual mass per baryon is lighter than $m_u$ and that the pressure contribution to the enthalpy is
negligible. 

We note that the material with the lowest specific enthalpy $h_{\rm min}$ is composed mainly of heavy nuclei in the EOSs employed in this study, in which nuclear statistical equilibrium (NSE) is assumed for the nuclear composition.
Thus, in Eq.~\eqref{eq:huth}, the energy released by the recombination of free nucleons into heavy nuclei is taken into account.
That is, Eq.~\eqref{eq:huth} is the condition for the material being unbound only for the case that the temperature of the material is low enough that free nucleons recombine into heavy nuclei.
One issue for using Eq.~\eqref{eq:huth} is that some fluid components, which stay in the central region and are strongly bounded by the central massive NS, can satisfy this condition.
Such components remain to be composed of free nucleons, and thus, even if the material satisfies Eq.~\eqref{eq:huth}, it is not ejected eventually.
To remove the contribution of this component, we in addition impose the condition of $r \geq 500$\,km to identify the real ejecta component when we employ Eq.~\eqref{eq:huth} as the condition.

Obviously, the total mass of the ejecta becomes largest when we employ Eq.~\eqref{eq:huth}.
Thus, in our previous studies, in particular for the case that we employed Eq.~\eqref{eq:ut1}, the total mass of the ejecta would be underestimated (see Table~\ref{tab:3dmodel}).
Also, the property of the ejecta such as average electron fraction depends weakly on the choice of the condition for the unbound object.

\subsection{Initial Condition}
As in our previous work \citep{fujibayashi2017a,fujibayashi2018a}, we use hydrodynamical configurations obtained by three-dimensional (3D) numerical relativity simulations for equal-mass binary NS mergers as the initial condition for our 2D simulations.
At several tens of milliseconds after the merger, the system, which is composed of a massive NS and an accretion disk, becomes nearly axisymmetric.
We averaged the 3D hydrodynamical quantities at $\approx 50$ ms after the merger over the azimuthal angle around the rotational axis to obtain a 2D axisymmetric hydrodynamical configuration used in our simulation.
For the viscous tensor, we set $\tau^0{}_{\alpha\beta}=0$ initially.
That is, we assume that the fluid is ideal initially.
The initial condition for the geometrical variables are determined in the assumption of the conformal flatness for $\gamma_{ij}$ and by solving the Hamiltonian and momentum constraints for given matter configuration.

In this work, we set the origin of the time $t=0$ to the beginning of our 2D simulation, which is $\approx 50$ ms after the onset of the merger.

\subsection{3D models} \label{subsec:3d}

\begin{table*}[t]
\caption{List of the models for three-dimensional merger simulation }
\begin{center}
\begin{tabular}{lllllllllll}

\hline \hline

Model & EOS & $M_\mathrm{NS}$ & $R_\mathrm{cold}$ & $t_\mathrm{end}-t_\mathrm{merge}$ & $M_\mathrm{ej}$ & $\langle V_\mathrm{ej}\rangle$ & $\langle s_\mathrm{ej}\rangle$ &  $\langle Y_\mathrm{e,ej}\rangle$ & $N_\mathrm{tp}$ & $X_\mathrm{la}$ \\
 & & $(M_\odot)$ & (km) & (ms) & ($10^{-2}M_\odot$) & ($c$) & ($k_\mathrm{B}$) &\\
\hline
DD2-125  & DD2  & 1.25 & 13.14 & 48.5 & 0.10 & 0.19 & 27 & 0.20 &2203 & 0.17\\
SFHo-125 & SFHo & 1.25 & 11.96 & 50.8 & 0.13 & 0.18 & 21 & 0.27 &2511 & 0.072\\
DD2-135 & DD2   & 1.35 & 13.18 & 61.7 & 0.15 & 0.19 & 25 & 0.23 &2761 & 0.093\\
\hline
SFHo-135 & SFHo & 1.35 & 11.92 & 66.6 & 0.83 & 0.27 & 16 &  0.24 & 2197 & 0.031\\

\hline
\end{tabular}
\end{center}
{\bf Note.} $M_\mathrm{NS}$, $R_\mathrm{cold}$, $t_\mathrm{end}-t_\mathrm{merge}$, $M_\mathrm{ej}$, $\langle V_\mathrm{ej} \rangle$, $\langle s_\mathrm{ej} \rangle$, $\langle Y_\mathrm{e,ej} \rangle$, $N_\mathrm{tp}$, and $X_\mathrm{la}$ are the mass of each NS, radius of the cold spherical NS with the mass of $M_\mathrm{NS}$, time after merger at which the simulation is terminated, ejecta mass, their average velocity, entropy, and electron fraction, the number of tracer particles we used, and the mass fraction of lanthanides, respectively.
The ejecta is defined as the material that satisfies Eq.~\eqref{eq:ut1}.

\label{tab:3dmodel}
\end{table*}

In this subsection we briefly describe models for the 3D merger simulation.
The 3D simulations are performed using our numerical relativity code with an approximate neutrino radiation transport \citep{Sekiguchi2015a,Sekiguchi2016a}.
We employ the puncture-BSSN formalism for the evolution of the geometrical quantities, and the leakage-based scheme described in \S \ref{subsec:rad} for the neutrino transport.
In these 3D simulations, the neutrino absorption is taken into account as the heating source, but not the other processes such as the neutrino pair-annihilation.

In this paper we explore three models with different nuclear EOSs and total masses of the system as shown in Table~\ref{tab:3dmodel}.
As mentioned in \S \ref{sec:intro}, we use two finite-temperature EOSs for nuclear matter, referred to as SFHo \citep{steiner2013a} and DD2 \citep{banik2014a}, which predict the maximum mass of cold spherical NSs of $2.06\,M_\odot$ and $2.42\,M_\odot$, respectively.
We extend the tables of these EOSs to low-density and low-temperature ranges down to $\approx0.16\,\mathrm{g/cm^3}$ and $10^{-3}\,$MeV, respectively, using the EOS of ideal gas with arbitrarily relativistic and degenerate electrons \citep{timmes2000a}.
We consider the equal-mass binaries with two sets of NS masses, 1.25 and 1.35 $M_\odot$.

In Table~\ref{tab:3dmodel}, the properties of the dynamical ejecta (defined by Eq.~\eqref{eq:ut1}) are shown.
They depend on the total mass of the binary and adopted EOS \citep{Bauswein2013a,Hotokezaka2013a}.
The SFHo EOS is ``softer" than DD2, in the sense that if we compare the radius of the NS with the same gravitational mass, the radius in the SFHo EOS is smaller than that in the DD2.
The softer EOS generally enhances the shock heating efficiency in the early evolution stage of the massive NS, which plays a role to increase the fraction of the unbound material.
This leads to the larger ejecta masses for SFHo models than those for DD2 models for equal-mass binary mergers.
In addition, the positron capture is enhanced in the shocked ejecta due to its high temperature, and thus the average values of electron fraction for SFHo models are higher than that for DD2 models.
For DD2 models, the mass ejection is driven predominantly by the tidal effect, which leads to the lower electron fraction of the ejecta.
The dynamical mass ejection occurs primarily toward the direction of the equatorial plane, in particular for the low-mass cases, because the tidal effect is the main mass ejection channel.

We note that the ejecta mass depends on its definition (see \S~\ref{subsec:ubound}).
For example, the ejecta mass for model DD2-125 evaluated using the condition Eq.~\eqref{eq:huth} with $r\geq 500$\,km is $\approx 0.002\,M_\odot$, which is approximately twice as large as that using the condition Eq.~\eqref{eq:ut1} ($\approx 0.001\,M_\odot$; see Table~\ref{tab:3dmodel}).
On the other hand, the average electron fraction of the ejecta changes only slightly (by less than 5\,\%).

\begin{table*}[t]
\caption{List of the models for two-dimensional post-merger simulation}
\begin{center}
\begin{tabular}{lllllllllllllll}

\hline \hline
Model & EOS & $M_\mathrm{NS}$ & $\Delta x_0$ & $N_\mathrm{grid}$ & $L$ & $\alpha_\mathrm{vis}$ & $t_\mathrm{end}$ & $M_\mathrm{disk}$ & $M_\mathrm{ej,tot}$ & $\langle V_\mathrm{ej}\rangle$& $\langle s_\mathrm{ej}\rangle$& $\langle Y_\mathrm{e,ej}\rangle$ &$N_\mathrm{tp}$ & $X_\mathrm{la}$\\
 & & $(M_\odot)$ & (m) &  & (km) && (s) & ($10^{-2}M_\odot$) & ($10^{-2}M_\odot$) & $(c)$ & $(k_\mathrm{B})$ & & \\
\hline
DD2-125M & DD2 & 1.25 & 200 & 921 & 8411 & 0.04 & 6.2 & 32.8 (26.6)  & 11.4 & 0.089 & 15.0 & 0.322 & 39808 & 0.00441\\
SFHo-125H & SFHo & 1.25 & 160 & 993 & 8745 & 0.04 & 3.8 & 14.8 (12.7)  & 6.1 & 0.108 &18.7& 0.342 & 24576 & 0.00261 \\
DD2-135M & DD2 & 1.35 & 200 & 921 & 8411 & 0.04 & 6.1 & 24.8 (21.2) & 8.6 & 0.098 &17.3& 0.337& 39168 & 0.00276\\
\hline
DD2-125L$^\mathrm{a}$ & DD2 & 1.25 & 250 & 857 & 8148 & 0.04& 7.6 &32.6 (26.3) & 10.6 &0.089&15.5& 0.326 & 44416 & 0.00441 \\
DD2-135L$^\mathrm{a}$ & DD2 & 1.35 & 250 & 857 & 8148 & 0.04& 6.9 & 24.3 (21.2) & 8.5 &0.095&18.8& 0.346 & 48768 & 0.00378 \\
DD2-125M-h$^\mathrm{b}$ & DD2 & 1.25 & 200 & 921 & 8411& 0.10 & 1.5 & 35.4 (31.6)  & 19.8 & 0.113 & 13.8 & 0.282 & 9856 & 0.0298\\
DD2-135M-v14$^\mathrm{c}$ & DD2 & 1.35 & 200 & 921 & 8411 & 0.04& 2.8 & 27.9 (23.7) & 8.7 & 0.113 & 20.4 &0.341&--- & ---\\
DD2-135M-irr $^\mathrm{d}$ & DD2 & 1.35 & 200 & 921 & 8411 & 0.04& 5.6 &  24.7 (21.1) &  6.9 & 0.106 & 16.3 & 0.287 &--- & ---\\
DD2-135M-v0$^\mathrm{f}$ & DD2 & 1.35 & 200 & 921 & 8411 & 0.00 & 1.1 &  40.9 (11.0) &  0.3 & 0.164 & 21.1 & 0.262 &--- & ---\\
\hline
\end{tabular}
\end{center}
{\bf Note.} $M_\mathrm{NS}$, $\Delta x_0$, $N_\mathrm{grid}$, $L$, $\alpha_\mathrm{vis}$, $t_\mathrm{end}$, $M_\mathrm{disk}$, $M_\mathrm{ej,tot}$, $\langle V_\mathrm{ej} \rangle$, $\langle s_\mathrm{ej} \rangle$, $\langle Y_\mathrm{e,ej} \rangle$, $N_\mathrm{tp}$, and $X_\mathrm{la}$ are the mass of each NS, the grid spacing of inner computational region, grid number in each direction, size of the computational domain, viscous parameter, time at which the simulation is terminated, initial disk mass, ejecta mass, its average velocity, entropy, electron fraction, the number of tracer particles we used, and the mass fraction of lanthanides, respectively.
The initial disk mass is defined as the mass outside the NS radius $R_\mathrm{NS}$ (see \S \ref{subsec:nsdisk} for its definition) at $t=10$\,ms.
In the parenthesis the mass of the material with $\rho<10^{12}\,\mathrm{g/cm^3}$ is shown.
For $M_\mathrm{ej,tot}$, $\langle V_\mathrm{ej} \rangle$, $\langle s_\mathrm{ej} \rangle$, and $\langle Y_\mathrm{e,ej} \rangle$, the values at $t=t_\mathrm{end}$ are shown.
$^\mathrm{a}$ The model with a lower resolution.
$^\mathrm{b}$ The model with a higher viscous parameter.
$^\mathrm{c}$ The model without viscosity in the region with $\rho > 10^{14}\,\mathrm{g/cm^3}$.
$^\mathrm{d}$ The model without neutrino irradiation.
$^\mathrm{f}$ The model without viscosity.
\label{tab:model}
\end{table*}

\subsection{2D Models}

The models for our 2D simulations in this study are listed in Table~\ref{tab:model}.
For each model, the same EOS tables as those employed in the 3D simulation are used.
To evolve the radiation viscous-hydrodynamics equations in the cylindrical coordinates $(x,z)$, we employ the same nonuniform grid as that in our previous work \citep{fujibayashi2017a,fujibayashi2018a}, in which the grid structure is determined by the uniform grid spacing for the inner region ($\Delta x_0$), the range of the inner region ($R_\mathrm{star}$), the increase rate of the grid spacing for the outer region ($1+\delta$), and the total grid number in each direction ($N_\mathrm{grid}$).
For all the models in this study, we adopt $R_\mathrm{star}=30$\,km and $\delta=0.0075$.
We list $\Delta x$ and $N_\mathrm{grid}$ for each model, together with the size of the computational domain, $L$, in Table~\ref{tab:model}.

Our fiducial models are DD2-125M, SFHo-125H, and DD2-135M.
For DD2-125M and DD2-135M, we set $\Delta x_0 = 200$\,m.
On the other hand, for SFHo-125H, we employ a finer grid resolution with $\Delta x_0 = 160$\,m than those in other models, because the stellar radii for the SFHo EOS are smaller than those for the DD2 EOS, and thus, the conservation of the total energy is less satisfied in a lower resolution model.
We also perform several other simulations; lower resolution ones (DD2-125L and DD2-135L) with $\Delta x_0 = 250$\,m, in order to confirm the convergence of the numerical results with the adopted grid resolutions; one without the viscosity in the region for which the density is higher than $10^{14}\,\mathrm{g/cm^3}$ (DD2-135M-v14) to clarify that the viscosity in the innermost region of the NS does not play an important role for the early viscosity-driven ejecta (\S~\ref{subsec:dyn}); one without the viscosity in the whole region (DD2-135M-v0) to examine the effect of the viscosity on the evolution of the remnant and the ejecta; and one without neutrino heating/irradiation in the optically-thin region (DD2-135M-irr) to understand the importance of neutrino absorption on the evolution of the remnant and the ejecta.

A number of magnetohydrodynamics simulations have shown that the effective value of $\alpha_\mathrm{vis}$ becomes $O$(0.01) for the disks rotating in the vicinity of central remnants \citep[e.g.,][]{kiuchi2018a,Fernandez2019a}.
In addition, in our previous work \citep{fujibayashi2018a}, we found that (if $\alpha_\mathrm{vis} \leq 0.04$) features of the post-merger ejecta do not depend strongly on the viscous parameter except for the mass ejection timescale, which is approximately proportional to $\alpha_\mathrm{vis}{}^{-1}$.
Therefore, we set $\alpha_\mathrm{vis}=0.04$ for our fiducial models in this work.
For the DD2-125 model, we also perform a simulation with a very high viscosity $\alpha_\mathrm{vis}=0.10$ (DD2-125M-h), to explore the dependence of the result on the viscosity strength (i.e., the timescale of the viscous evolution of the system).

\subsection{Particle Trace Method}
\label{subsubsec:tracer}
The evolutions of our grid-based hydrodynamical quantities should be converted into Lagrangian particle motions for the nucleosynthesis calculations (\S~\ref{sec:nucleosynthesis}).
For this purpose, we adopt a tracer particle method.
As initial positions of the tracer particles, we select 128 points with polar angles in the range of $\theta=0$--$\pi/2$ on the arc with the radius of 8000\,km.
We repeat the same procedure every 0.05\,s to set tracer particles.
We then evolve them backward in time.
The numerical method of tracing particles is basically the same as that in \cite{nishimura2015a} except for the difference of the coordinate system employed.
The time integration is based on the predictor-collector method with second-order accuracy.
We use the bilinear interpolation to convert grid-based hydrodynamical quantities into those of the tracer particles.
The particle position $(x^{(n)},z^{(n)})$ at time $t^{(n)}$ is obtained by
\begin{eqnarray}
x^{(n)} &=& x^{(n+1)} + v^{x*} \Delta t,\\
z^{(n)} &=& z^{(n+1)} + v^{z*} \Delta t,
\end{eqnarray}
where $v^{x*}$ and $v^{z*}$ denote the modified velocity components of the particle estimated by the predictor-collector method and $\Delta t = t^{(n)}-t^{(n+1)} (<0)$.

\section{Simulation Result} \label{sec:hydro}
\subsection{Dynamics of the System} \label{subsec:dyn}
\begin{figure*}[t]
\epsscale{1.17}
\plottwo{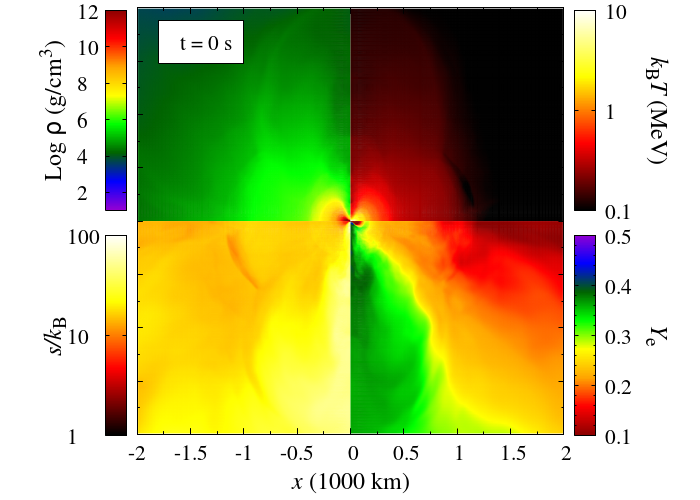}{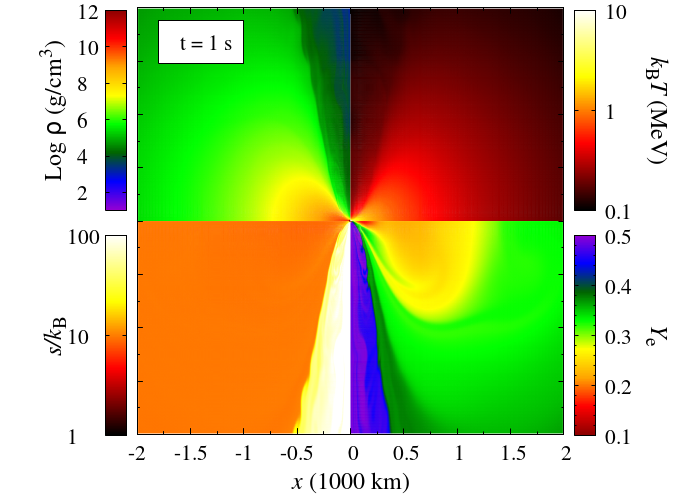}
\plottwo{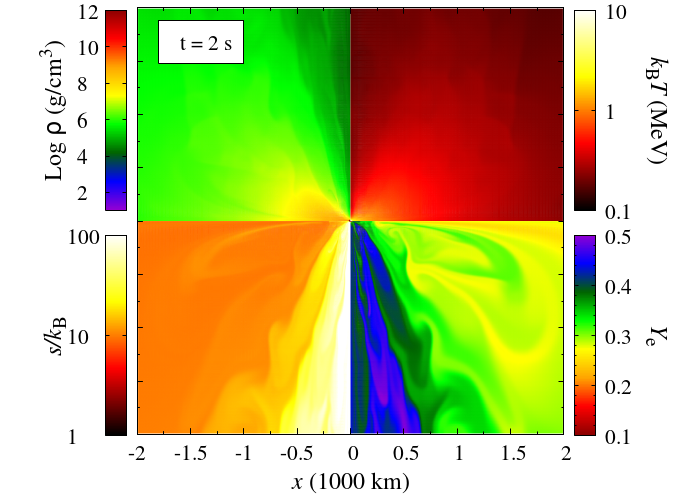}{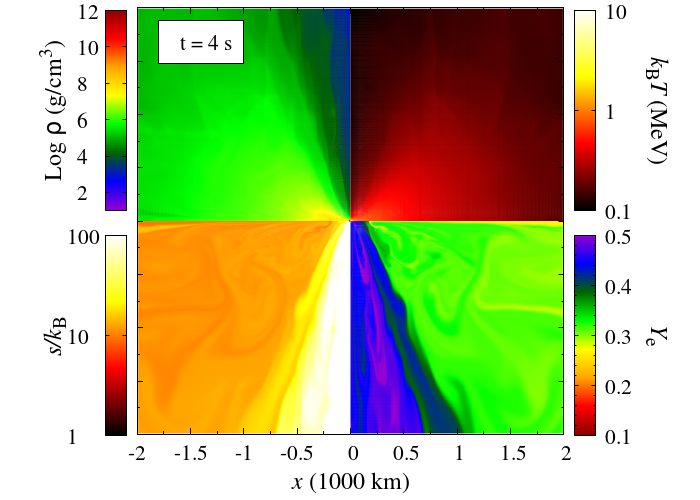}
\caption{Snapshots for model DD2-125M at $t=0,$ 1, 2, and 4\,s.
Each panel with the color bars displays the profiles of the rest-mass density (top-left), temperature (top-right), entropy (bottom-left), and electron fraction (bottom-right).
}
\label{fig:dyn}
\end{figure*}

Although the qualitative evolution process of the post-merger remnant composed of a massive NS and an accretion disk was already described in our previous paper \citep{fujibayashi2018a}, we briefly summarize it again in this subsection.
Figure~\ref{fig:dyn} shows snapshots of the rest-mass density, electron fraction, temperature, and entropy per baryon for model DD2-125M.
The top-left panel of Fig.~\ref{fig:dyn} shows the initial profile of the simulation.
There is a dense disk extending to $\sim 100$~km around the massive NS.
The disk has high density ($\rho\sim 10^{12}$ g/cm$^3$), high temperature ($k_\mathrm{B}T\sim 10$~MeV), and low electron fraction ($Y_\mathrm{e}\lesssim 0.1$).
The material of low electron fraction is located in the region with the density of $\sim10^6\,\mathrm{g/cm^3}$ around the equatorial plane.
This material was expelled during the dynamical merger phase without experiencing positron capture by neutrons because of its low temperature and resulting in poor electron-positron pair creation.
A part of the material is the dynamical ejecta, and the rest is still mechanically bound.

The massive NS formed as a remnant after the binary NS merger is differentially rotating and it is first evolved by the viscous effect.
The timescale is estimated by
\begin{align}
\frac{R_\mathrm{eq}{}^2}{\nu} &\approx 5\,\mathrm{ms}\,\biggr( \frac{\alpha_\mathrm{vis}}{0.04} \biggl)^{-1}\biggr( \frac{c_\mathrm{s}}{c/3} \biggl)^{-1} \notag\\
&\times \biggr( \frac{R_\mathrm{eq}}{15\,\mathrm{km}} \biggl)^2 \biggr( \frac{H_\mathrm{tur}}{10\,\mathrm{km}} \biggl)^{-1}, \label{eq:vis}
\end{align}
where $R_\mathrm{eq}$ is the equatorial radius of the NS and we use Eq.~(\ref{eq:nudef}) for viscous coefficient $\nu$.
In this timescale, the angular momentum distribution of the massive NS is modified due to the viscous angular momentum transport effect.
This fast viscous evolution drives the {\it early viscosity-driven ejecta}, which is the ejecta component swept by the shock wave powered by extracting the rotational kinetic energy of the NS, in particular its outer region.\footnote{In our previous work~\citep{fujibayashi2018a}, we described that this mass ejection was powered by the rotational kinetic energy of the entire region in the NS.
However, as found in Appendix~\ref{app:mod}, the inner region of the NS contributes only slightly to the mass ejection and the viscous effect on the surface of the NS, which has the largest angular velocity, drives the mass ejection.
We also find that the mass of this ejecta component depends weakly on the magnitude of $\zeta$.
This indicates that this early component is partially due to the transitional effect in the timescale of $\zeta^{-1}$, in which the viscous tensor $\tau^0{}_{\alpha\beta}$ approaches $\sigma_{\alpha\beta}$.
}

The range of the electron fraction of this component is approximately the same as that of dynamical ejecta (see \S~\ref{subsubsec:fiducial}).
The amount of this component depends on the viscous parameter $\alpha_\mathrm{vis}$ as found in \S~\ref{subsec:highv}.

In the polar direction, the high-entropy material is ejected mainly due to the heating by neutrinos emitted from the remnant (panels for $t>0$\,s of Fig.~\ref{fig:dyn}).
This polar mass ejection is efficient in particular for $t \lesssim 0.1$\,s during which neutrinos are copiously emitted from the disk and continues throughout our simulation time (i.e., at least for seconds), because the massive NS is present and continuously emits copious neutrinos (cf. Fig~\ref{fig:MNS}).

The disk gradually expands through the outward viscous angular momentum transport and the viscous heating.
Due to the expansion and the neutrino cooling, the temperature of the disk decreases.
When the neutrino cooling rate becomes lower than the viscous heating rate, the disk material begins to be ejected because most part of the viscous heating is used for the disk expansion.
This {\it late-time viscosity-driven ejecta} consists mainly of the material with $Y_\mathrm{e} = 0.3$--0.35 and $s/k_\mathrm{B} \approx 10$ (panels for $t>0$~s of Fig.~\ref{fig:dyn}; see also \S~\ref{subsubsec:fiducial}).
The dynamics of the system among the explored models is quite similar to each other, although the properties of the ejecta depend quantitatively on the EOS, mass of the massive NS, and the magnitude of $\alpha_\mathrm{vis}$.

\subsection{Remnant Massive Neutron Star and Disk} 
\label{subsec:nsdisk}

\subsubsection{Radius of the NS Surface}

\begin{figure}
\epsscale{1.17}
\plotone{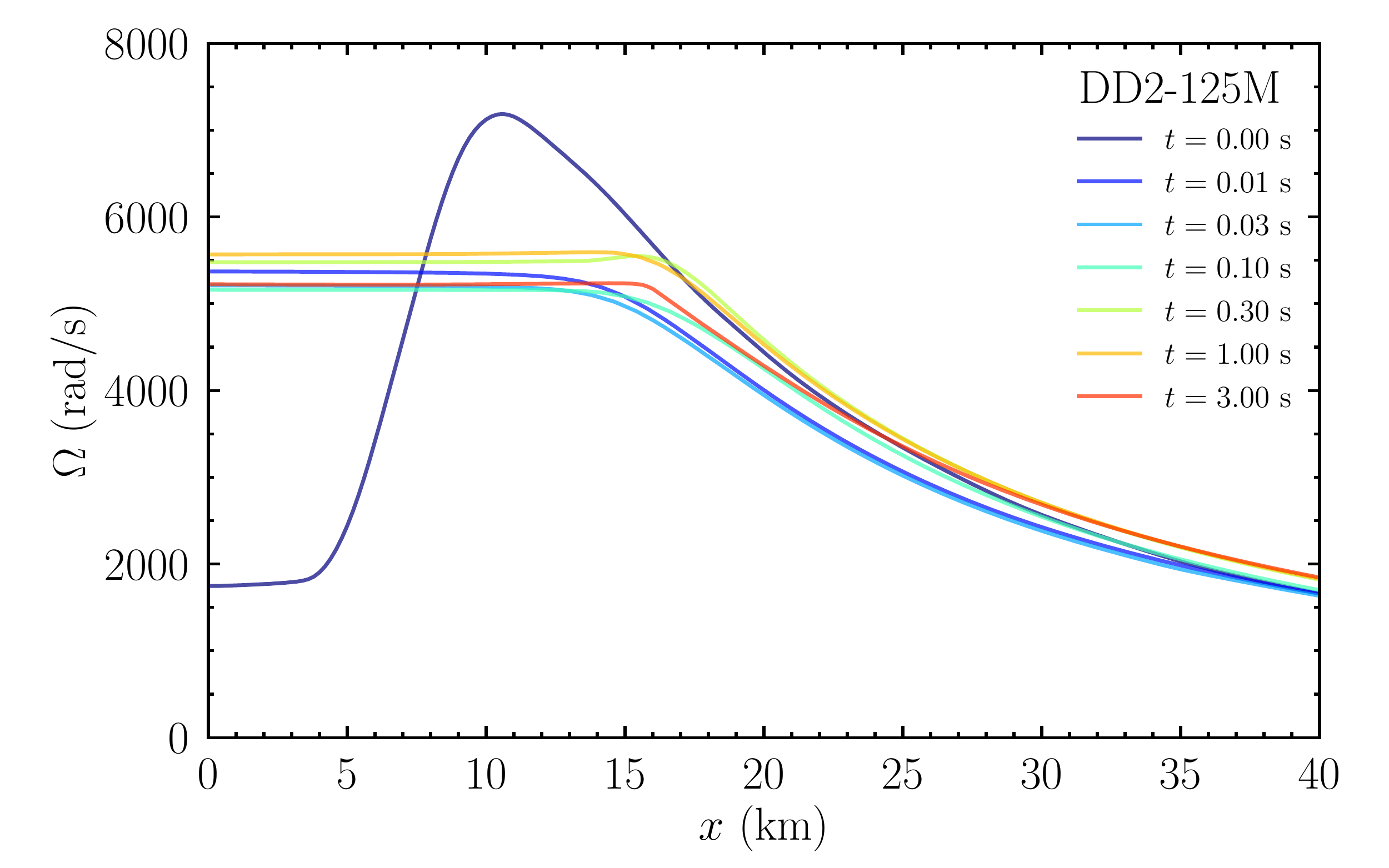}
\plotone{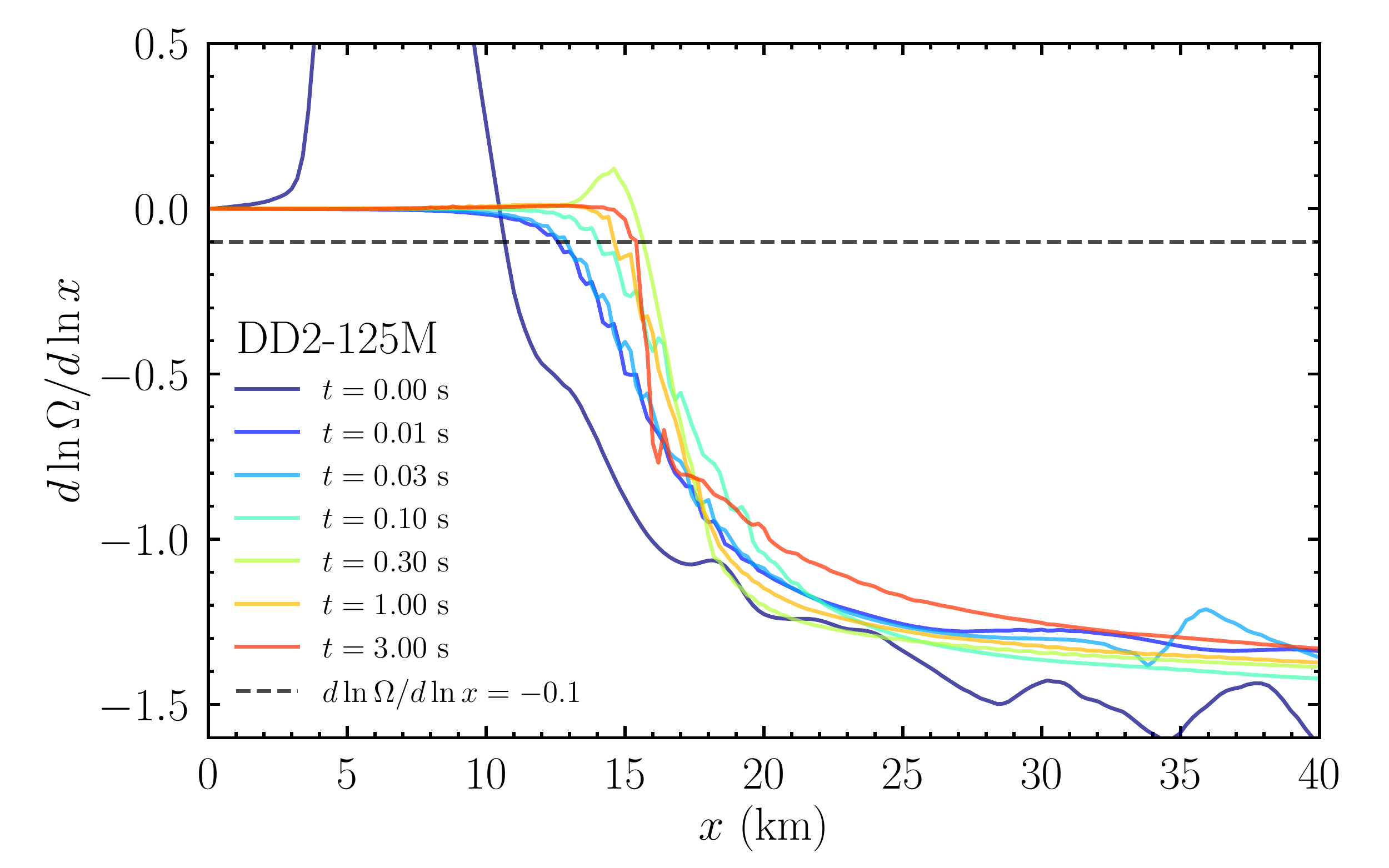}
\plotone{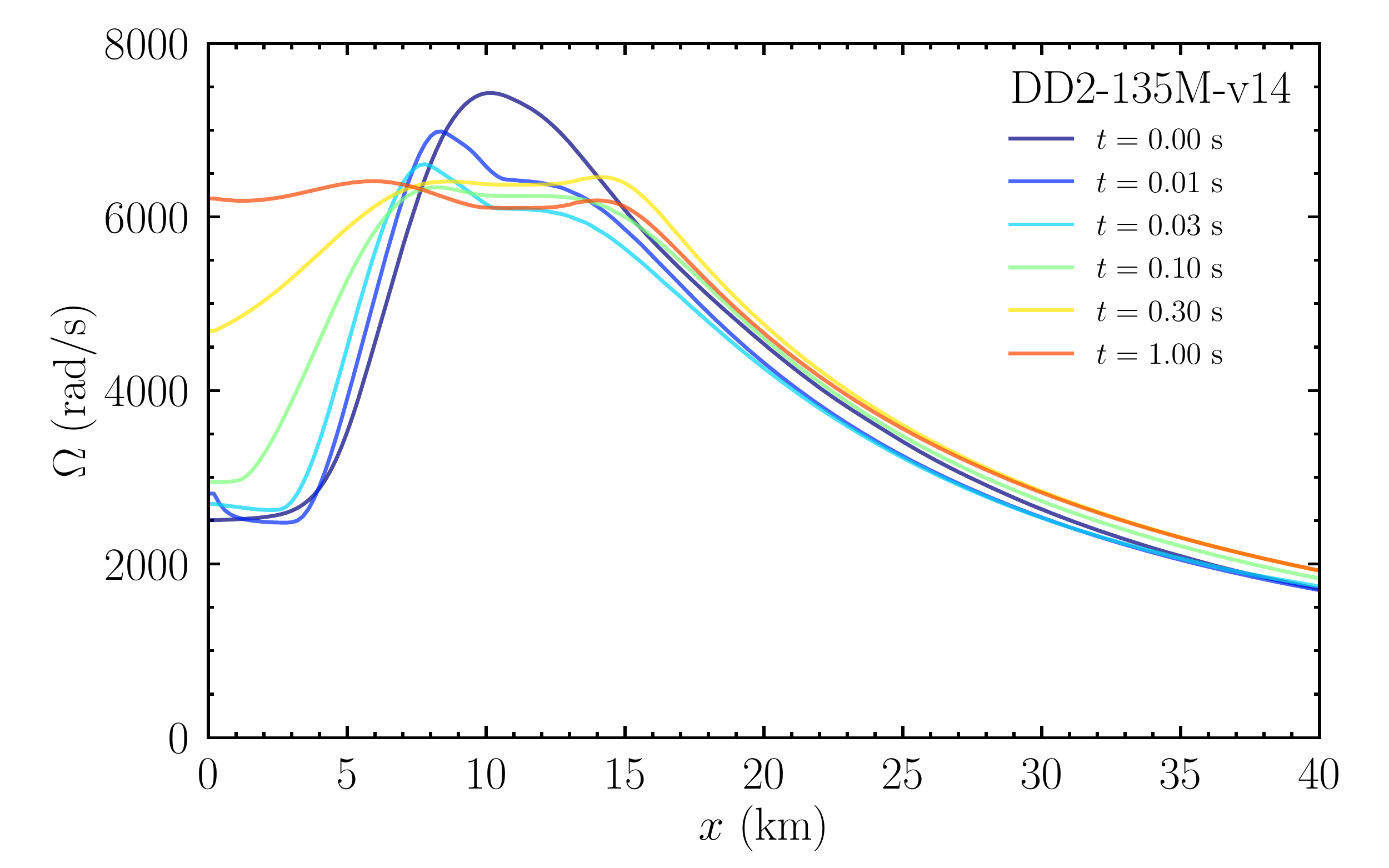}
\caption{
Angular velocity (top) and its logarithmic derivative with respect to the cylindrical radius (middle) along the equatorial plane for model DD2-125M.
Bottom panel shows the angular velocity evolution for model DD2-135M-v14.
}
\label{fig:dodx}
\end{figure}
In this section, we present the time evolution of quantities related to the massive NS and the disk surrounding it for each model.
First, we describe how we distinguish the disk from the outer part of the massive NS.

It is not trivial to define the boundary that divides these two components, because the material of the massive NS is continuously connected to the disk (i.e., the density varies continuously).
We use the angular velocity profile to define the NS and disk components, respectively.
The top panel of Fig.~\ref{fig:dodx} shows the angular velocity profile along the equatorial plane at selected time slices for DD2-125M.
At $t=0$, the system including the high-density region of the remnant ($x\lesssim 15$\,km) is entirely in a differentially rotating state.
The angular velocity is small in the central region, peaks at $\approx$10~km, and then decreases with the radius.
This is the typical rotational profile of a merger remnant \citep[e.g.,][]{Shibata2005a}.
Within a few milliseconds after the simulation begins, the rotational profile of this region changes to be rigid due to a viscous angular momentum transport process, the timescale of which is estimated by Eq.~\eqref{eq:vis}.
This is clearly illustrated in the middle panel of Fig.~\ref{fig:dodx} that shows the logarithmic derivative of angular velocity with respect to the cylindrical radius, i.e., $d\ln\Omega/d\ln x$.
For $t>0.01$\,s, the innermost region ($x\lesssim 15$\,km) settles into a rigidly rotating state, i.e., $d\ln\Omega/d\ln x\approx 0$.
On the other hand, in the outer region with $x\gtrsim 15$ km, its value gradually approaches the Keplerian value $-1.5$, which is the expected feature of the disk.
We define the NS radius, $R_\mathrm{NS}$, as the innermost cylindrical radius that satisfies $d\ln\Omega/d\ln x = -0.1$ on the equatorial plane.
Although the value is defined along the equatorial direction, we employ this to all the polar angle directions to define the NS region.
By this definition, we regard the NS as an object with a nearly rigid-rotation profile.

We note that defining the NS surface in terms of density is nontrivial particularly in an early stage of the evolution ($t\lesssim 0.3$\,s), in which the density gradient at the NS surface is not sufficiently large.
However, for $t\gtrsim 0.3$\,s, the density can also be used to separate the two regions because the density gap between those regions becomes very large.

The bottom panel of Fig.~\ref{fig:dodx} shows the angular velocity evolution for model DD2-135M-v14 (for this model $d\ln\Omega/dx$ is positive or close to zero inside the NS).
In this model, $d \ln \Omega/d \ln x$ steeply decreases also for $x \agt 15$\,km and thus we can define the NS surface in the method mentioned above. On the other hand, the viscous angular momentum transport does not occur inside a radius of $\approx 10$\,km, in which the rest-mass density is higher than $10^{14}\,\mathrm{g/cm^3}$.
However, as illustrated in Appendix~\ref{app:mod}, the properties of the early viscosity-driven ejecta for this model are not significantly different from those of DD2-135M.
This shows that the early ejecta component is driven by releasing the rotational kinetic energy near the NS surface (not the main body of the NS).
We find that the slow increase of the angular velocity for $t\gtrsim 0.1$\,s in this model is due to the convective motion inside the stellar body (see Appendix~\ref{app:mod}).

\subsubsection{Diagnostics}
\label{subsubsec:rem-diag}
With the radius of the NS surface determined above, we define the baryon rest mass, angular momentum, and neutrino luminosity of the NS as
\begin{align}
M_\mathrm{NS} &= \int_{r\le R_\mathrm{NS}} d^3x\,\rho_*,\\
J_\mathrm{NS} &= \int_{r\le R_\mathrm{NS}} d^3x\,\rho_*h u_\phi,\\ 
L_{\nu,\mathrm{NS}} &= \int_{r\le R_\mathrm{NS}} d^3x\,\alpha W^{-3} Q_{\mathrm{(leak)}t},
\end{align}
where $\rho_* = W^{-3} \alpha u^t \rho$, and $Q_{\mathrm{(leak)}t}$ is the time component of the leakage rate (see \S \ref{subsec:rad}).
Here we multiply the lapse function $\alpha$ to take the gravitational redshift into account for the neutrino luminosity, and thus, this luminosity corresponds to the value observed at infinity.
The quantities for the disk ($M_\mathrm{disk}$, $J_\mathrm{disk}$, and $L_{\nu,\mathrm{disk}}$) are also defined by integrating the same quantities in the region of $r>R_\mathrm{NS}$.
We also define the average electron fraction, average entropy, and total viscous heating rate of the disk by
\begin{align}
Y_\mathrm{e,disk} &= \frac{1}{M_\mathrm{disk}}\int_{r > R_\mathrm{NS}} d^3x\,\rho_*Y_\mathrm{e},\\
s_\mathrm{disk} &= \frac{1}{M_\mathrm{disk}}\int_{r > R_\mathrm{NS}} d^3x\,\rho_*s,\\
L_\mathrm{vis,disk} &= \int_{r > R_\mathrm{NS}} d^3x\, \alpha W^{-3} \frac{1}{2} \rho h\nu \tau^{0\alpha\beta}\tau^0{}_{\alpha\beta} u_t,
\end{align}
where $(1/2) \rho h\nu \tau^{0\alpha\beta}\tau^0{}_{\alpha\beta}$ is the viscous heating rate in the fluid rest frame \citep[see \S 2.6 in][]{fujibayashi2018a}.

\subsubsection{Properties of the NS}
\label{subsubsec:ns}

\begin{figure}
\epsscale{1.17}
\plotone{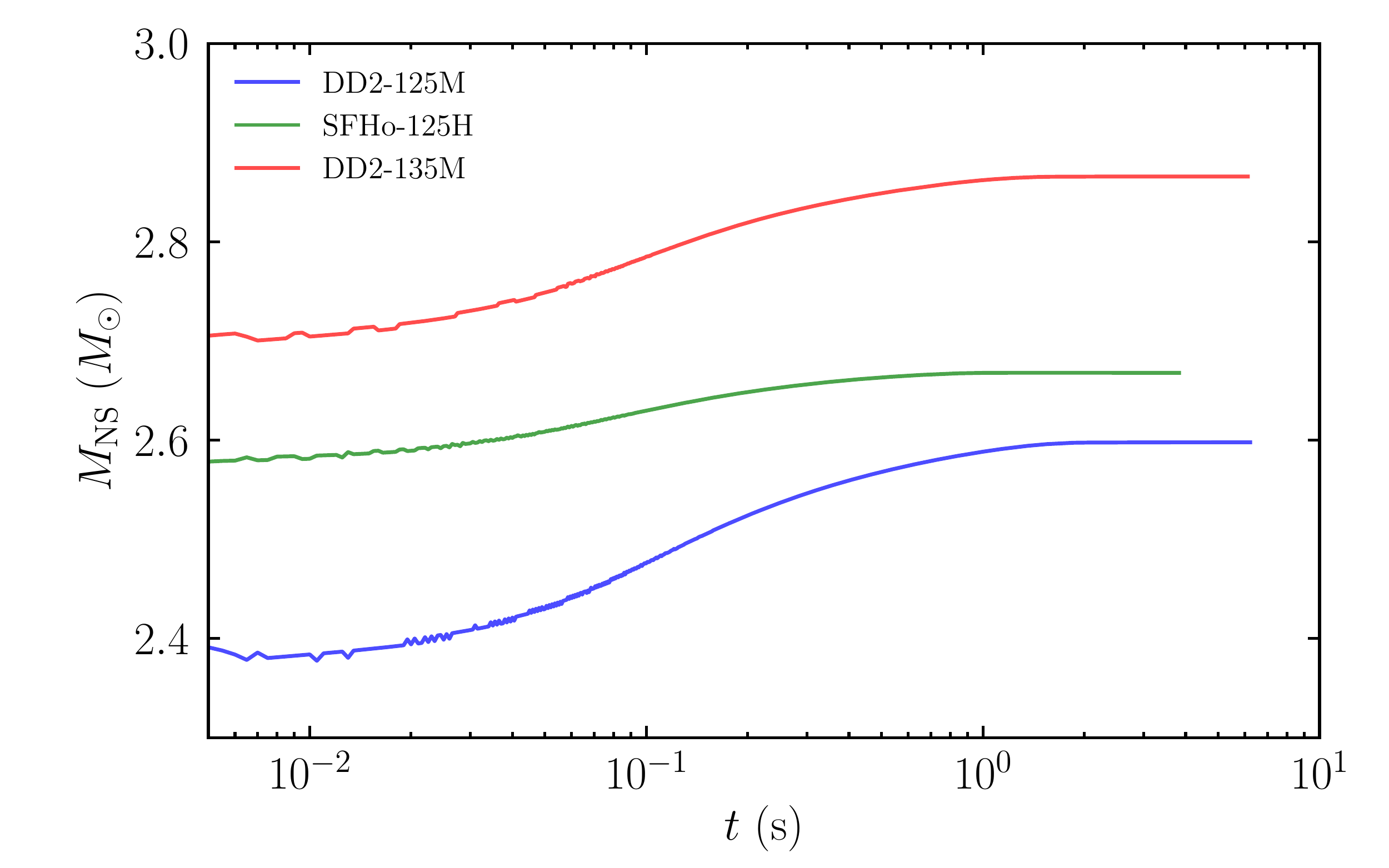}
\plotone{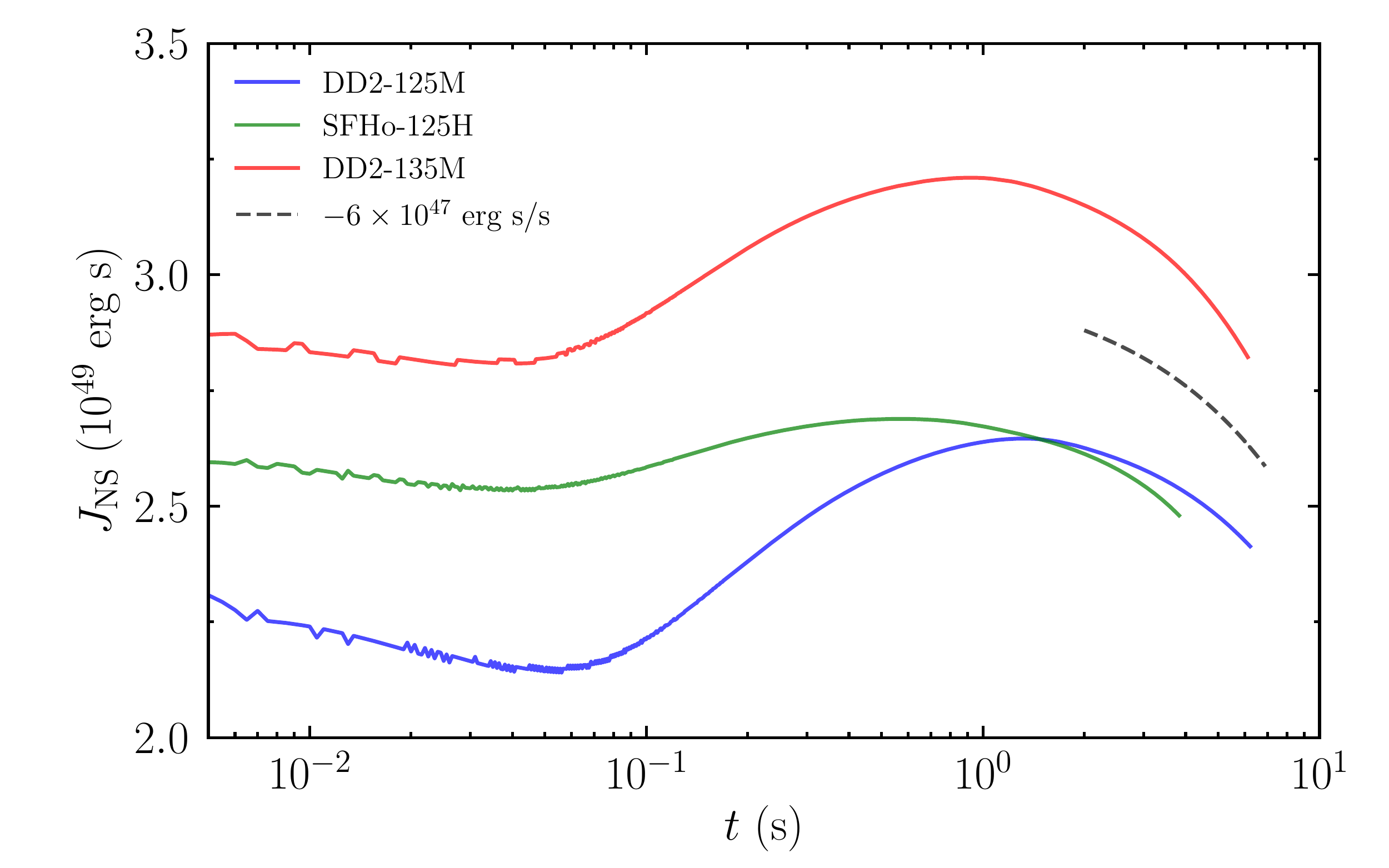}
\plotone{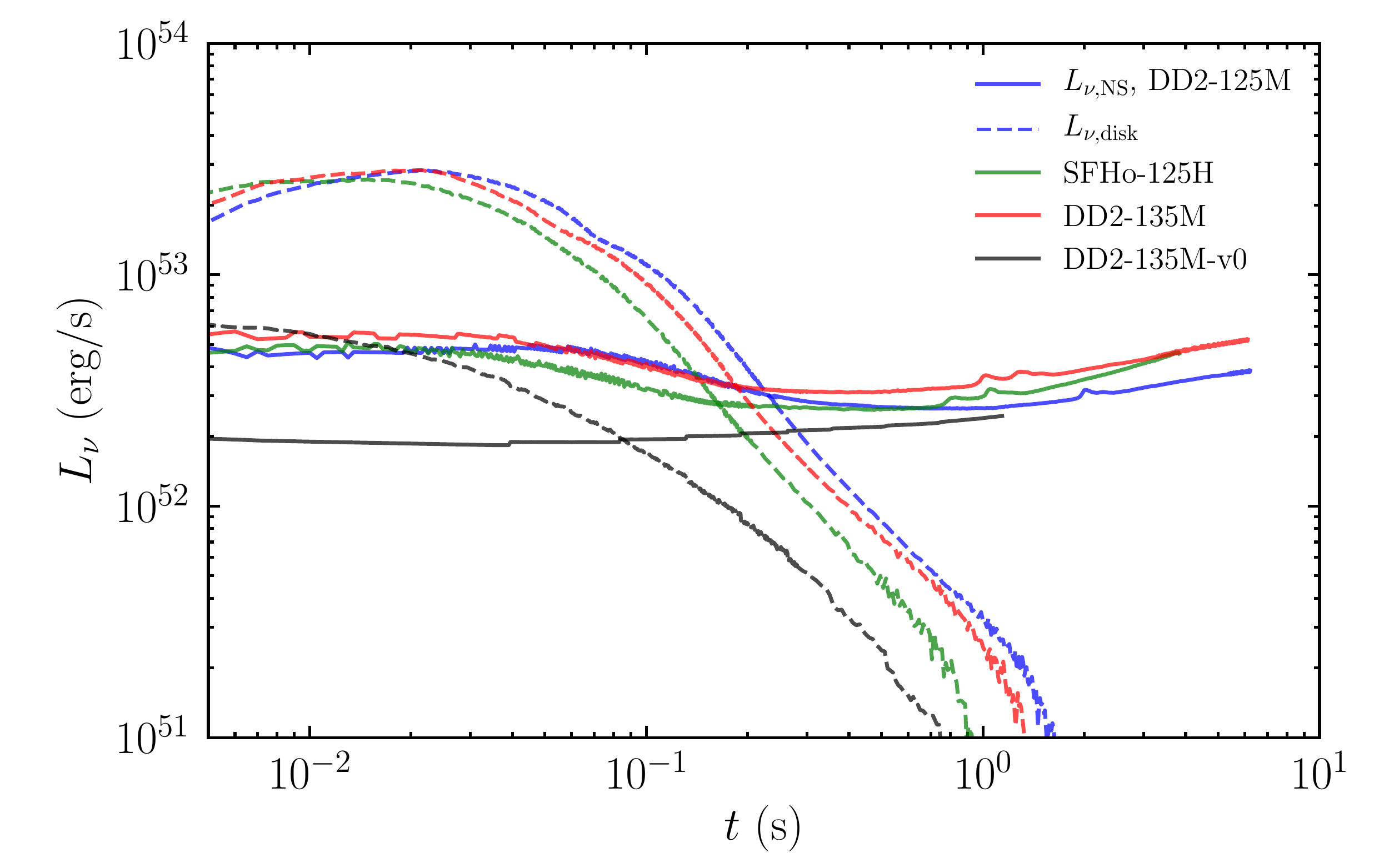}
\caption{
Baryon mass (top), angular momentum (middle), and total neutrino luminosity (bottom) of the NSs for models DD2-125M (blue), SFHo-125H (green), and DD2-135M (red).
The dashed curve in the middle panel corresponds to the case of the constant angular momentum loss rate of $6\times 10^{47}$ erg s/s (see Appendix~\S~\ref{app:angloss}).
In the bottom panel, the neutrino luminosities of the disks are also plotted by the dashed curves, and those for model DD2-135M-v0 are also shown.
}
\label{fig:MNS}
\end{figure}

The top and middle panels of Fig.~\ref{fig:MNS} show the baryon mass and angular momentum of the remnant massive NSs ($M_\mathrm{NS}$, $J_\mathrm{NS}$) for models DD2-125M, SFHo-125H, and DD2-135M.
$M_\mathrm{NS}$ and $J_\mathrm{NS}$ increase with time in an early stage of its evolution, $t\lesssim 1$\,s, in response to mass accretion.
This mass accretion is caused primarily by the cooling of the disk material:
The material located near the NS surface does not have the Keplerian angular momentum and is sustained dominantly by the thermal pressure (cf. Fig~\ref{fig:dodx}).
By the neutrino cooling, the thermal pressure decreases and then the material falls onto the NS.
The increase of the baryon mass is smaller for the SFHo case than for the DD2 cases simply due to the smaller disk mass for the SFHo case (see Table~\ref{tab:model}).
For $t \gtrsim$ 0.5\,s, the total angular momentum of the NS begins to decrease, while the increase of the NS mass saturates.
This is a consequence of the fact that neutrinos emitted from the rotating NS carry its angular momentum away (see Appendix~\ref{app:angloss}).
The angular momentum loss rate is $\sim 10^{48}\,\mathrm{erg\,s/s}$, which is consistent with an analytic estimation presented in Appendix~\ref{app:angloss}.

The bottom panel of Fig.~\ref{fig:MNS} shows the neutrino luminosity of the massive NS and the disk for each model.
The luminosity of the NS is $\sim$ (5--7)$\times 10^{52}$ erg/s for $t\lesssim 0.1$\,s and gradually decreases for $t>0.1$\,s.
The slight increase of the luminosity for $t\gtrsim 0.5$\,s is partly due to the finite grid resolution (see Appendix~\ref{app:mod}).
The neutrino luminosity of the disk is (1--3)$\times 10^{53}$ erg/s for $t\lesssim 0.1$\,s for each model, which is several times higher than that of the NS.
The high luminosity is sustained for $t\lesssim 0.1$\,s by the viscous heating in the disk.
However, as the temperature of the disk decreases due to the neutrino cooling and the viscous expansion, the luminosity decreases for $t\gtrsim 0.1$\,s.
By comparing models DD2-135M and DD2-135M-v0, we find that the viscous effect enhances the neutrino luminosity of the NS and disk by a factor of 3--5 for $t\lesssim 0.1$\,s (for our choice of the viscous parameter, $\alpha_\mathrm{vis}=0.04$).
This enhancement of the neutrino luminosity results in the larger amount of the neutrino-driven ejecta (see \S~\ref{subsubsec:decomp}).

\begin{figure*}
\epsscale{1.17}
\plottwo{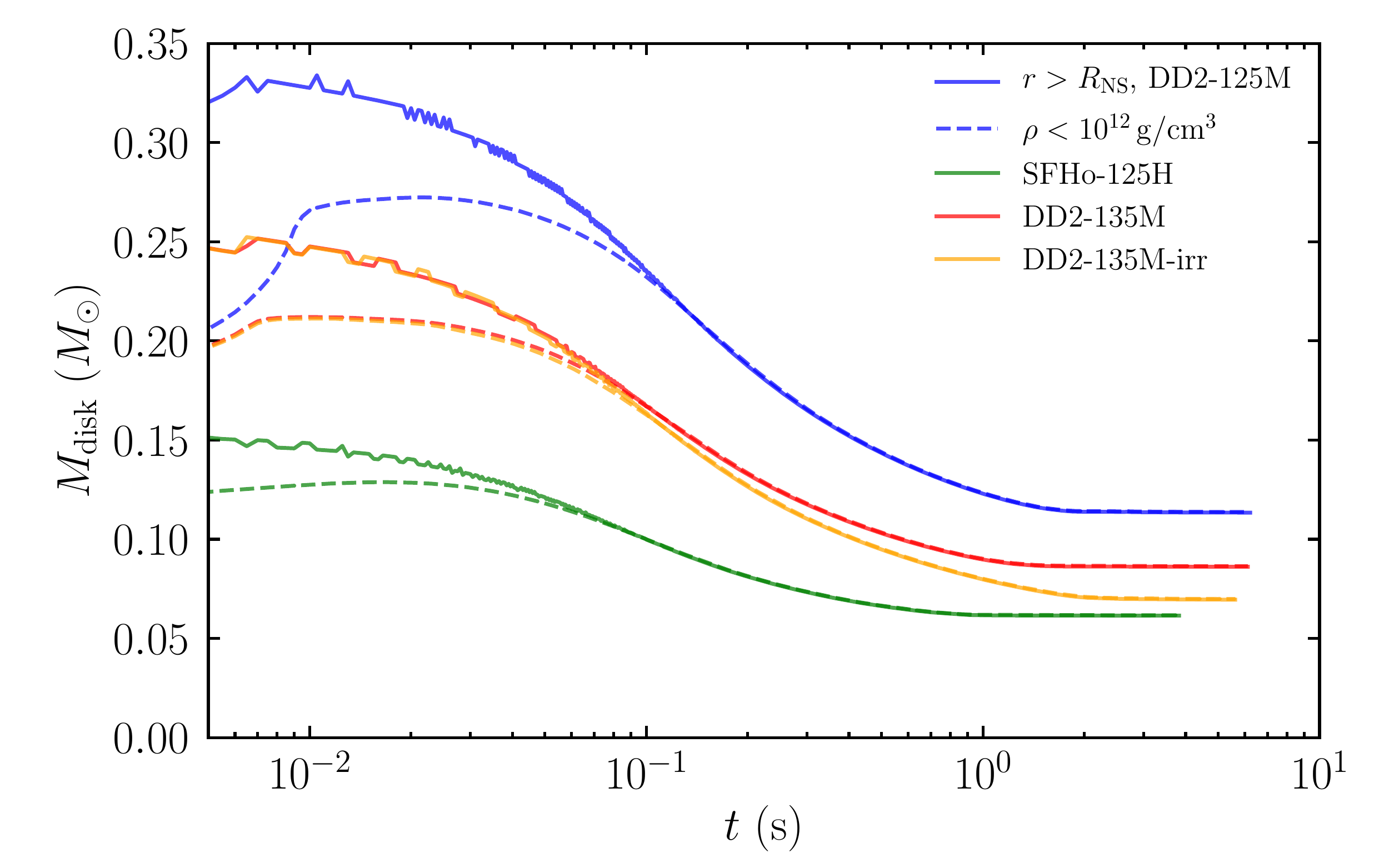}{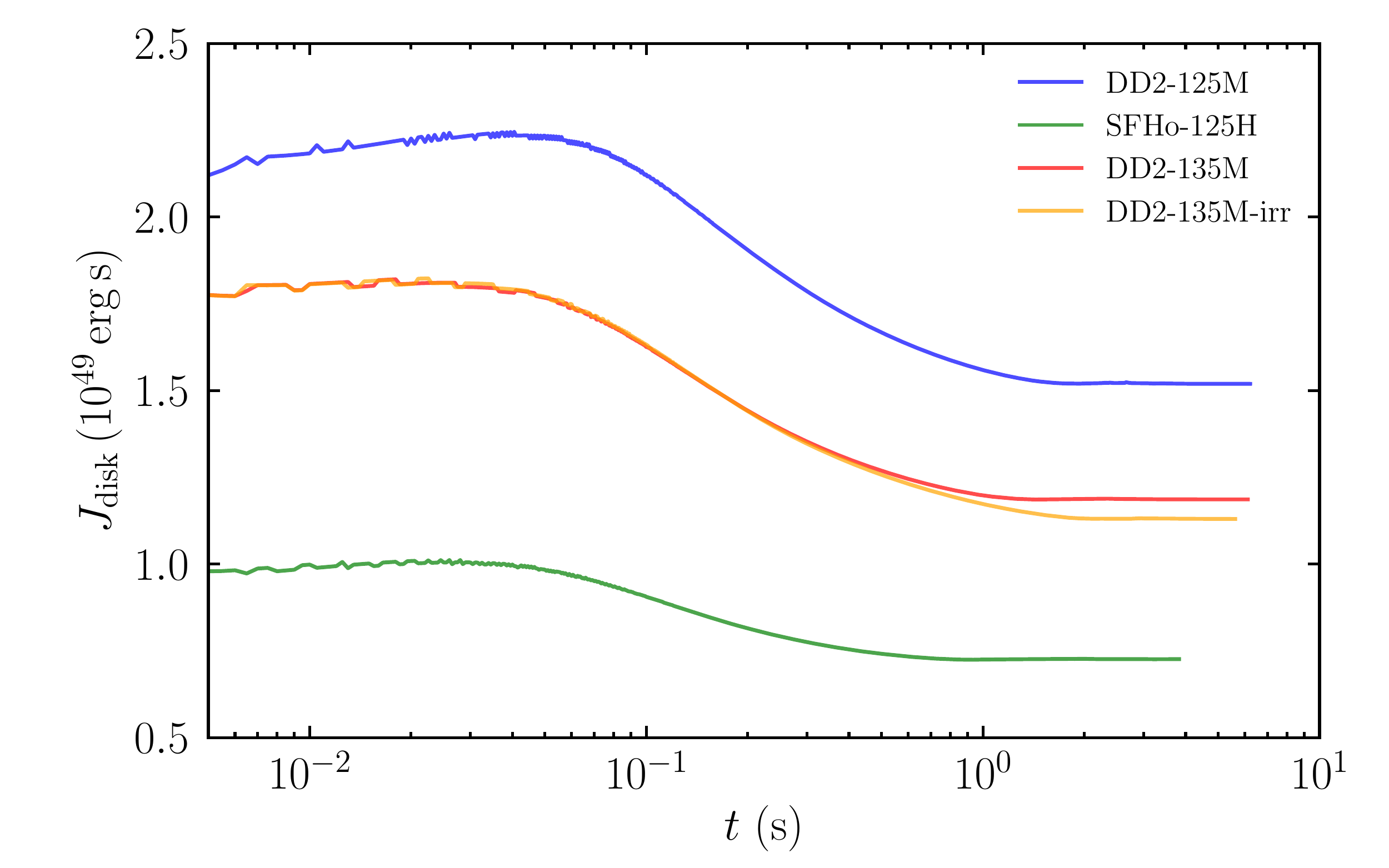}
\plottwo{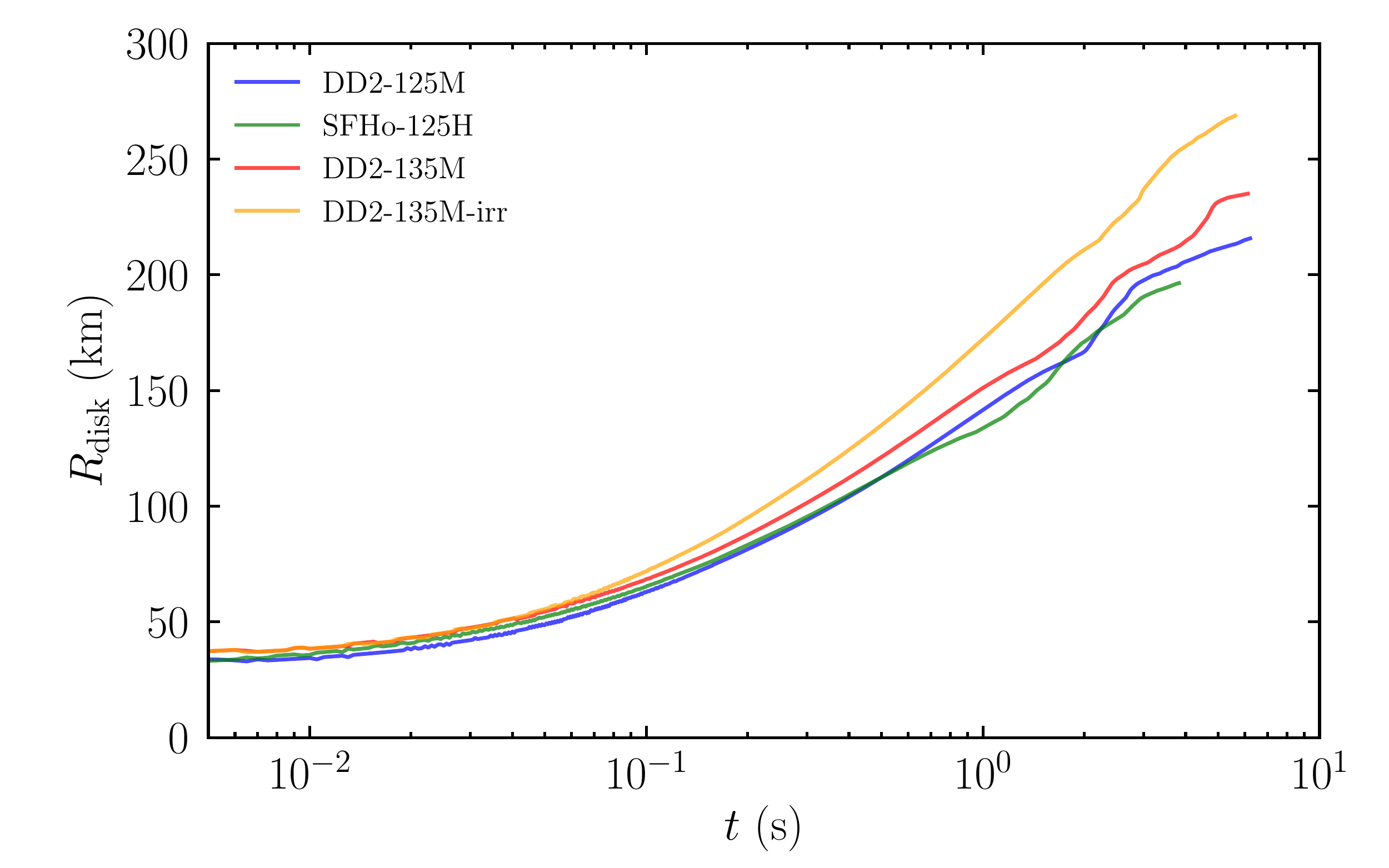}{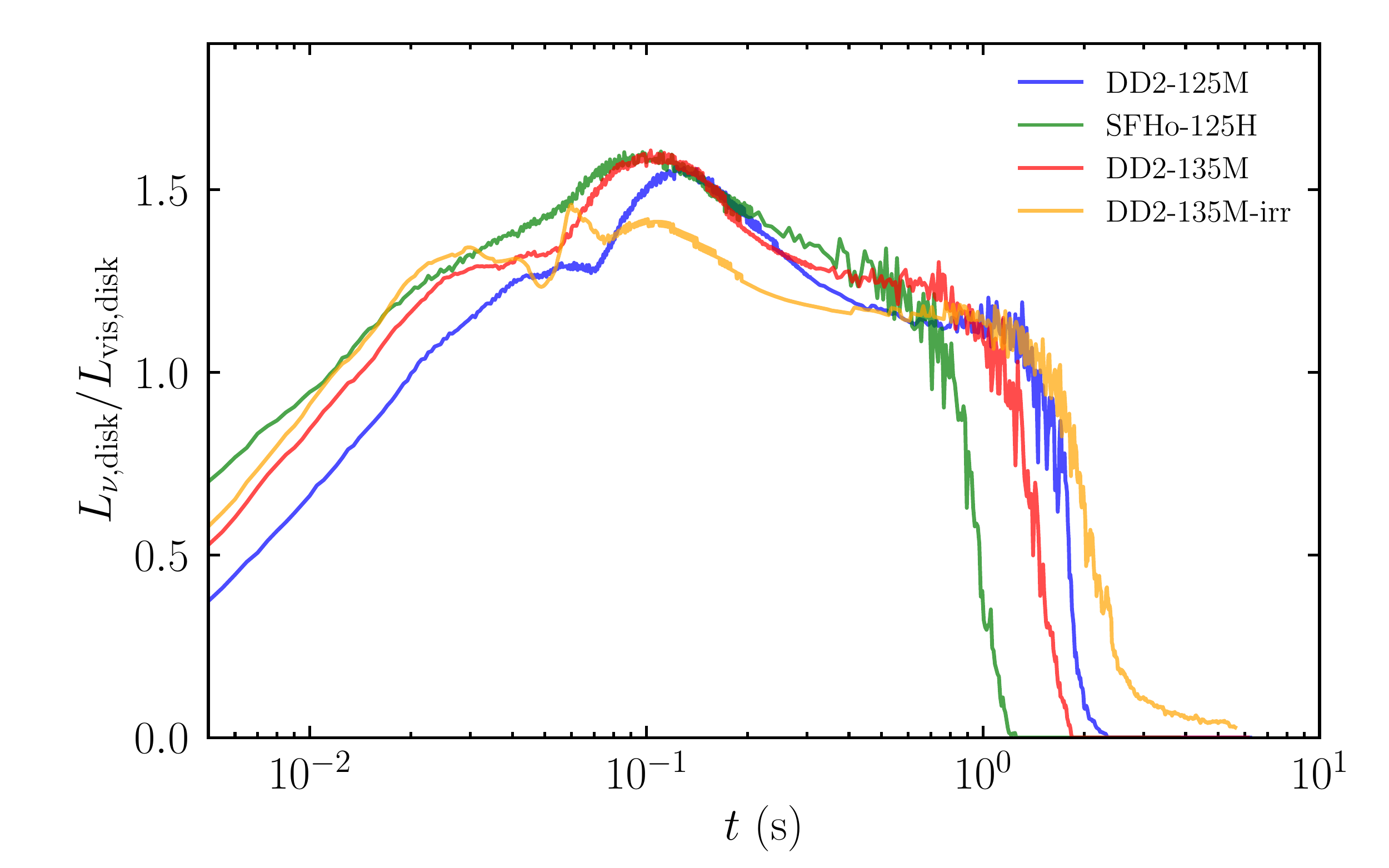}
\plottwo{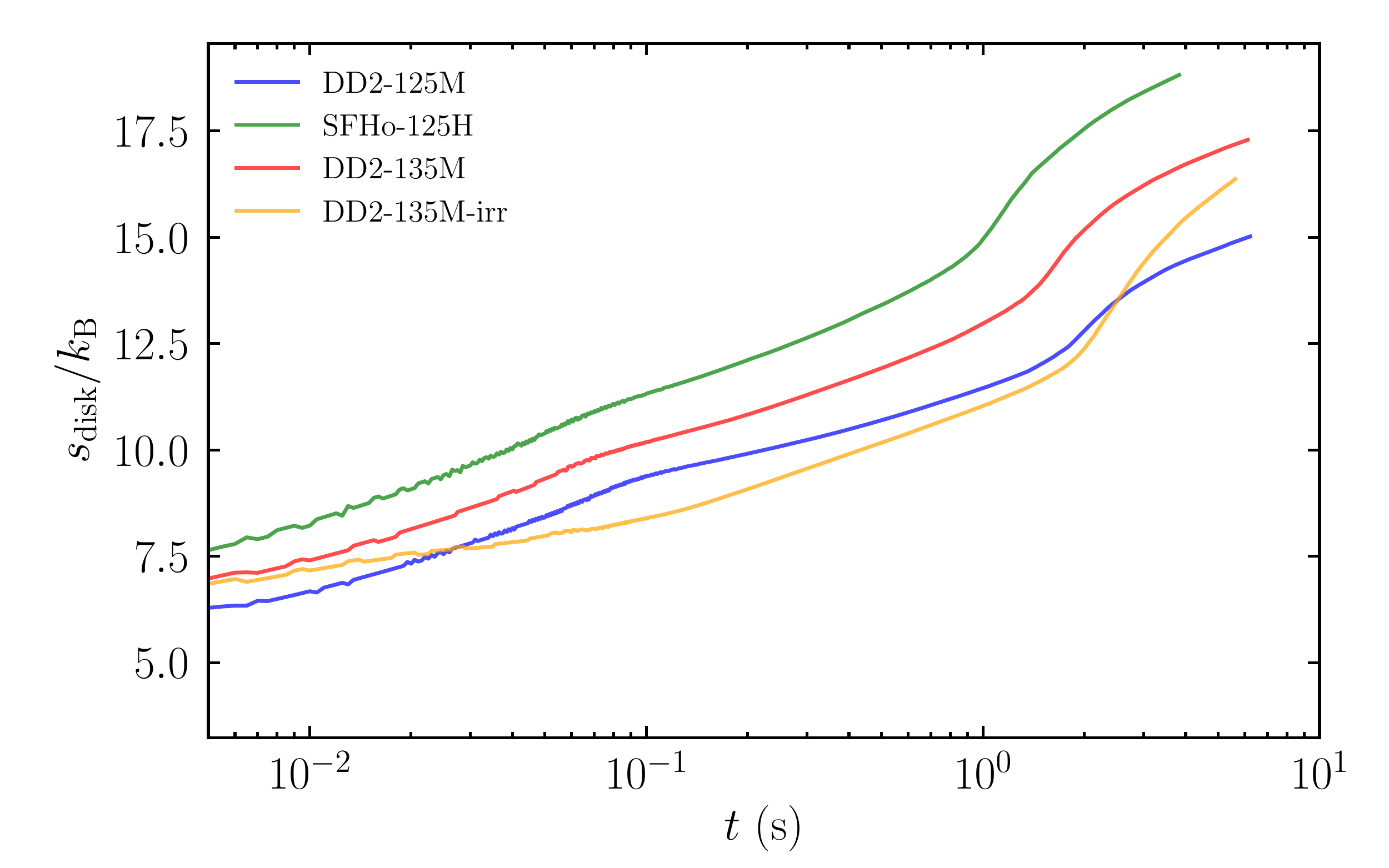}{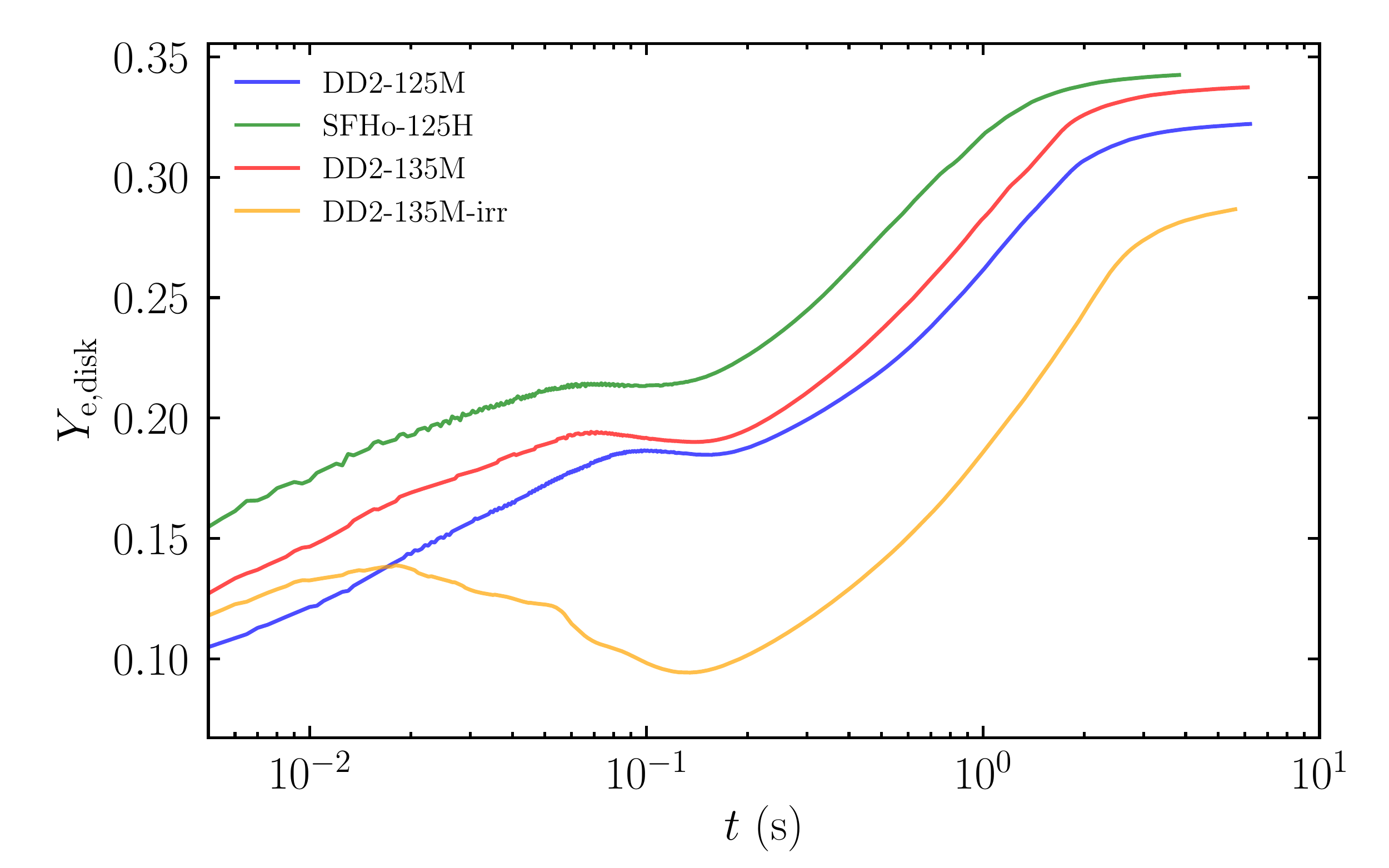}
\caption{
Mass (top-left), angular momentum (top-right), average radius (middle-left), the ratio of neutrino luminosity to viscous heating rate (middle-right), average entropy (bottom-left) and electron fraction (bottom-right) of the disks for models DD2-125M, SFHo-125H, DD2-135M, and DD2-135M-irr.
In the top-left panel, the mass in the region with $\rho < 10^{12}\,\mathrm{g/cm^3}$ is also plotted by the dashed curve for each model.
In the top two panels, the saturated values show the total ejecta mass ($M_\mathrm{ej,tot}$) and angular momentum.
We note the disk material here implies the material located outside the massive NS and contains the unbound component.
}
\label{fig:disk}
\end{figure*}

\subsubsection{Properties of the Disk}
The quantities relevant to the disks for models DD2-125M, SFHo-125H, and DD2-135M are shown in Figure~\ref{fig:disk}.
Note that this ``disk" component includes unbound material (see the definition in \S~\ref{subsubsec:rem-diag}).
The baryon mass and angular momentum of the disk ($M_\mathrm{disk}$ and $J_\mathrm{disk}$) rapidly drop during an early time of $t \lesssim 1$\,s due to the mass accretion onto the NS caused by the neutrino cooling (cf. \S~\ref{subsubsec:ns}).
The timescale of this accretion (i.e., the neutrino cooling timescale) is estimated by
\begin{eqnarray}
\frac{M_\mathrm{g,NS}M_\mathrm{disk}}{R_\mathrm{NS}L_{\nu,\mathrm{disk}}} &\approx& 0.64\,\mathrm{s}\nonumber \\
&\times& \biggl(\frac{M_\mathrm{g,NS}}{2.5M_\odot}\biggr) \biggl(\frac{M_\mathrm{disk}}{0.1M_\odot}\biggr)\nonumber\\
&&\biggl(\frac{R_\mathrm{NS}}{10\,\mathrm{km}}\biggr)^{-1}
\biggl(\frac{L_{\nu,\mathrm{disk}}}{10^{53}\,\mathrm{erg/s}}\biggr)^{-1}, \label{eq:acctime}
\end{eqnarray}
where $M_\mathrm{g,NS}$ is the gravitational mass of the NS.
The numerical result is consistent with this timescale.

The baryon mass and angular momentum of the disk for the SFHo case are smaller than those for the DD2 cases because of the smaller tidal effect on the merging NSs for the former.
The disk mass is slightly larger for model DD2-125M than that for DD2-135M, because the former has a smaller compactness and thus the self-gravity of the central massive NS is weaker.
We note that the disk component has a non-negligible fraction of the angular momentum of the remnant.
The fraction of the disk angular momentum is 25--30\% of the total angular momentum of the remnant depending weakly on the EOS and the total mass of the system.
Thus only less than 75\% of the total angular momentum of binary NSs at merger goes into the remnant massive NSs.

The disk mass is often defined by a certain density criterion \citep[e.g.,][]{Shibata2017a,Radice2018a}.
In the top-left panel of Fig.~\ref{fig:disk}, we also show the mass of the material with $\rho < 10^{12}\,\mathrm{g/cm^3}$, which agrees approximately with the mass using our criterion, $r>R_\mathrm{NS}$, at late times $t\gtrsim 0.1$\,s.

We define the average radius of the disk $R_\mathrm{disk}$ by
\begin{eqnarray}
R_\mathrm{disk} = \frac{1}{M_\mathrm{g,NS}}\frac{J_\mathrm{disk}^2}{M_\mathrm{disk}^2}, \label{eq:typrad}
\end{eqnarray}
where we assumed that $J_\mathrm{disk}$ can be approximated by $M_\mathrm{disk}\sqrt{M_\mathrm{g,NS}R_\mathrm{disk}}$, and for simplicity, we use the Arnowitt-Deser-Misner mass \citep[ADM mass;][]{arnowitt1960a} obtained initially in the simulation as the gravitational mass of the NS.
We note that $R_\mathrm{disk}$ depends only on the average specific angular momentum, $J_\mathrm{disk}/M_\mathrm{disk}$, and does not change by heating or cooling of the disk.
Moreover, this is a good indicator of the disk radius only for the case that the disk has a Kepler motion and does not appropriately indicate the disk radius when the material in the disk starts to be ejected from the system.

The middle-left panel of Fig.~\ref{fig:disk} shows that this radius gradually increases, indicating that the average specific angular momentum of the disk increases.
For the earlier phase $t\lesssim 1$\,s, this occurs by the accretion of the disk material with small specific angular momentum onto the NS due to the neutrino cooling.
After the disk mass saturates for $t\gtrsim 1$\,s, it is due to the viscous angular momentum transport from the outer envelope of  NS to the disk.

In the middle-right panel of Fig.~\ref{fig:disk}, we plot the ratios of the total (net) neutrino luminosity of the disk, $L_{\nu,\mathrm{disk}}$, which includes the heating by neutrino interaction as well, to the viscous heating rate in the disk, $L_\mathrm{vis,disk}$, for each model. 
The ratio increases with time for $t\lesssim 0.1$\,s because the optical depth of the disk, which is $\sim 10$--$100$ at the beginning of the simulation, decreases due to the accretion and expansion of the disk.
It peaks at $t\sim 0.1$\,s, when the optical depth of the disk decreases to $\sim 1$ and neutrinos can carry away the thermal energy generated by the viscous heating and stored in the opaque disk.
Then the ratio decreases to $\sim 1$ for $0.1$\,s $\lesssim t \lesssim 1$\,s, when the disk is optically thin and the viscously-generated thermal energy can be emitted immediately.
For $t\lesssim 1$\,s, the ratio is close to or larger than unity, indicating that the viscous heating does not contribute substantially to the disk expansion.
However, after the accretion timescale of the disk ($\sim 1$\,s, see Eq.~\eqref{eq:acctime}), the neutrino emission and the disk accretion become weaker.
Then for $t\gtrsim 1$\,s, this ratio drops steeply.
This implies that the neutrino cooling is no longer efficient and most of the viscous heating is used for the disk expansion.
We note that $t\sim 1$\,s also corresponds to the time at which the weak interaction timescale becomes longer than the timescale of the viscous expansion of the disk (i.e., the weak interaction freezes out; see \S \ref{subsec:ejecta}).

In Fig.~\ref{fig:disk} we also plot the evolution of the average entropy (bottom-left) and electron fraction (bottom-right) of the disk, which increase gradually with time.
The increase of the entropy is due to the viscous and neutrino heating.
The electron fraction of the disk is determined by an equilibrium condition of electron/positron capture, which is enhanced during the disk expansion because the electron degeneracy becomes weak (the neutrino irradiation plays a minor role for the increase of the electron fraction; see \S~\ref{subsec:ejecta}).
Its increase stagnates at $t\sim 0.1$\,s because of the feedback effect of the efficient neutrino emission \citep{Kawanaka2007a,Chen2007a}: As the middle-right panel of Fig.~\ref{fig:disk} shows, the efficiency of the neutrino emission at $t\sim 0.1$\,s is quite high.
In such a situation, the increase of the specific entropy by the viscous heating is suppressed (bottom-left panel of Fig.~\ref{fig:disk}).
As a result, the electron degeneracy is kept to be strong, suppressing the increase of $Y_\mathrm{e}$ (bottom-right panel of Fig.~\ref{fig:disk}).
Due to the neutrino irradiation, the value of the electron fraction at $t \sim 0.1$\,s in our models ($\sim 0.2$) is higher by $\sim 0.1$ than that found in the context of the disk around a BH \citep{Siegel2018a}.
In the absence of the neutrino irradiation, the electron fraction becomes lower (see \S~\ref{subsubsec:diskirr}).
After the neutrino cooling becomes inefficient, the entropy increases continuously, while the electron fraction saturates after the gradual increase until $t\sim 1$\,s, because the timescale of electron/positron capture in the disk becomes longer than the viscous evolution timescale due to the decrease of the temperature (see \S \ref{subsec:ejecta}).
The final average value of $Y_\mathrm{e}$ is 0.30--0.35.

\subsubsection{Effect of the Neutrino Irradiation on the Properties of the Disk}
\label{subsubsec:diskirr}
Figure~\ref{fig:disk} also shows the quantities for model DD2-135M-irr, in which the neutrino irradiation is switched off.
Here, we explore the effect of the neutrino irradiation on the quantities of the disk by comparing models DD2-135M and DD2-135M-irr.
For model DD2-135M-irr, the neutrino cooling of the inner part of the disk, which has smaller specific angular momentum than the outer part, is more efficient than that for model DD2-135M because of the absence of the neutrino irradiation.
As a result, the average specific angular momentum of the disk ($J_\mathrm{disk}/M_\mathrm{disk}$), and thus, the average disk radius ($R_\mathrm{disk}$) increase more quickly than those for model DD2-135M.

In addition, the average electron fraction of the disk is lower by 0.05--0.1 than that for model DD2-135M throughout the evolution because of the absence of the neutrino irradiation, which usually increases the electron fraction.
At $t\sim0.1$\,s, due to the feedback effect, the average electron fraction of the disk is maintained to be $\sim 0.1$, which is consistent with that found in the disk around a BH without the neutrino irradiation \citep{Kawanaka2007a,Chen2007a,Siegel2018a}.
Because the electron fraction of the disk becomes lower, the electron fraction of the late-time viscosity-driven ejecta becomes also lower (see \S~\ref{subsubsec:decomp}).

\subsection{Ejecta} \label{subsec:ejecta}
\subsubsection{Diagnostics}
In this subsection, we describe the properties of the ejecta.
First of all, we summarize the method to calculate the ejecta quantities.
We define the mass of the ejecta that escaped from a sphere, $M_\mathrm{ej,esc}$, by integrating the unbound material outflowing through the surface of a radius $r_\mathrm{esc}$,
\begin{eqnarray}
M_\mathrm{ej,esc} = \int^t dt \frac{dM_\mathrm{ej,esc}}{dt},
\end{eqnarray}
where the mass ejection rate is defined by
\begin{eqnarray}
\frac{dM_\mathrm{ej,esc}}{dt} = \int dS_k v^k \rho_* H (-hu_t-h_\mathrm{min}), \label{eq:dmdt}
\end{eqnarray}
with the area element on the sphere $dS_k$, and the Heaviside function $H$.
Hereafter $r_\mathrm{esc}$ is set to 8000\,km, at which the temperature of the material is sufficiently low and the recombination of free nucleons has already occurred.
Thus, Eq.~\eqref{eq:huth} can be a suitable condition for the material being unbound.

We also define the rate of the total energy (sum of the rest-mass, internal, and kinetic energy) of the ejecta passing through the radius $r_\mathrm{esc}$ as
\begin{eqnarray}
\frac{dE_\mathrm{ej,esc}}{dt} = \int dS_k v^k \rho_* \hat{e} H(-hu_t-h_\mathrm{min}),
\end{eqnarray}
where $\hat{e}=h\alpha u^t -P/(\rho\alpha u^t)$ and the total energy of the ejecta is calculated by
\begin{eqnarray}
E_\mathrm{ej,esc} = \int^t dt \frac{dE_\mathrm{ej,esc}}{dt}.
\end{eqnarray}
This quantity is decomposed in the region far from the source of the gravity into
\begin{eqnarray}
E_\mathrm{ej,esc} \approx M_\mathrm{ej,esc} + T_\mathrm{kin} + U + \frac{M_\mathrm{ej,esc}M_\mathrm{g,NS}}{r_\mathrm{esc}},\label{eq:dec-eej}
\end{eqnarray}
where $T_\mathrm{kin}$ and $U$ are the kinetic and internal energies of the ejecta at a sufficient distant region and the last term approximately denotes the contribution from the gravitational binding energy.
For the gravitational mass of the NS, we again use the ADM mass with the value obtained at the beginning of the simulation.
From these expressions, we define the average ejecta velocity by
\begin{eqnarray}
\langle V_\mathrm{ej}\rangle = \sqrt{2\biggl(\frac{E_\mathrm{ej,esc}} {M_\mathrm{ej,esc}} -1 -\frac{M_\mathrm{g,NS}}{r_\mathrm{esc}}\biggr)},
\end{eqnarray}
where we assumed that the internal energy of the ejecta is eventually transformed into its kinetic energy during its expansion, i.e., $T_\mathrm{kin}+U = M_\mathrm{ej,esc}\langle V_\mathrm{ej}\rangle^2/2$ at infinity.
Because there are several ejecta components having different velocities, we define the velocity of the material passing through the sphere at a given time by 
\begin{eqnarray}
V_\mathrm{ej} = \sqrt{2 \biggl( \biggl(\frac{dM_\mathrm{ej,esc}}{dt}\biggr)^{-1} \frac{dE_\mathrm{ej,esc}}{dt} -1 -\frac{M_\mathrm{g,NS}}{r_\mathrm{esc}} \biggr) },
\end{eqnarray}
where we used differential form of Eq.~\eqref{eq:dec-eej} and assumed
\begin{align}
\frac{d}{dt} (T_\mathrm{kin} + U) = \frac{1}{2} \frac{dM_\mathrm{ej,esc}}{dt}V_\mathrm{ej}{}^2.
\end{align}
With this definition we can derive the velocity of the material ejected at different times.

\begin{figure*}
\epsscale{1.17}
\plottwo{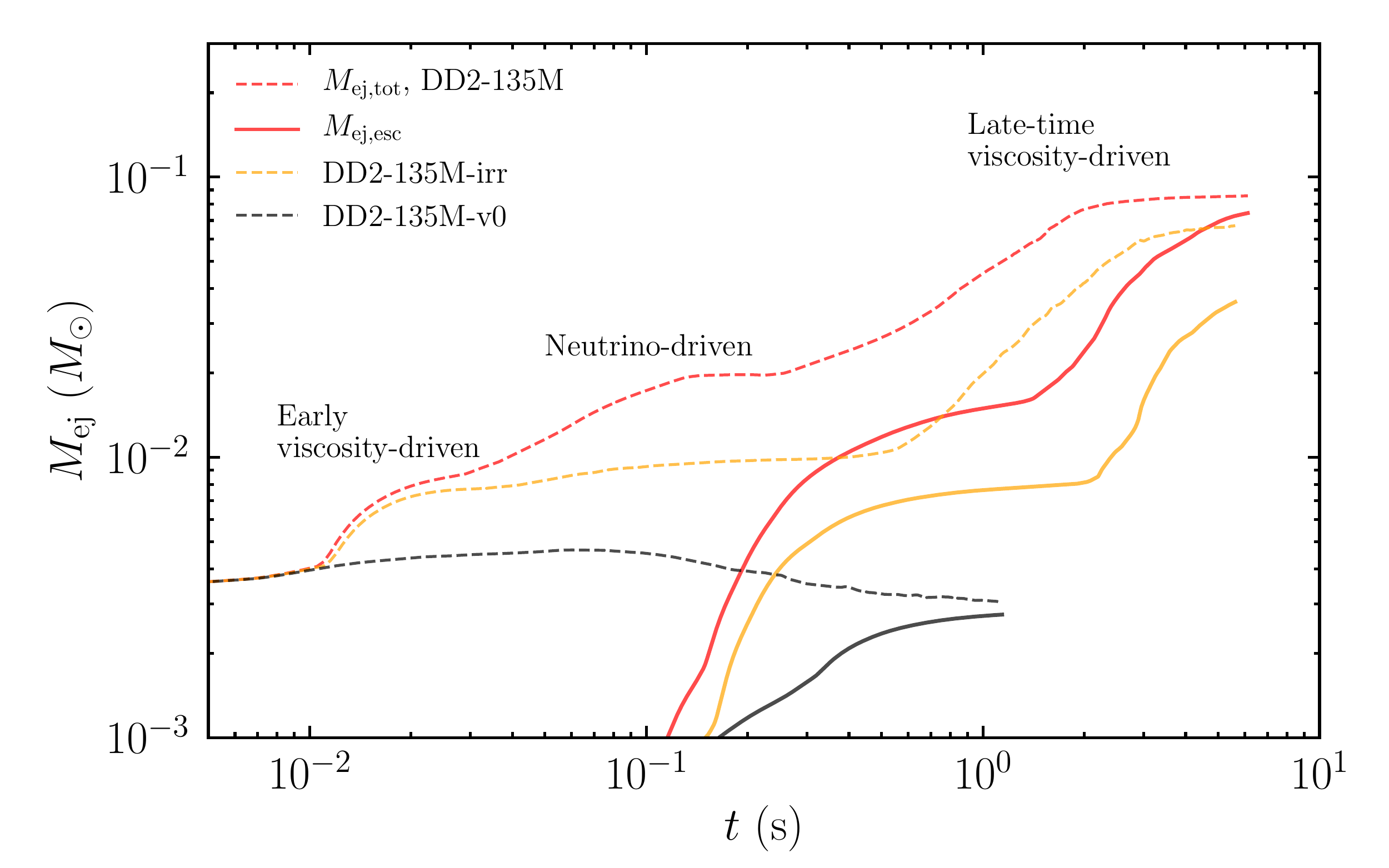}{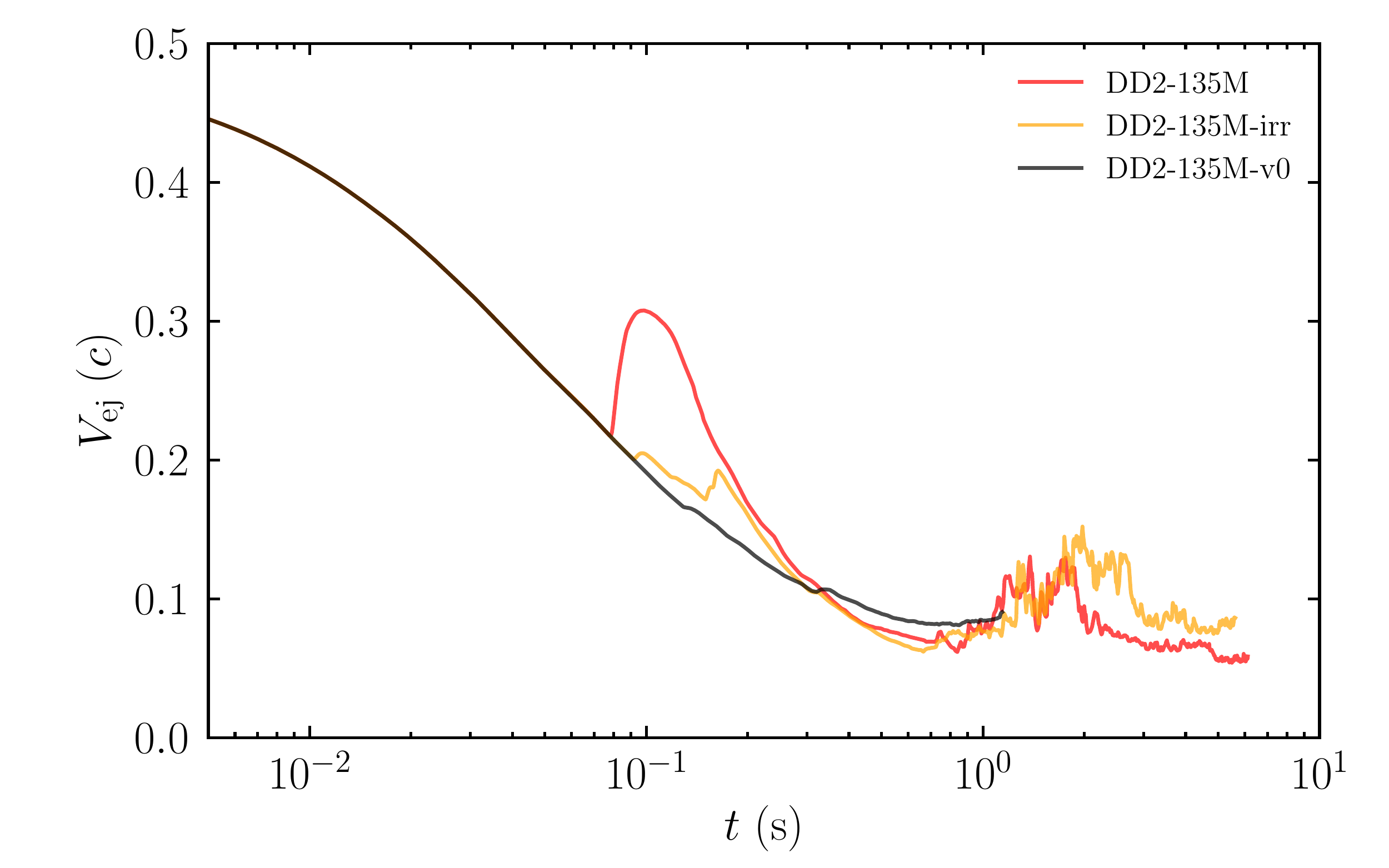}
\plottwo{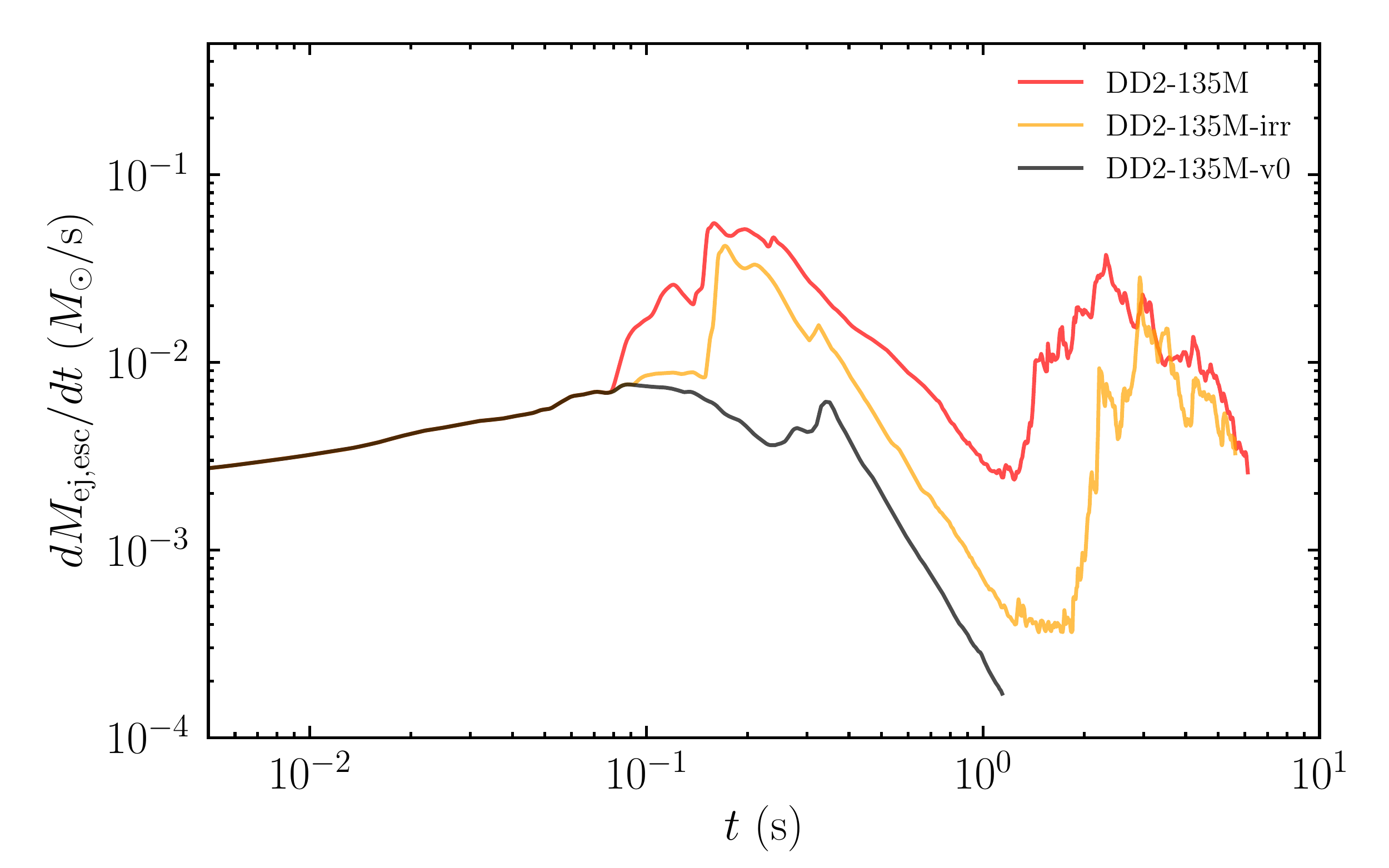}{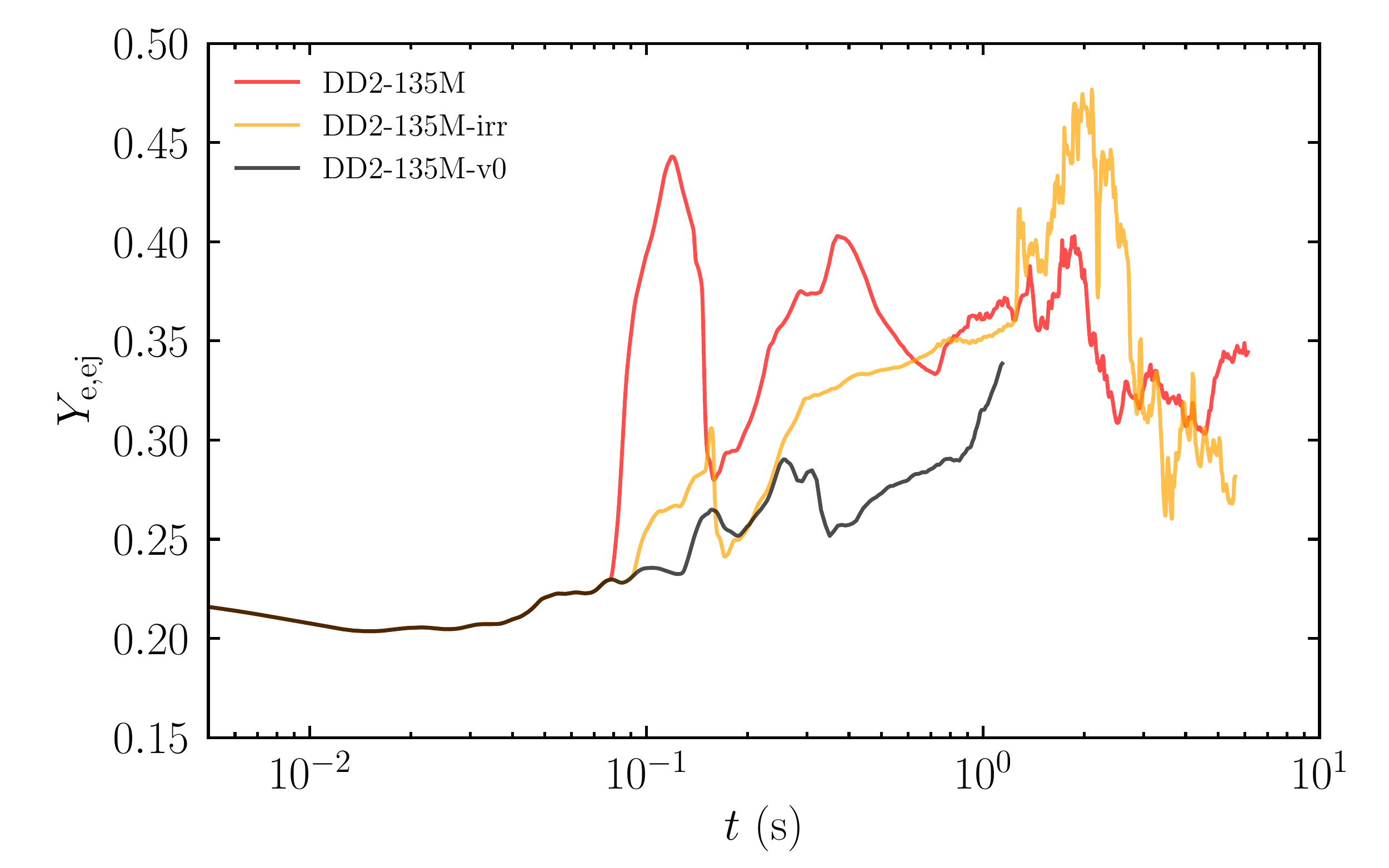}

\caption{Ejecta masses ($M_\mathrm{ej,tot}$ and $M_\mathrm{ej,esc}$; top-left), ejecta velocities ($V_\mathrm{ej}$; top-right), mass ejection rates ($dM_\mathrm{ej,esc}/dt$; bottom-left), and average electron fraction ($Y_\mathrm{e,ej}$; bottom-right) for models DD2-125M, DD2-135M-irr, and DD2-135M-v0.
In the top-left panel, the solid and dashed curves show $M_\mathrm{ej,tot}$ and $M_\mathrm{ej,esc}$, respectively.
}
\label{fig:ejecta-decom}
\end{figure*}

There is still the material satisfying Eq.~\eqref{eq:huth} inside the sphere with $r=r_\mathrm{esc}$ at the end of the simulations.
Thus, we define the total mass by
\begin{align}
M_\mathrm{ej,tot} &= M_\mathrm{ej,esc} \notag \\
&+ \int_{\mathrm{500\,km}\leq r<r_\mathrm{esc}} d^3x \rho_* H (-hu_t-h_\mathrm{min}). \label{eq:mejtot}
\end{align}
This value is interpreted as approximately the maximum mass that the merger remnant can eject.
We define the mass-weighted average electron fraction of this material by
\begin{align}
\langle Y_\mathrm{e,ej} \rangle = \frac{1}{M_\mathrm{ej,tot}} \biggl (\int^t dt \int dS_k v^k \rho_* Y_\mathrm{e}H (-hu_t-h_\mathrm{min})\notag \\
+ \int_{\mathrm{500\,km}\leq r<r_\mathrm{esc}} d^3x \rho_* Y_\mathrm{e} H (-hu_t-h_\mathrm{min}) \biggr),
\end{align}
which includes both of the ejecta that escaped from and stays inside the sphere of $r=r_\mathrm{esc}$.
In order to derive the electron fraction of the different ejecta components, we also define the flux-weighted average electron fraction of the material passing through the sphere at a given time by
\begin{align}
Y_\mathrm{e,ej} = \frac{1}{dM_\mathrm{ej,esc}/dt}\int dS_k v^k \rho_* Y_\mathrm{e} H(-hu_t-h_\mathrm{min}).
\end{align}

\subsubsection{Contribution of Different Ejecta Components}
\label{subsubsec:decomp}
In Fig.~\ref{fig:ejecta-decom}, the ejecta mass (top-left), mass ejection rate (bottom-left), ejecta velocity (top-right), and average electron fraction (bottom-left) for models DD2-135M, DD2-135M-irr, and DD2-135M-v0 are compared to explore the contribution of the effects of neutrino irradiation and viscosity.
Here, we focus on the models of the remnant of the merger with a large total gravitational mass of $2.7\,M_\odot$.
The following discussion is also held for models DD2-125M and SFHo-125M.
At the very beginning of the simulation, unbound material of $M_\mathrm{ej,tot} \approx 0.004 \,M_\odot$ was present reflecting the merger process for this model.
This amount is somewhat larger than that shown in Table~\ref{tab:3dmodel} because of the difference in the definition of the ejecta (see \S~\ref{subsec:ubound} and \S~\ref{subsec:3d}).
For the inviscid model DD2-135M-v0, $M_\mathrm{ej,tot}$ peaks at $t\sim 0.1$\,s and then decreases, indicating that a part of the ejecta was not actually the ejecta component and it falls back to the central region.
Its saturated value for $t \gtrsim 1$\,s, $\approx 0.003M_\odot$, shows the actual mass of the ejecta.
With the viscosity, on the other hand, a larger amount of the material is ejected as found below.

For model DD2-135M, it is found that $M_\mathrm{ej,tot}$ increases at three different times, $t\sim 0.01$\,s, $\sim$ 0.1\,s, and $\sim$ 1\,s (see the top-right panel of Fig.~\ref{fig:ejecta-decom}).
The first increase of $M_\mathrm{ej,tot}$ at $t\sim 0.01$\,s is also found for model DD2-135M-irr, but absent for model DD2-135M-v0, and therefore, this increase is due to the viscous effect.
This is the contribution of the early viscosity-driven ejecta.
By comparing models DD2-135M and DD2-135M-v0, the contribution of this component for our choice of the viscous parameter ($\alpha_\mathrm{vis}=0.04$) is found to be $\sim 0.01\,M_\odot$.
The second increase at $t\sim 0.1$\,s is absent for model DD2-135M-irr, and hence, we confirm that this ejecta is driven by the neutrino irradiation.
By comparing models DD2-135M and DD2-135M-irr, the contribution of this neutrino-driven component is found to be $\sim 0.01\,M_\odot$.
The contribution of the neutrino irradiation is fairly large compared to DD2-135M-v0 because of the large neutrino luminosity of the remnant due to the viscous heating (see Fig.~\ref{fig:MNS}).
The third one at $t\sim 1$\,s, which is found in both models DD2-135M and DD2-135M-irr, is the contribution of the late-time viscosity-driven ejecta.

In the bottom-left panel of Fig.~\ref{fig:ejecta-decom}, only two (not three) distinct mass ejection phases are found, one of which is found in an early phase of $t\lesssim 0.5$\,s and the other is in a late phase of $t\gtrsim 1$\,s.
The former is composed of the dynamical, early viscosity-driven, and neutrino-driven components.
The value of $dM_\mathrm{ej,esc}/dt$ for DD2-135M starts to deviate from that for model DD2-135M-irr at $t\approx 0.08$\,s, indicating that the neutrino-driven component starts to contribute to $M_\mathrm{ej,esc}$ at this time.
At $t\sim 0.1$\,s, the properties of the ejecta are determined by both the dynamical and neutrino-driven ejecta, the former of which has a lower electron fraction of $\approx 0.2$ and the latter has a higher value of $\approx 0.55$ (see Fig~\ref{fig:ye}).
$Y_\mathrm{e,ej}$ (the bottom-right panel of Fig.~\ref{fig:ejecta-decom}) then has an average value of the two components, which is $\approx$ 0.40--0.45.

After that, the early viscosity-driven ejecta begins to contribute dominantly to the properties of the ejecta at $t\approx 0.2$\,s, at which a peak in $dM_\mathrm{ej,esc}/dt$ is found in particular for model DD2-135M-irr.
This component has a low electron fraction of $\approx 0.25$ as found for model DD2-135M-irr.
$Y_\mathrm{e,ej}$ also decreases to $\approx 0.3$ at this time for model DD2-135M as a result of the large contribution of the early viscosity-driven ejecta to $dM_\mathrm{ej,esc}/dt$.

In the top-right panel of Fig.~\ref{fig:ejecta-decom}, a peak of $V_\mathrm{ej}$ at $t\sim 0.1$\,s is found for model DD2-135M, but absent for DD2-135M-irr, indicating that the neutrino-driven component has a velocity of $>$\,0.3\,$c$, which is larger than those of the dynamical and early viscosity-driven components at $t\approx 0.2$\,s.
This results in the simultaneous contribution of these ejecta components to $dM_\mathrm{ej,esc}/dt$ at $t\sim 0.1$\,s, although they are ejected at different times as found in the top-left panel of Fig.~\ref{fig:ejecta-decom}.

Finally, the late-time viscosity-driven ejecta associated with the viscous effect in the disk sets in.
$dM_\mathrm{ej,esc}/dt$ has a large value of $\gtrsim 0.01\,M_\odot$/s for $t=1$--3\,s, and after that, it decreases with time.
This component has smaller velocity of $V_\mathrm{ej}\sim 0.05$--0.10$\,c$ and a range of the average electron fraction of $Y_\mathrm{e,ej}\approx$ 0.30--0.35 reflecting the electron fraction of the disk ($Y_\mathrm{e,disk}$; see Fig.~\ref{fig:disk}).

\begin{figure*}
\epsscale{1.17}
\plottwo{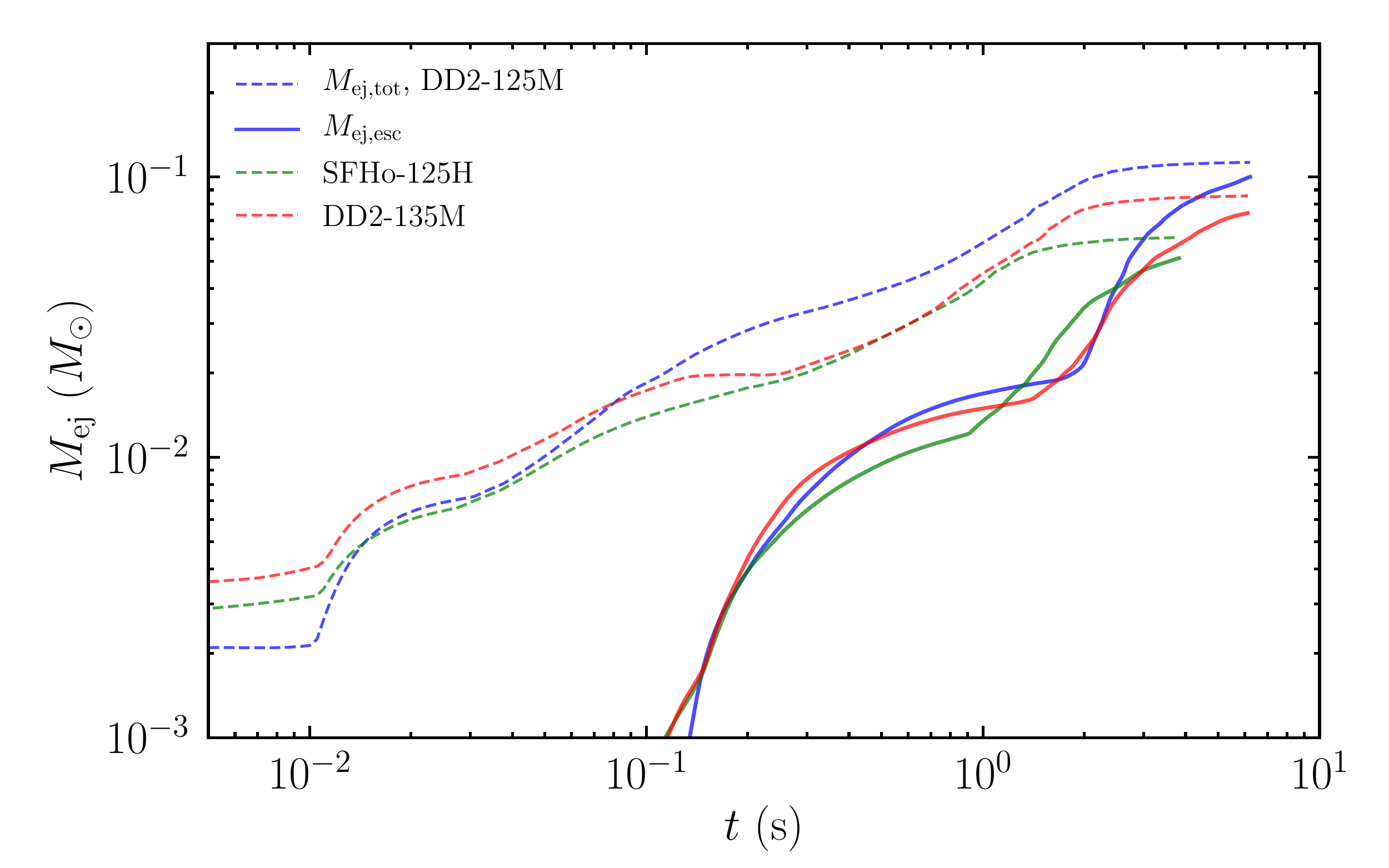}{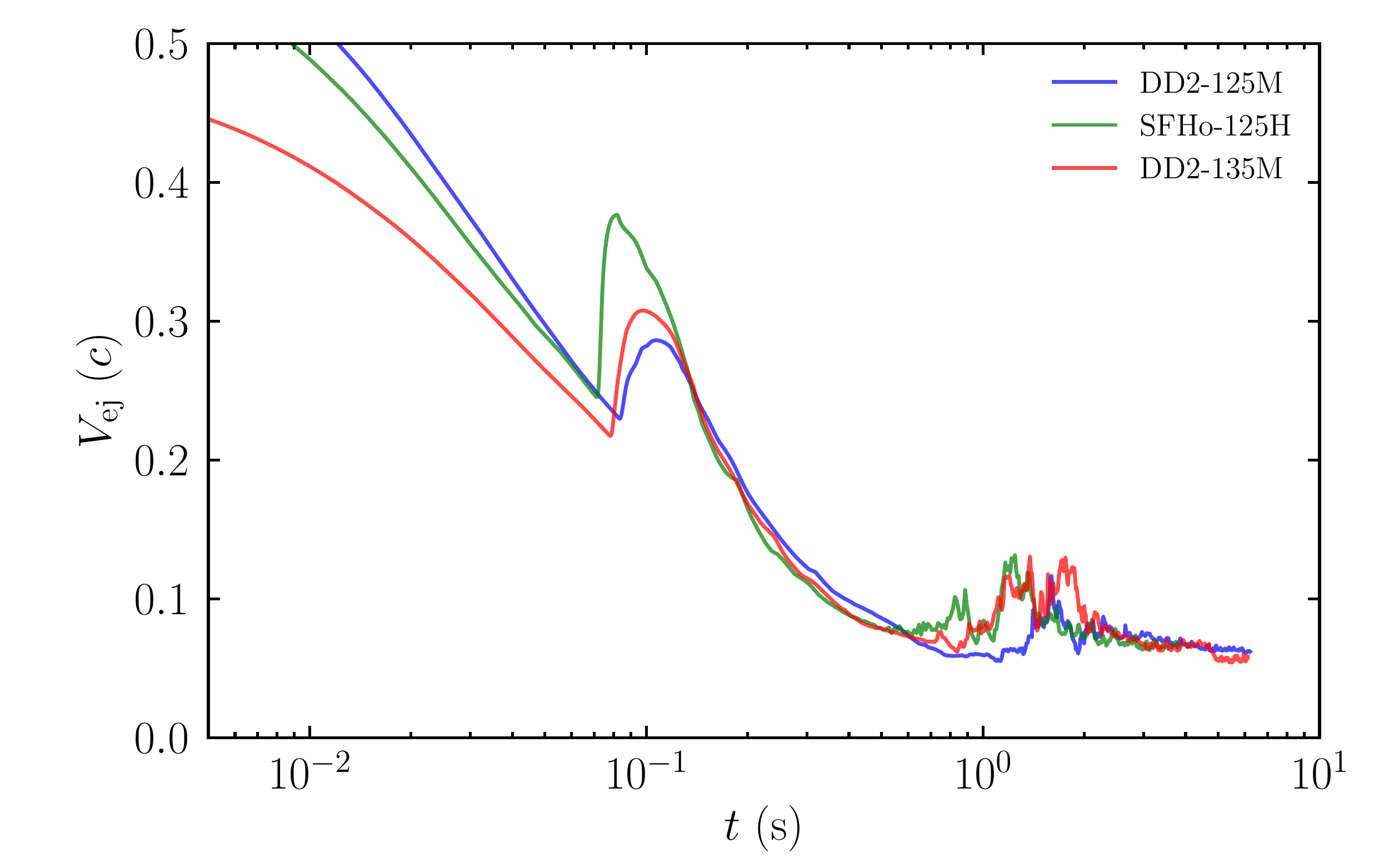}
\plottwo{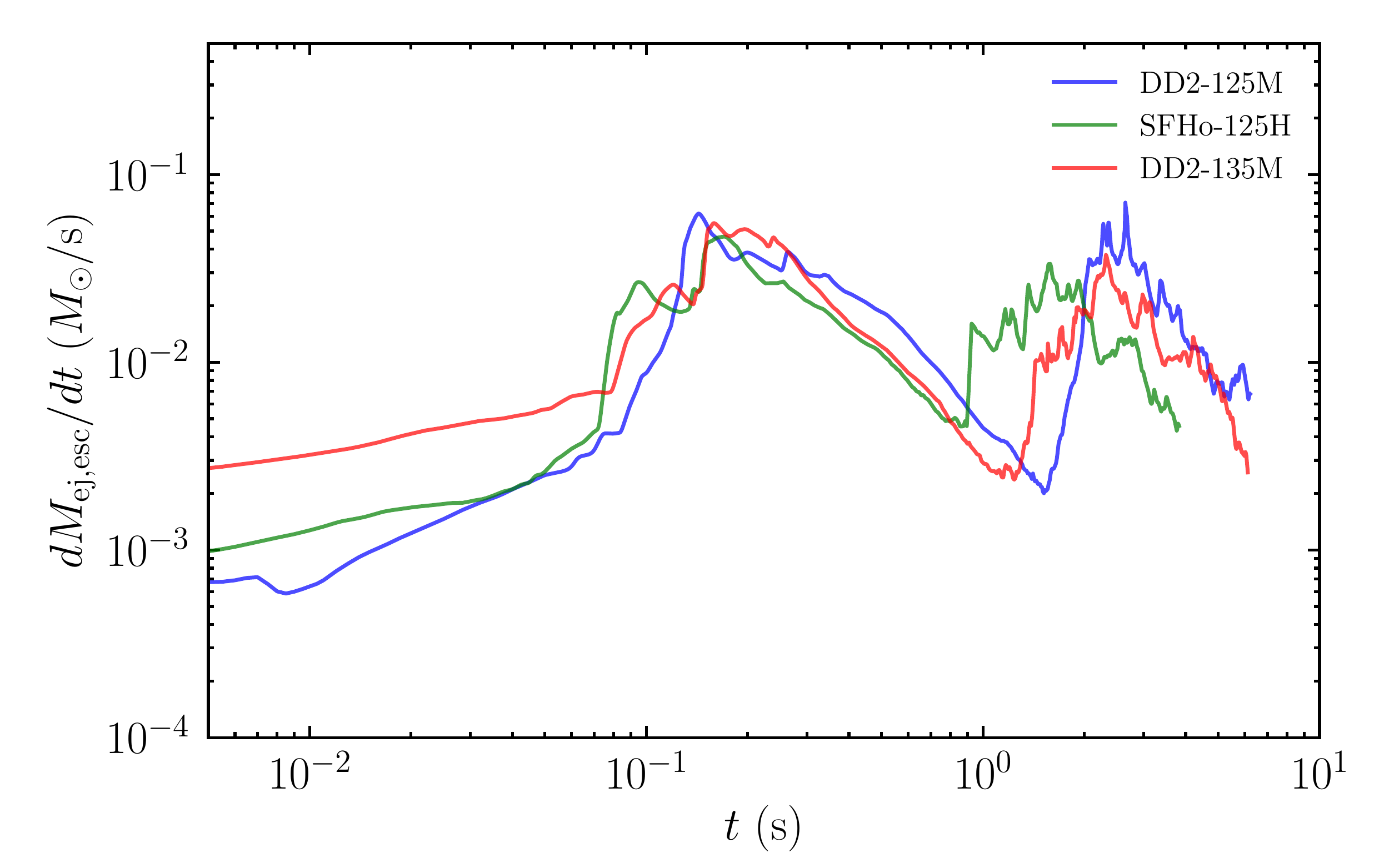}{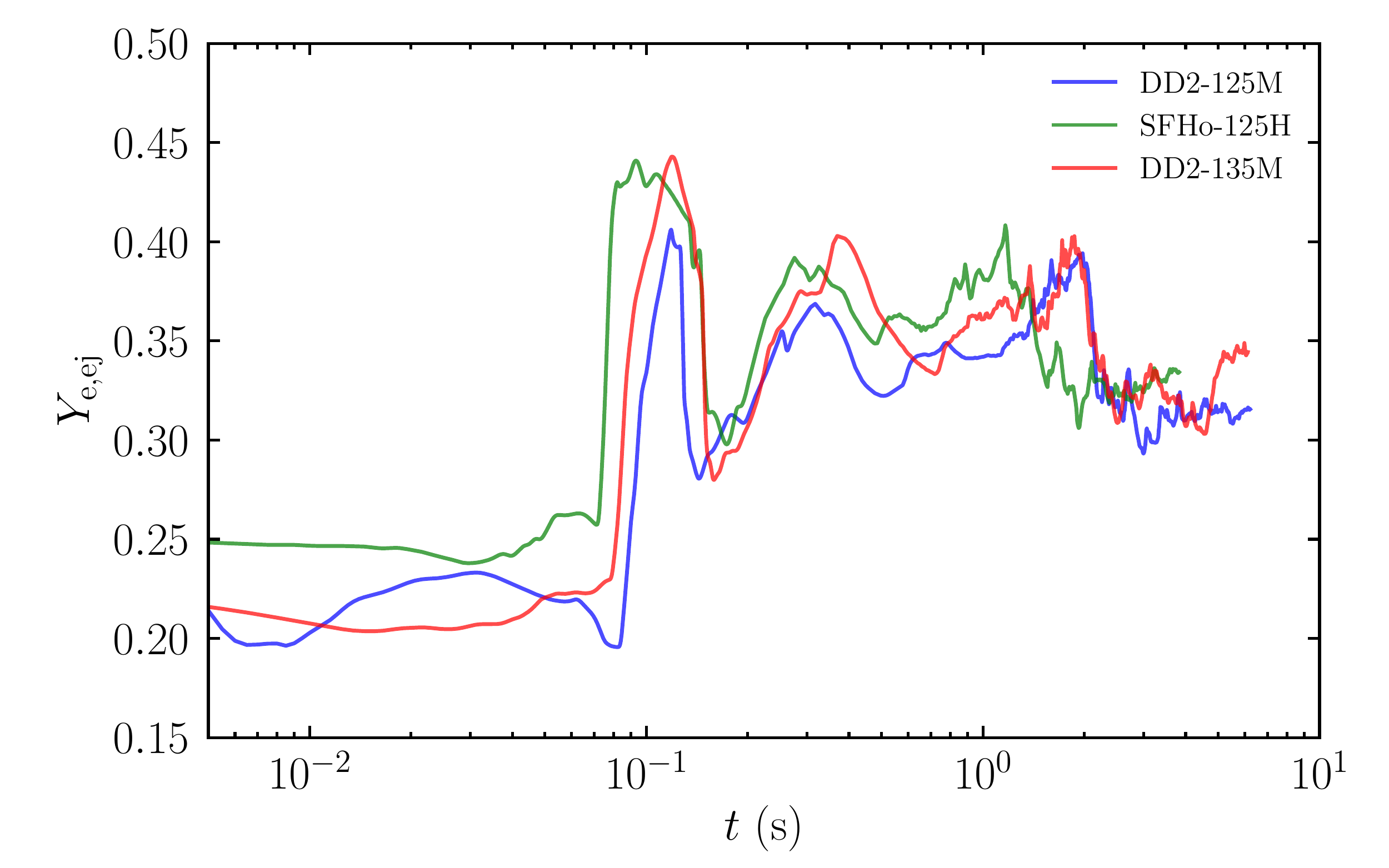}

\caption{
Same as Fig.~\ref{fig:ejecta-decom}, but for models DD2-125M, SFHo-125H, and DD2-135M.
}
\label{fig:ejecta}
\end{figure*}

For model DD2-135M, the mass-weighted average electron fraction over all the ejecta components is $\langle Y_\mathrm{e,ej}\rangle\approx0.34$ at the end of the simulation due to the domination of the late-time ejecta (see Table~\ref{tab:model}).
By comparing models DD2-135M and DD2-135M-irr in Table~\ref{tab:model}, $\langle Y_\mathrm{e,ej}\rangle$ is lower by 0.05 in the absence of the neutrino irradiation (reflected by the lower electron fraction of the disk; see \S~\ref{subsubsec:diskirr}).

Worthy to mention is that, even in the absence of the neutrino irradiation, the electron fraction of the ejecta increases to $\approx 0.3$ due to the positron capture after the decrease of the electron degeneracy in the disk (see \S~\ref{subsubsec:ejectaye}).
This indicates that, even in the absence of the long-lived massive NS, the electron fraction of the ejecta would not be very low as pointed out in our recent work for BH-disk systems \citep{Fujibayashi2020a}.

\subsubsection{Ejecta Properties for Fiducial models}
\label{subsubsec:fiducial}

Figure~\ref{fig:ejecta} compares the ejecta properties for models DD2-125M, SFHo-125H, and DD2-135M.
The curves are qualitatively similar to each other, showing that there are three mass ejection mechanisms as described in \S~\ref{subsubsec:decomp} irrespective of the adopted EOS and the total mass of the model.
The baryon mass of the unbound material for $t\lesssim0.03$\,s, composed of dynamical and early viscosity-driven ejecta, depends on models.
The baryon mass of these early ejecta components correlates with the baryon mass of the dynamical ejecta (see Table~\ref{tab:3dmodel}).
The amount of the subsequent neutrino-driven and late-time viscosity-driven ejecta, on the other hand, correlates with the disk mass (see the top-left panel of Fig.~\ref{fig:disk}).
The value of $M_\mathrm{ej,tot}$ saturates at $t= 2$--3\,s and by comparing with the top-left panel of the Fig.~\ref{fig:disk}, the saturated values indicate that approximately 30\,\% of the initial disk mass becomes the ejecta.

For all the fiducial models, the average ejecta velocity is $\langle V_\mathrm{ej} \rangle \sim 0.1\,c$ (see Table~\ref{tab:model}), which is somewhat larger than that of the late-time ejecta (see the top-left panel of Fig.~\ref{fig:ejecta}) because of the large kinetic energy of the dynamical, early viscosity-driven, and neutrino-driven ejecta components.

\begin{figure}
\epsscale{1.17}
\plotone{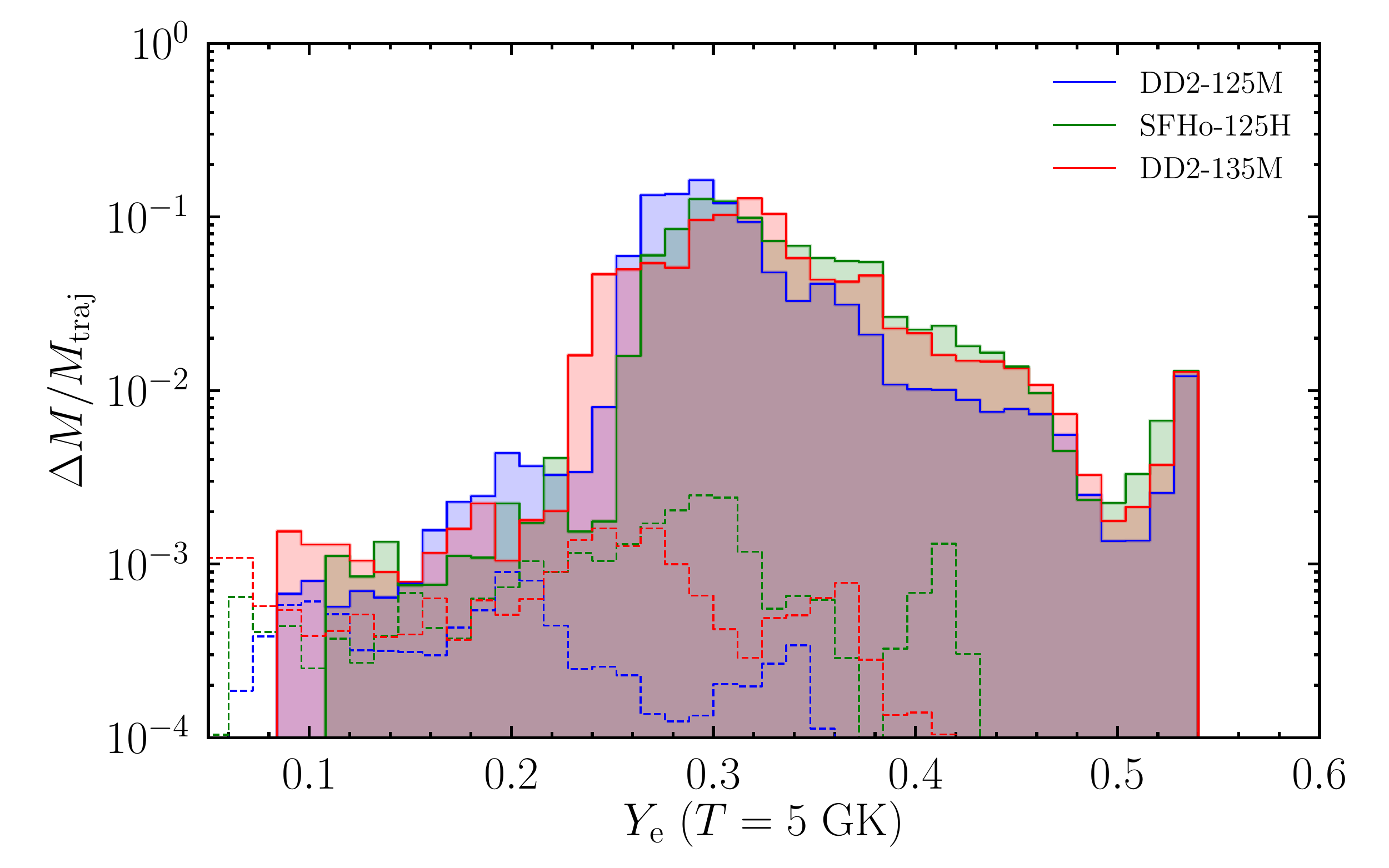}
\plotone{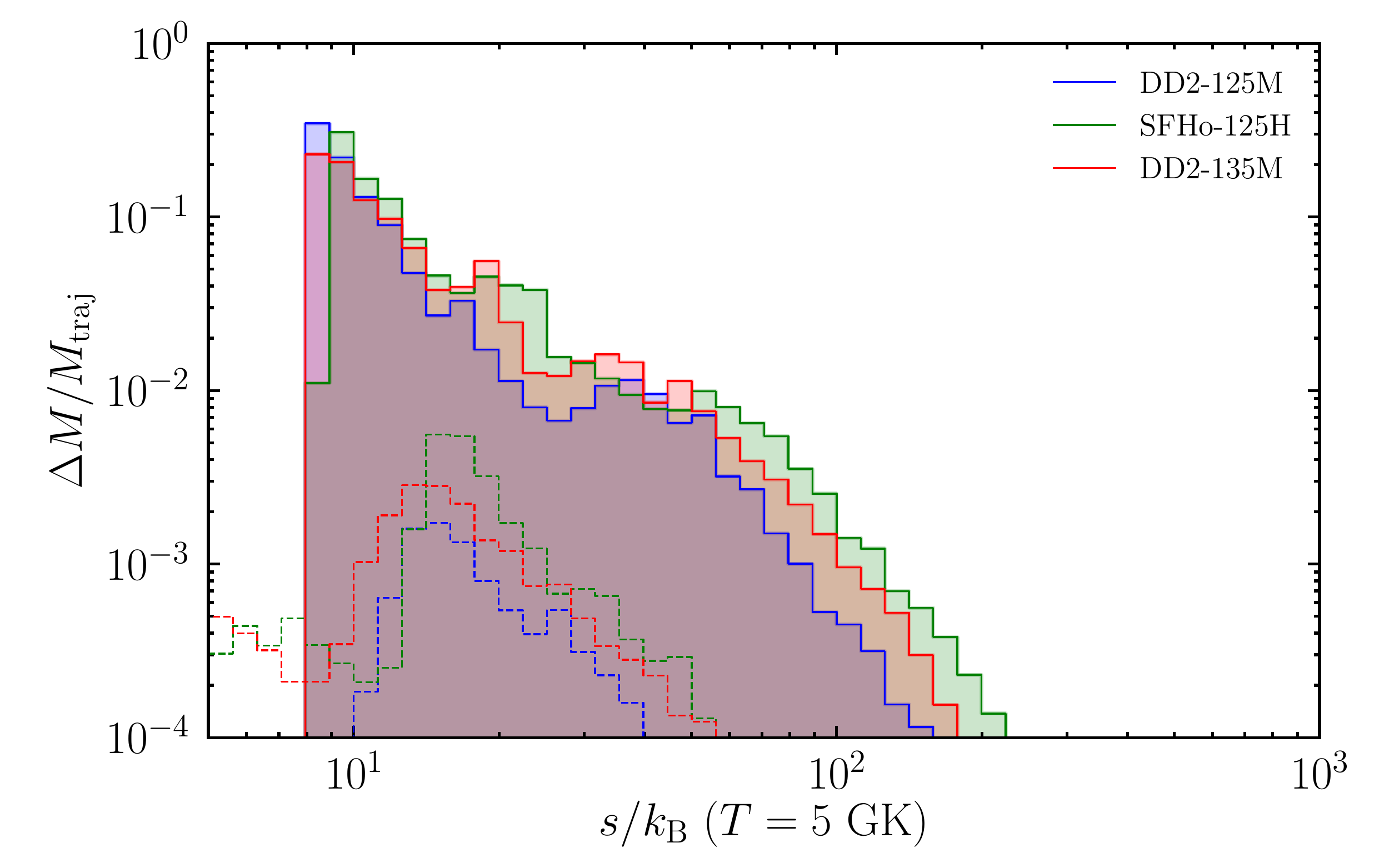}
\plotone{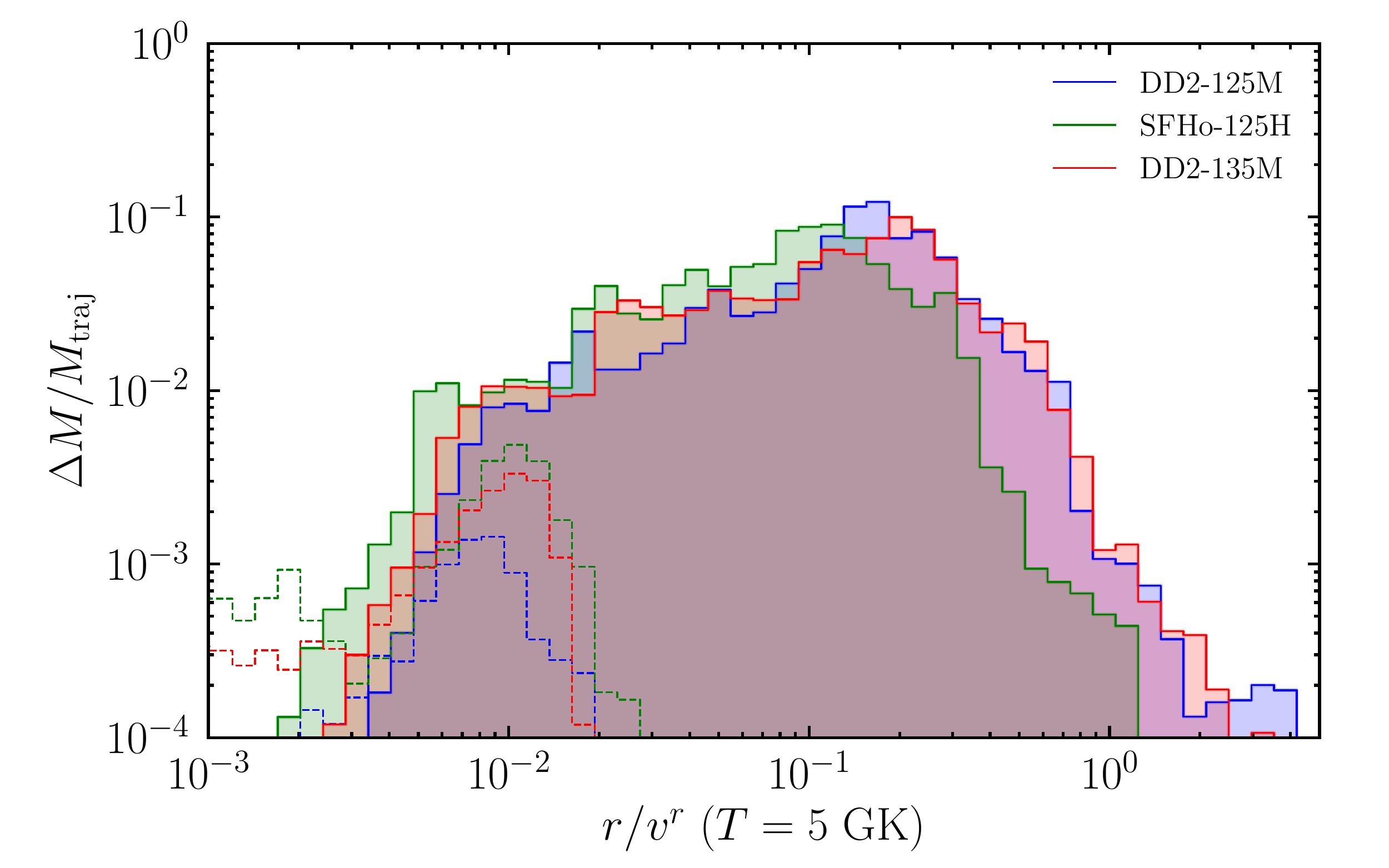}

\caption{
Mass histograms of the electron fraction (top), entropy (middle), and the expansion timescale (bottom) for tracer particles at $T=5$ GK for models DD2-125M (blue), SFHo-125H (green), and DD2-135M (red).
The dashed curves show the histograms obtained in the corresponding 3D models.
Each histogram is normalized by the total mass of the tracer particles.
}
\label{fig:yesv}
\end{figure}

Figure~\ref{fig:yesv} shows the mass histograms of electron fraction (top), entropy (middle), and expansion timescale (bottom; $t_\mathrm{exp} = r/v^r$ with $ v^r= dr/dt$) of the tracer particles at $T=5$\,GK for the fiducial models.
The distributions for all the models are quite similar and peak at $Y_\mathrm{e}\approx 0.3$, $s/k_\mathrm{B}\approx 10$, and $t_\mathrm{exp}\approx 0.1$\,s.
These peaks for each model are produced by the late-time viscosity-driven ejecta.
In this figure, the histogram in corresponding 3D simulations are also shown.
This illustrates that the low electron fraction and short expansion timescale sides of distributions are determined by the dynamical ejecta.\footnote{In Fig.~\ref{fig:yesv}, the lowest electron fraction, lowest entropy, and shortest expansion timescale ends found in 3D merger simulations are missing in 2D models.
Due to the small mass fraction of such components, they disappear owling that the 3D fluid profiles are averaged with respect to the azimuthal angle and mapped to the 2D initial conditions.}

\begin{figure}
\epsscale{1.17}
\plotone{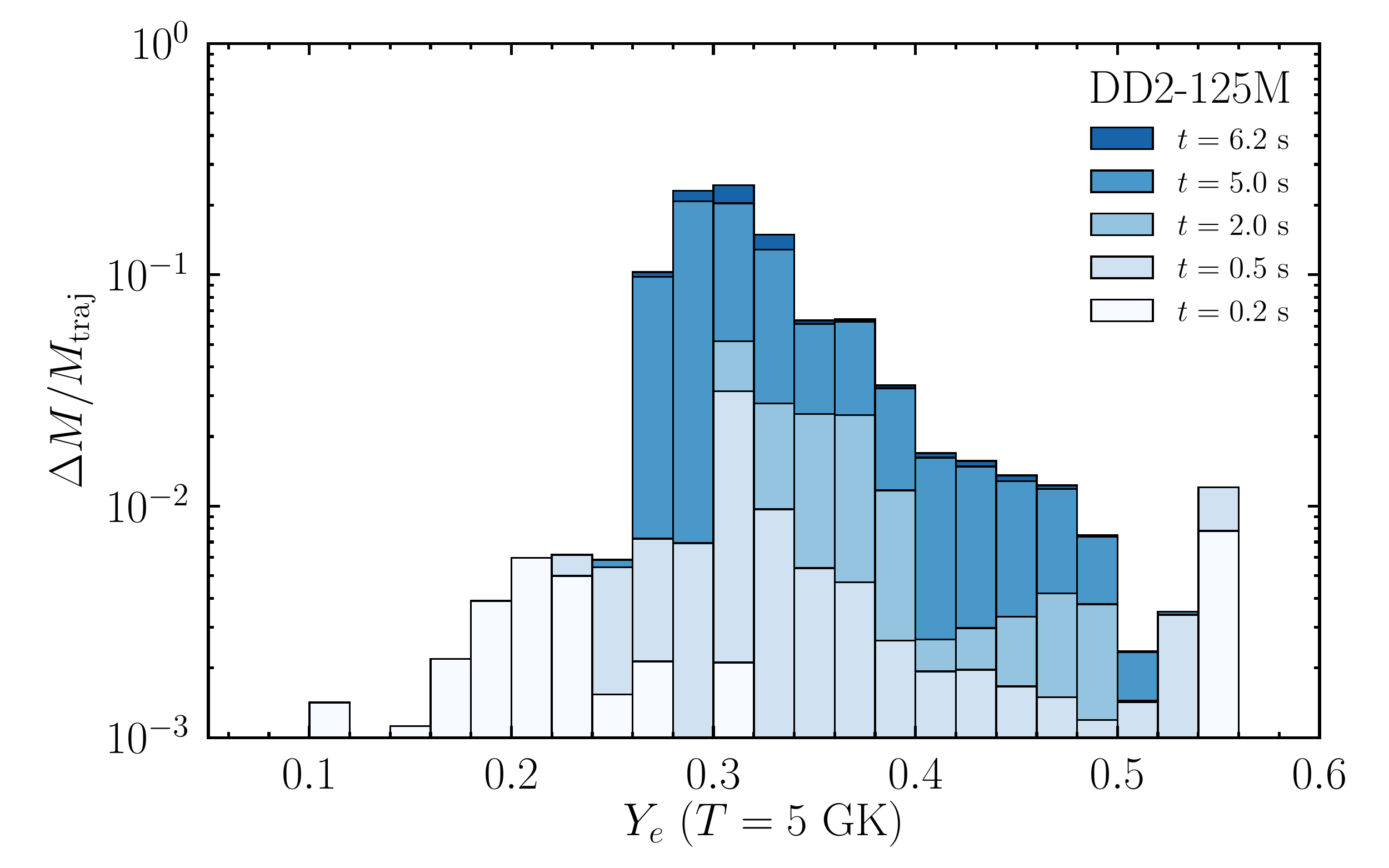}
\plotone{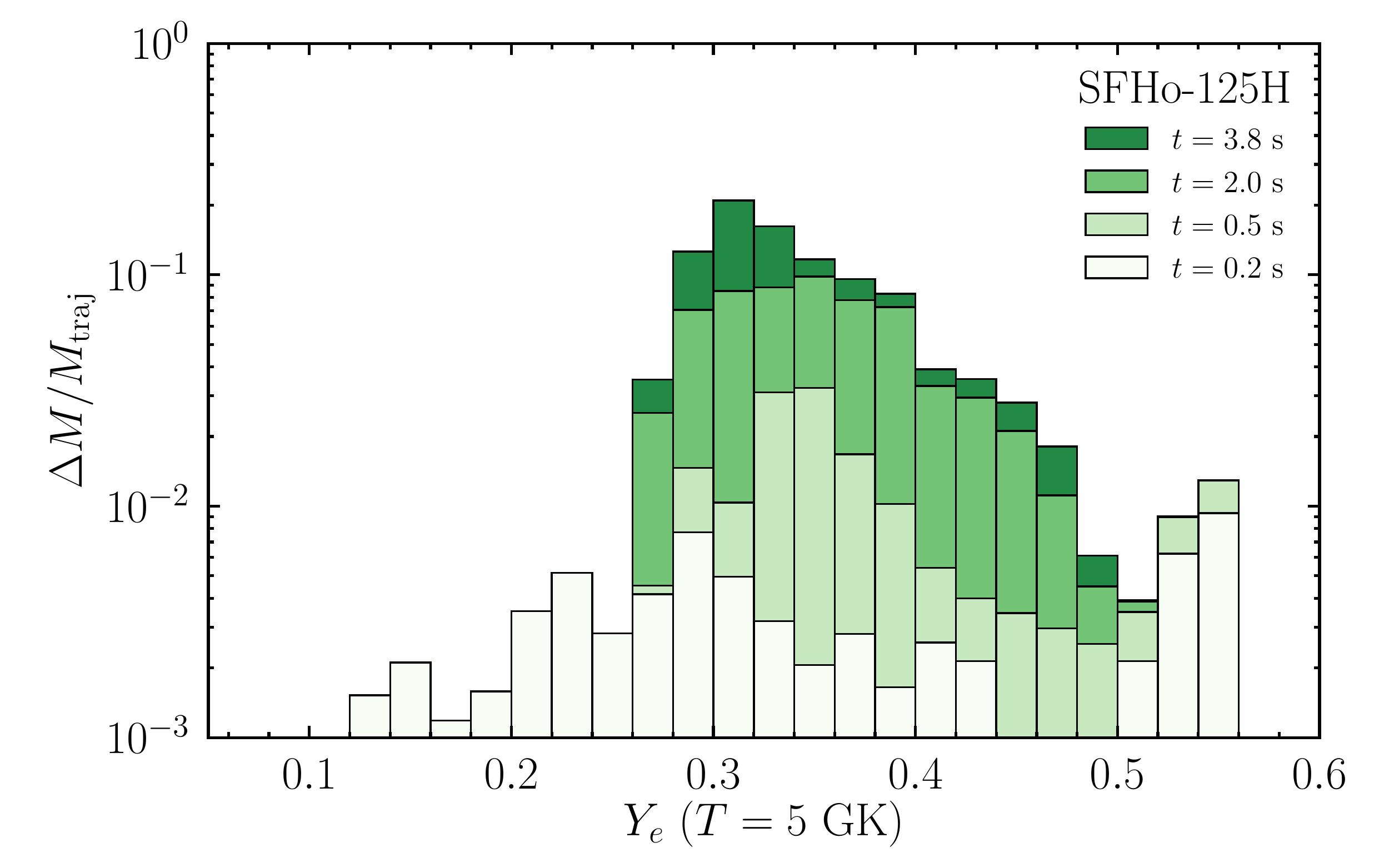}
\plotone{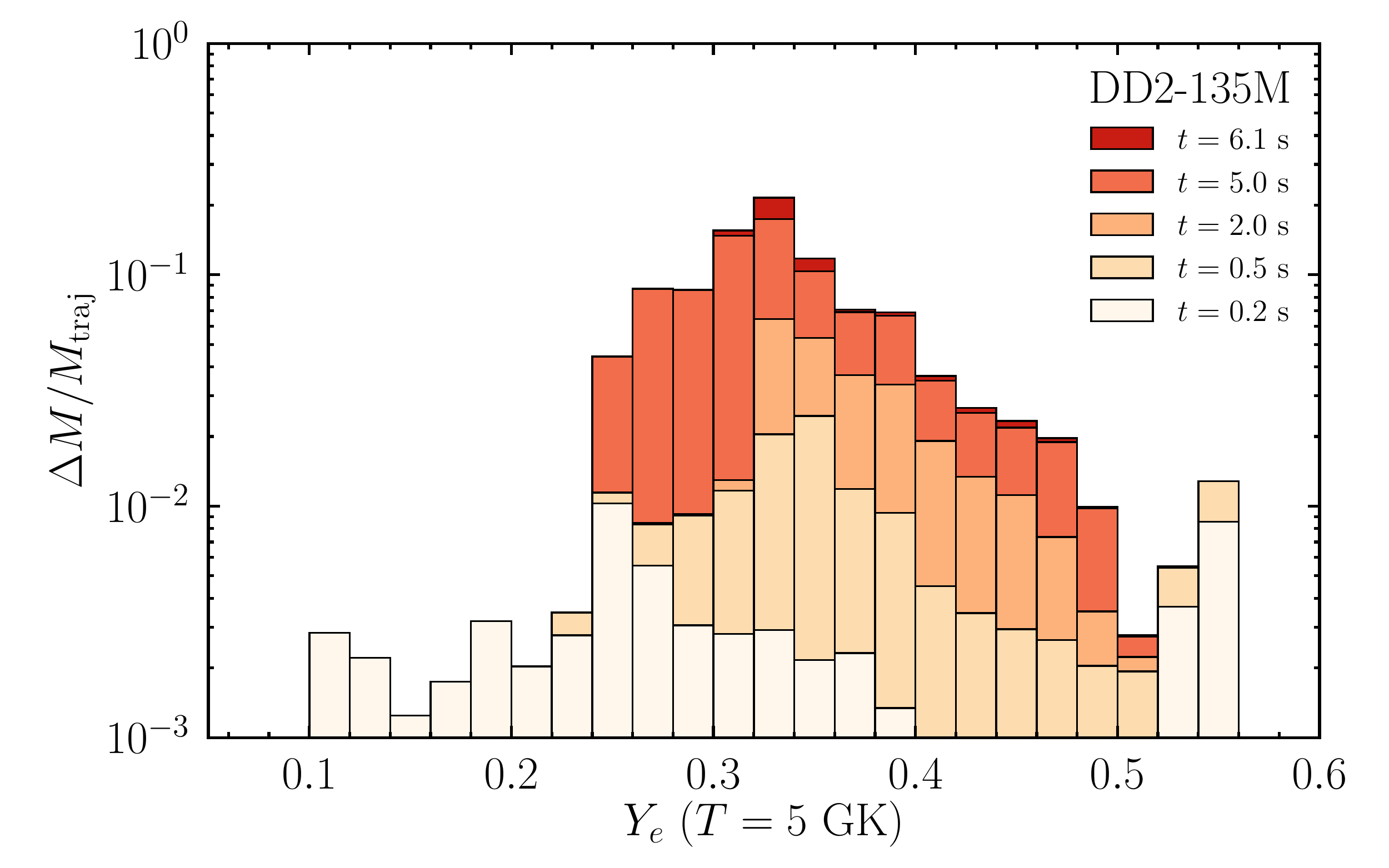}

\caption{
Temporal mass histograms of $Y_\mathrm{e}$ for the tracer particles that reach the radius $r_\mathrm{esc}$ by a given time for models DD2-125M (top), SFHo-125H (middle), and DD2-135M (bottom).
Each histogram is normalized by the total mass of the tracer particles.
Note that the range of the vertical axis for these panels differs from that of Fig.~\ref{fig:yesv}.
}
\label{fig:ye}
\end{figure}

\begin{figure*}
\epsscale{1.17}
\plottwo{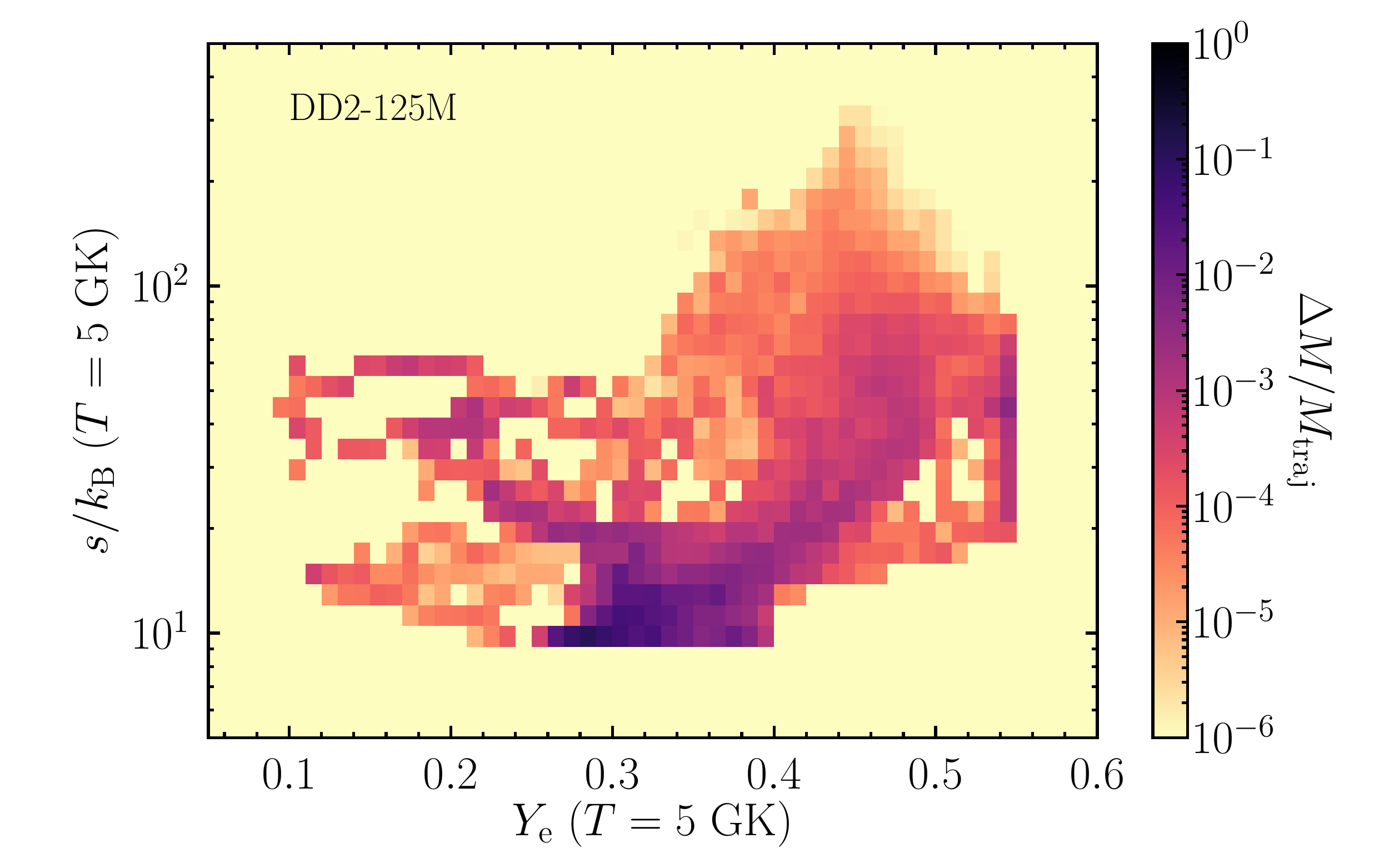}{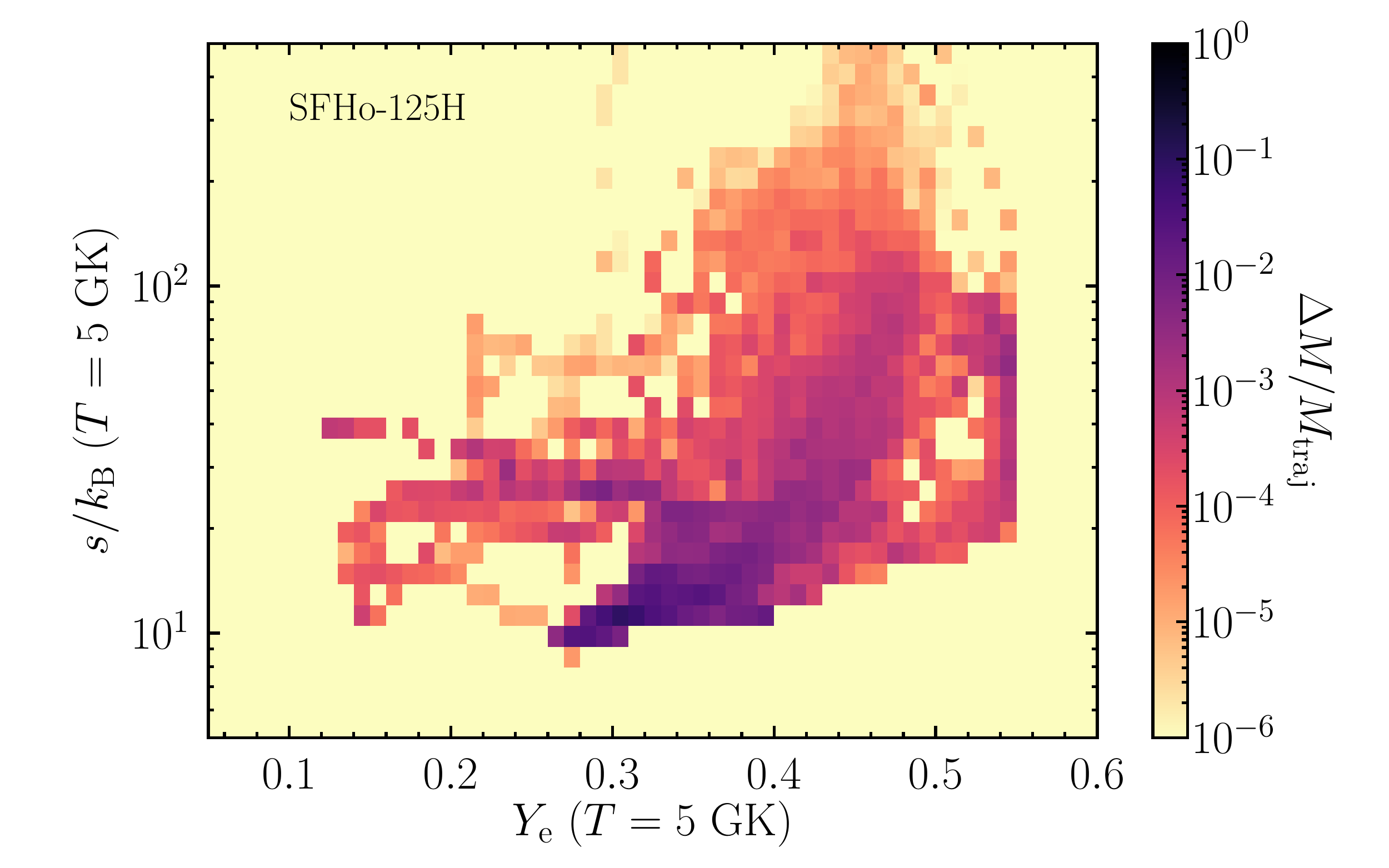}
\plottwo{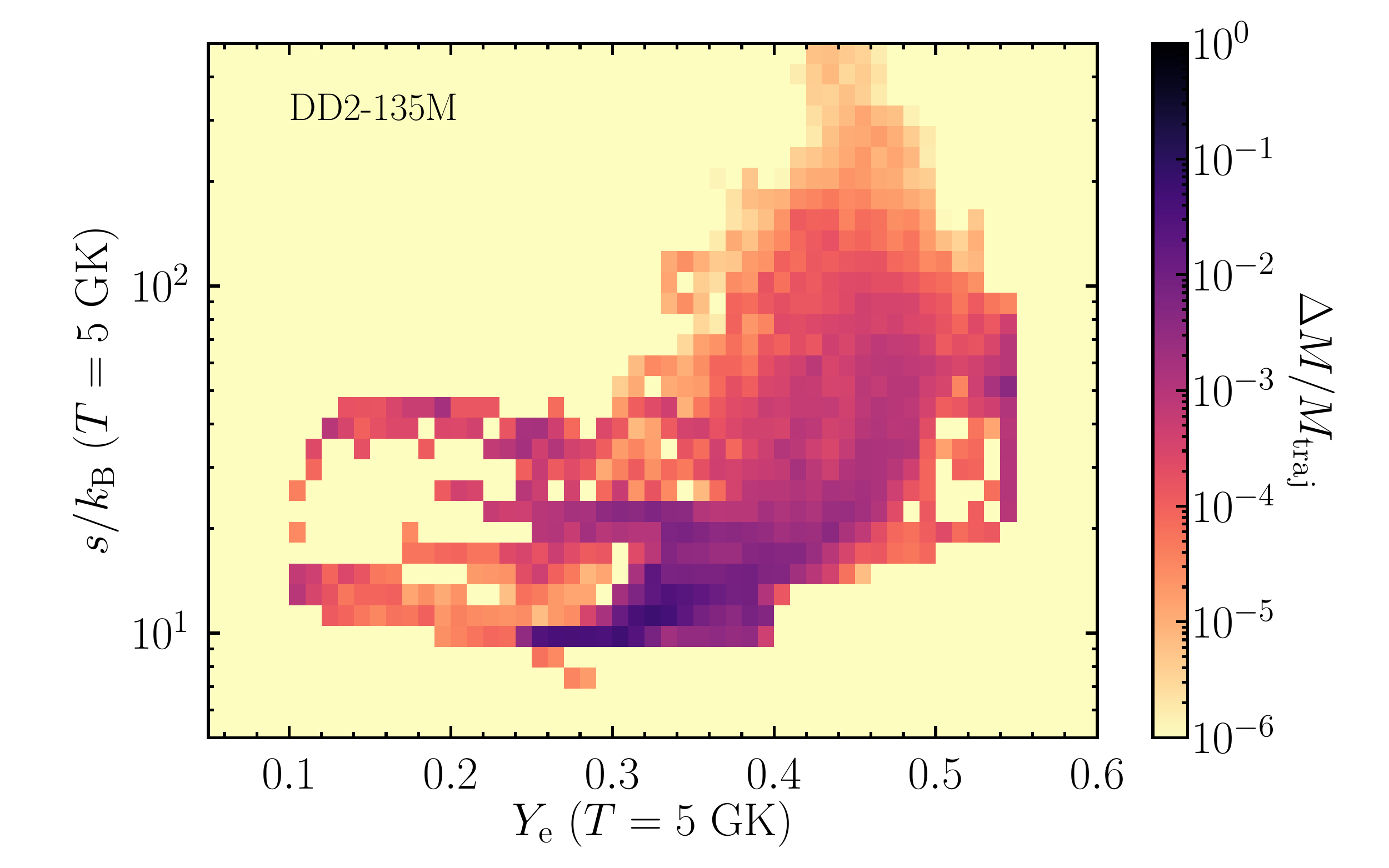}{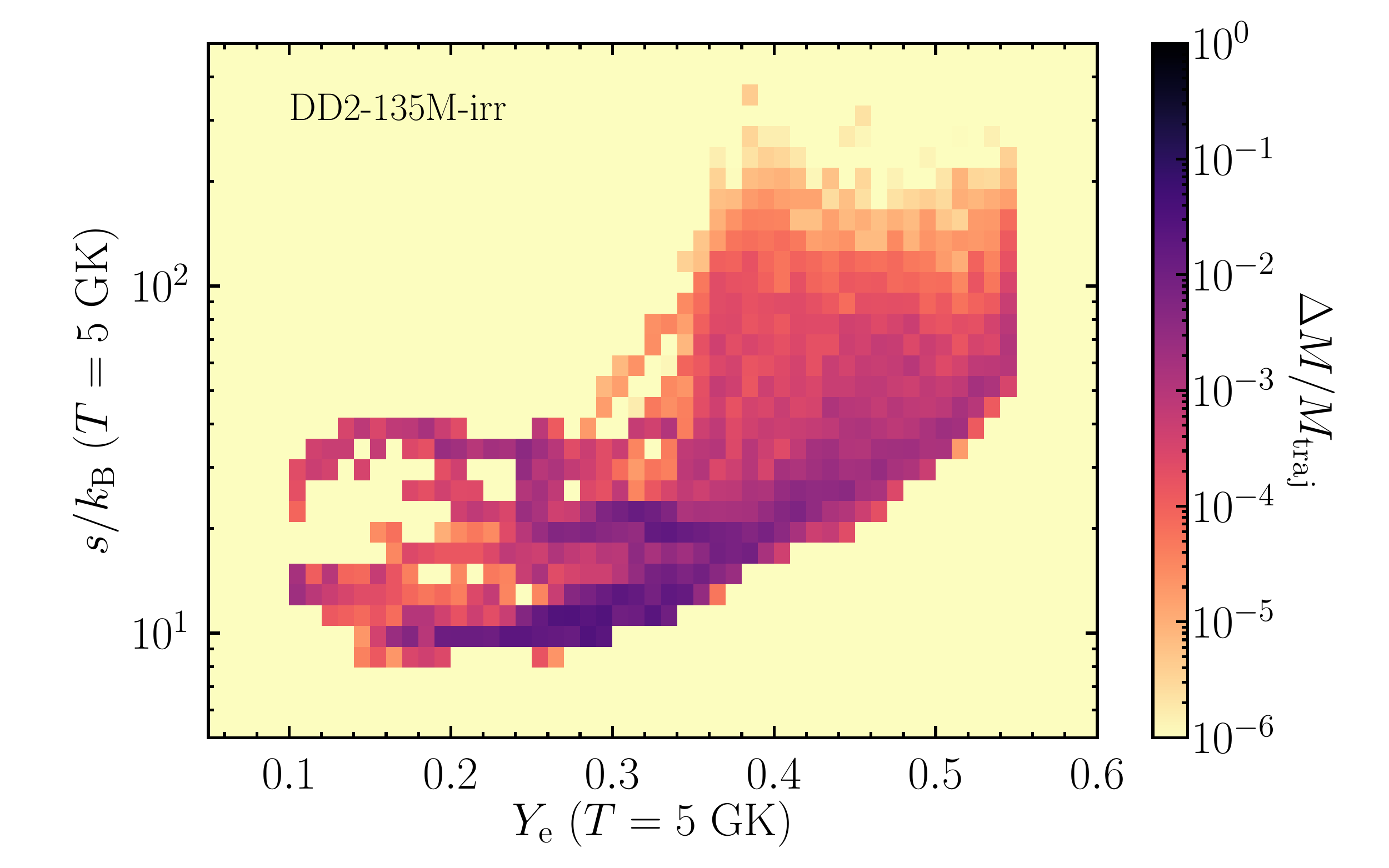}
\caption{
Mass distribution of the ejecta in the $s$--$Y_\mathrm{e}$ plane for models DD2-125M (top-left), SFHo-125H (top-right), DD2-135M (bottom-left), and DD2-135M-irr (bottom-right).
The color code shows the distribution of the mass normalized by the total mass for the tracer particles of each model.
}
\label{fig:sye}

\end{figure*}

The temporal $Y_\mathrm{e}$ distributions for the tracer particles that reach $r_\mathrm{esc}$ by a given time are shown in Fig.~\ref{fig:ye} for models DD2-125M, SFHo-125H, and DD2-135M.
At $t=0.2$\,s, there are two peaks at $Y_\mathrm{e} \sim 0.2$ and 0.55, which are determined by the dynamical plus early viscosity driven ejecta and neutrino-driven ejecta, respectively.
The subsequent change of the histograms clearly illustrates that the peak of $Y_\mathrm{e}$ in the final distribution is developed by the late-time viscosity-driven ejecta for $t>1$\,s.

The distributions of the ejecta in the $s$--$Y_\mathrm{e}$ plane are shown in Fig.~\ref{fig:sye} for models DD2-125M, SFHo-125H, and DD2-135M.
It is found that the distribution profiles depend very weakly on the EOSs and binary total mass.
We also find a positive correlation between $s$ and $Y_\mathrm{e}$.
The ejecta with $Y_\mathrm{e}\approx 0.3$ has entropy of $s/k_\mathrm{B}\lesssim 10$, while the ejecta with higher values of $Y_\mathrm{e}$ has higher entropy.
The electron fraction for most of the material ejected toward the non-polar direction is determined predominantly by the equilibrium between electron/positron capture (see \S~\ref{subsubsec:ejectaye}), and the equilibrium value becomes higher for weaker electron degeneracy.
Because the entropy negatively correlates with the electron degeneracy, the material with the higher entropy has the higher equilibrium value of the electron fraction.
This results in the positive correlation between $s$ and $Y_\mathrm{e}$.
In other words, the material experiencing the longer-time viscous heating has the higher electron fraction reflecting its higher entropy.
By comparing bottom two panels of Fig.~\ref{fig:sye}, it is found that the low-$Y_\mathrm{e}$ component is more abundant in the absence of the neutrino irradiation.

In Fig.~\ref{fig:sye}, there is a component that has a low electron fraction of $Y_\mathrm{e}=0.1$--0.2 and a moderate entropy of $s/k_\mathrm{B}\sim 50$.
This component is a part of the dynamical ejecta, which is initially expelled by the tidal effect and then heated up by a shock formed in spiral arms.
Because this component experiences the shock heating in a distant region from the center, the increase of the temperature due to the shock is not enough for the weak interaction (positron capture) to work in a sufficiently short timescale.
Thus, the electron fraction of this component remains low.

\subsubsection{Processes that Determines the Electron Fraction of the Ejecta}
\label{subsubsec:ejectaye}
\begin{figure}
\epsscale{1.17}
\plotone{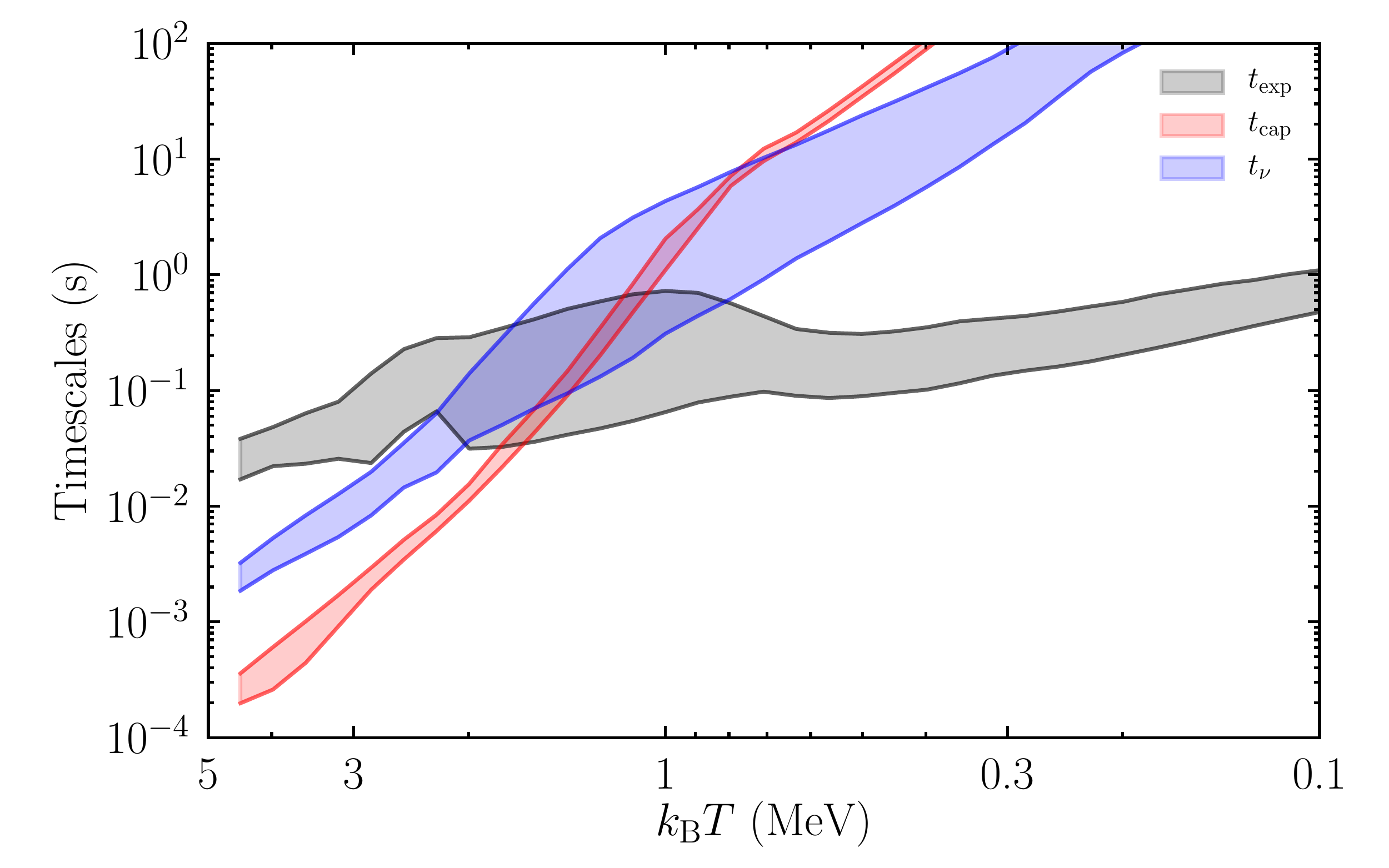}
\plotone{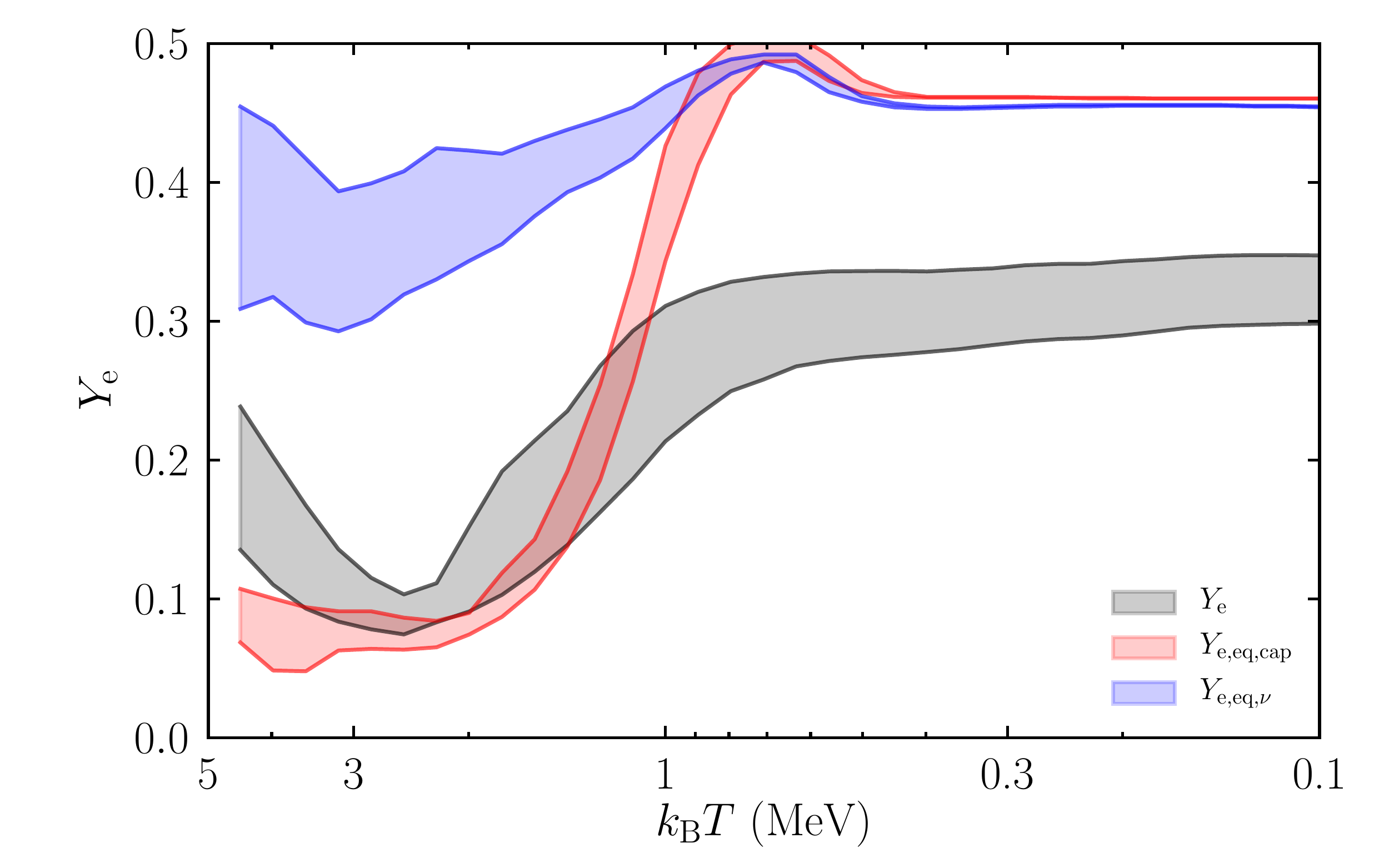}
\plotone{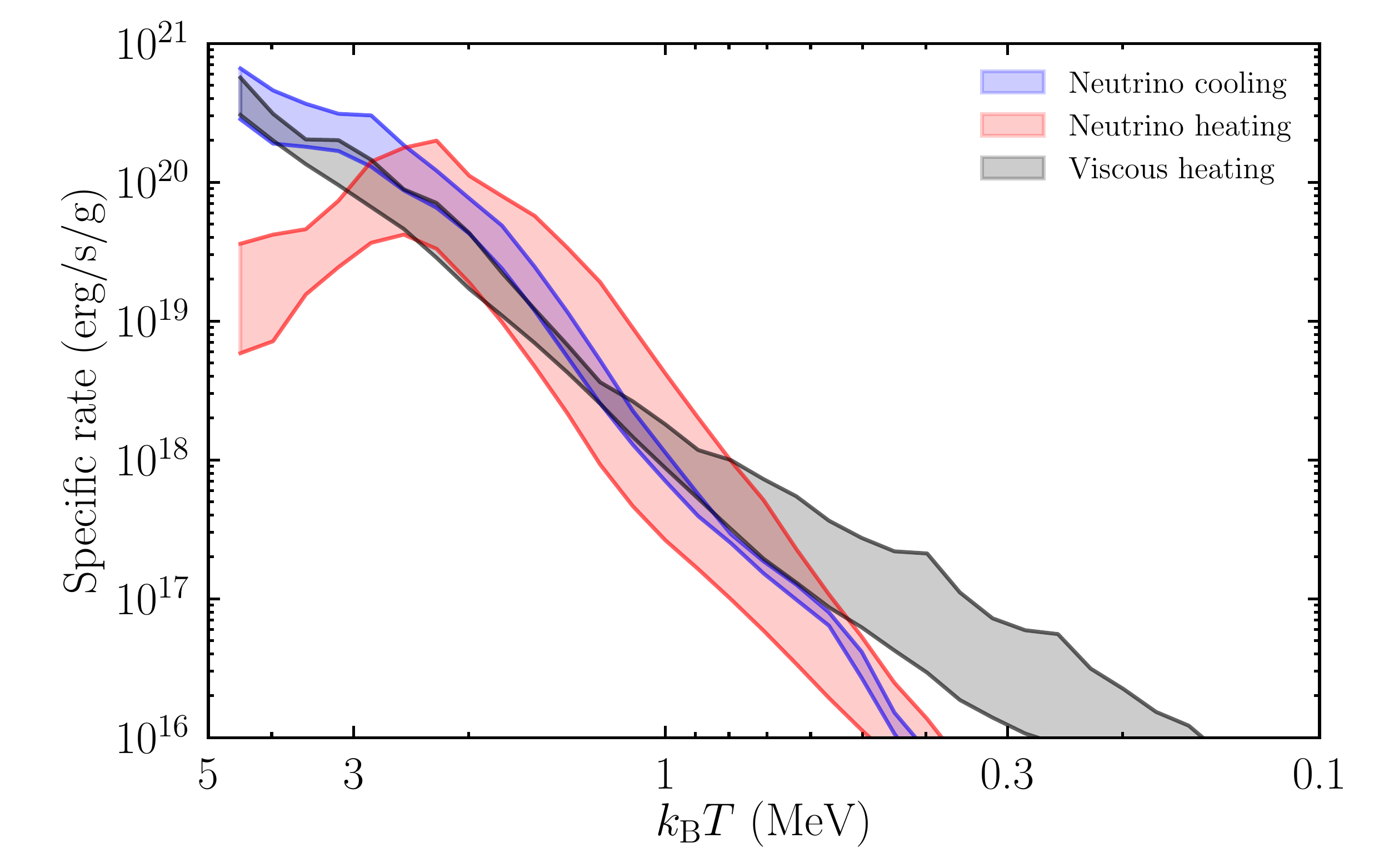}
\caption{
Top: the evolution of the timescales of the tracer particles for the electron (anti)neutrino absorption (blue), electron/positron capture (red), and the expansion of the ejecta (gray) with decreasing temperature (i.e., with the time evolution) for model DD2-125M (see text for their definitions).
Middle: the evolution of the electron fraction (gray), the equilibrium values for neutrino and antineutrino absorption (blue), and electron and positron captures (red) as functions of the decreasing temperature.
Bottom: the evolution of the specific heating rate due to the neutrino absorption (red) and the viscosity (gray) as well as the specific cooling rate due to electron and positron capture (blue) as functions of the decreasing temperature.
Shaded area indicates the region that contains 50\% of the trajectories which have the terminal entropy less than 30\,$k_\mathrm{B}$ around the median of each value.
}
\label{fig:timescale}
\end{figure}

The electron fraction of the material is determined by the the absorption of electron-type (anti)neutrinos and the electron/positron capture on free nucleons.
The time evolution of the electron fraction is described by
\begin{eqnarray}
\dot{Y}_\mathrm{e} = \dot{Y}_\mathrm{e,\nu} + \dot{Y}_\mathrm{e,pc} - \dot{Y}_\mathrm{e,\bar{\nu}} - \dot{Y}_\mathrm{e,ec},
\end{eqnarray}
where $\dot{Y}_\mathrm{e,\nu}$, $\dot{Y}_\mathrm{e,\bar{\nu}}$, $\dot{Y}_\mathrm{e,ec}$, and $\dot{Y}_\mathrm{e,pc}$ are the change rates of the electron fraction due to the absorption (or capture) of electron neutrinos, electron antineutrinos, electrons, and positrons, respectively, on free nucleons.
These reaction rates depend on the local density, temperature, electron fraction, and fluxes (and energies) of electron neutrinos and antineutrinos.
In the top panel of Fig.~\ref{fig:timescale}, the variations of the timescales of these reactions with decreasing temperature in tracer particles are shown together with their expansion timescales defined by $t_\mathrm{exp}=r/v^r$ with the coordinate radius $r$ and radial velocity $v^r(=dr/dt)$.
Here we define the timescale of the electron/positron capture by $t_\mathrm{cap}=\min(Y_\mathrm{e}/\dot{Y}_\mathrm{e,ec},Y_\mathrm{e}/\dot{Y}_\mathrm{e,pc})$ and that of the neutrino absorption by $t_\nu=\min(Y_\mathrm{e}/\dot{Y}_\mathrm{e,\nu},Y_\mathrm{e}/\dot{Y}_\mathrm{e,\bar{\nu}})$.
It is found that until the timescales of weak interaction processes become longer than the expansion timescale, the electron fraction of the tracer particle is determined mainly by the electron/positron capture (red color) and subdominantly by the neutrino absorption (blue color).

In the middle panel of Fig.~\ref{fig:timescale}, the variation of the electron fraction with decreasing temperature in the tracer particles is shown together with its equilibrium values for neutrino and antineutrino absorption, $Y_\mathrm{e,\nu}$, and electron and positron captures, $Y_\mathrm{e, \mathrm{cap}}$.
These equilibrium values are determined by the equations $\dot{Y}_\mathrm{e,\nu}=\dot{Y}_\mathrm{e,\bar{\nu}}$ and $\dot{Y}_\mathrm{e,ec}=\dot{Y}_\mathrm{e,pc}$.
Here, we assumed that the material is not optically thick to neutrinos.
That is, in our analysis, we pay attention only to the temperature range in which the beta equilibrium is not established.
This assumption is reasonable for low temperature of $k_\mathrm{B}T\lesssim 3$\,MeV.
It is found that the electron fraction of the tracer particle is determined approximately by the equilibrium value for electron and positron captures for $k_\mathrm{B}T\sim 1$--3\,MeV.
We note that the electron fraction deviates from $Y_\mathrm{e, \mathrm{cap}}$ for $k_\mathrm{B}T\gtrsim 3$\,MeV because the material is opaque to neutrinos and the electron fraction should be determined by the beta equilibrium; the electron fraction is determined not only by the electron/positron captures but also by electron (anti)neutrino absorption.
By contrast, at $k_\mathrm{B}T \sim 1$\,MeV, the electron fraction freezes out when the weak interaction timescale becomes longer than the expansion timescale (see the top panel of Fig.~\ref{fig:timescale}).
This illustrates how the electron fraction is determined in the material ejected towards the non-polar direction (cf. Fig.~\ref{fig:ye}).
The effect of the neutrino absorption on the electron fraction is minor for most of the material ejected toward the non-polar direction.
On the other hand, for the material ejected toward the polar direction, the electron fraction is determined predominantly by the neutrino absorption and it is higher than that of the ejecta toward the non-polar direction (see Fig.~\ref{fig:dyn}).

\begin{figure*}
\epsscale{1.17}
\plottwo{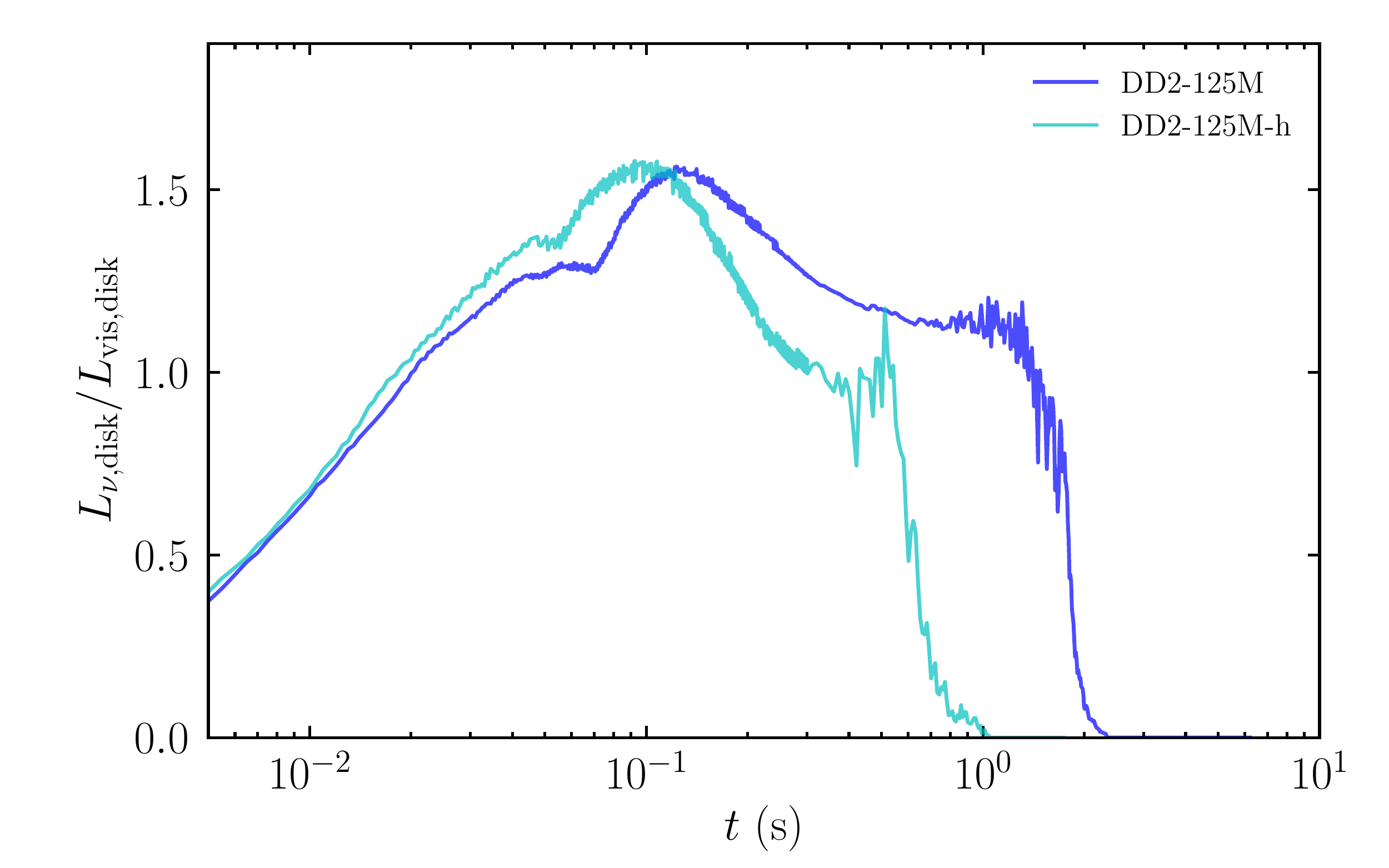}{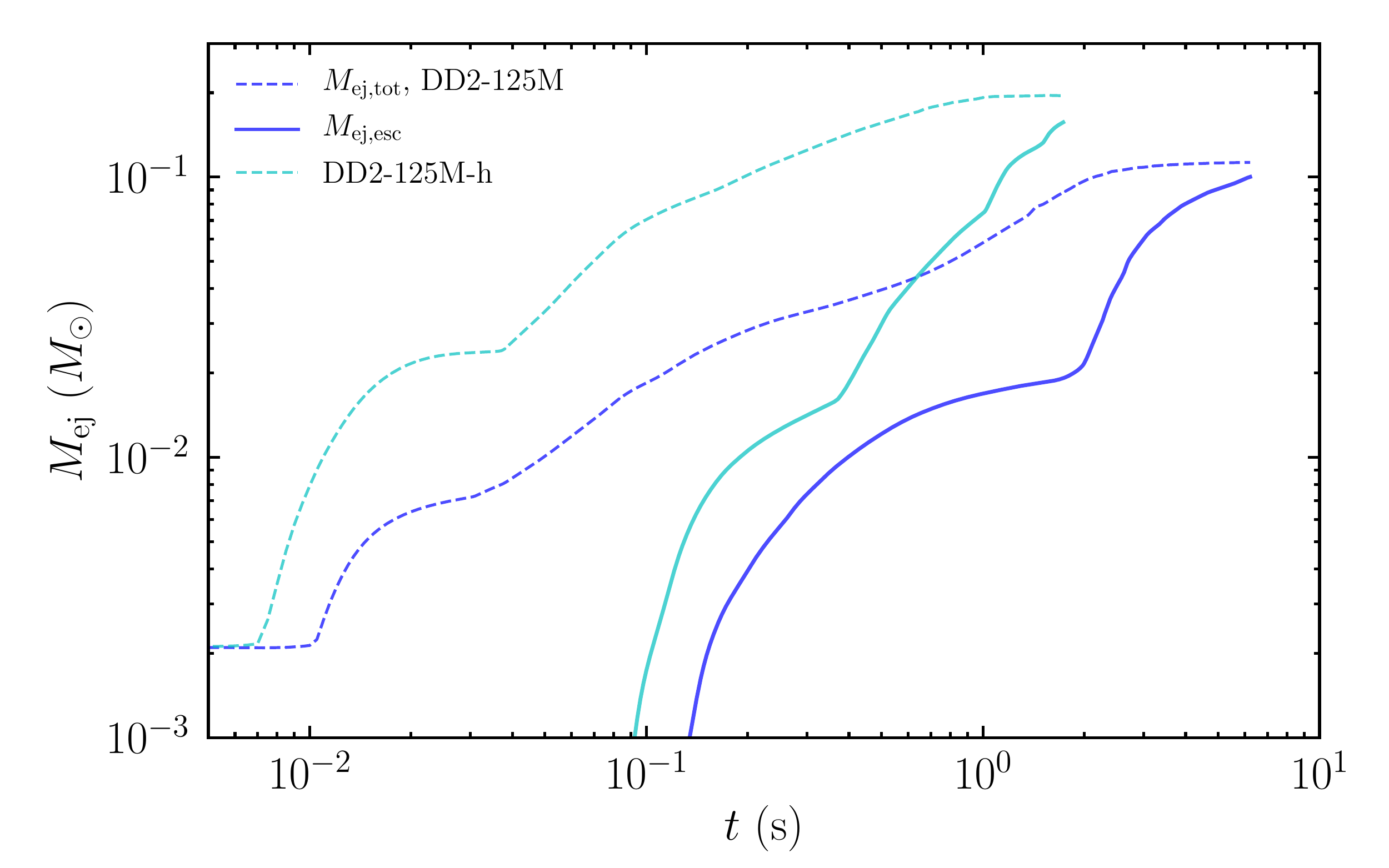}
\plottwo{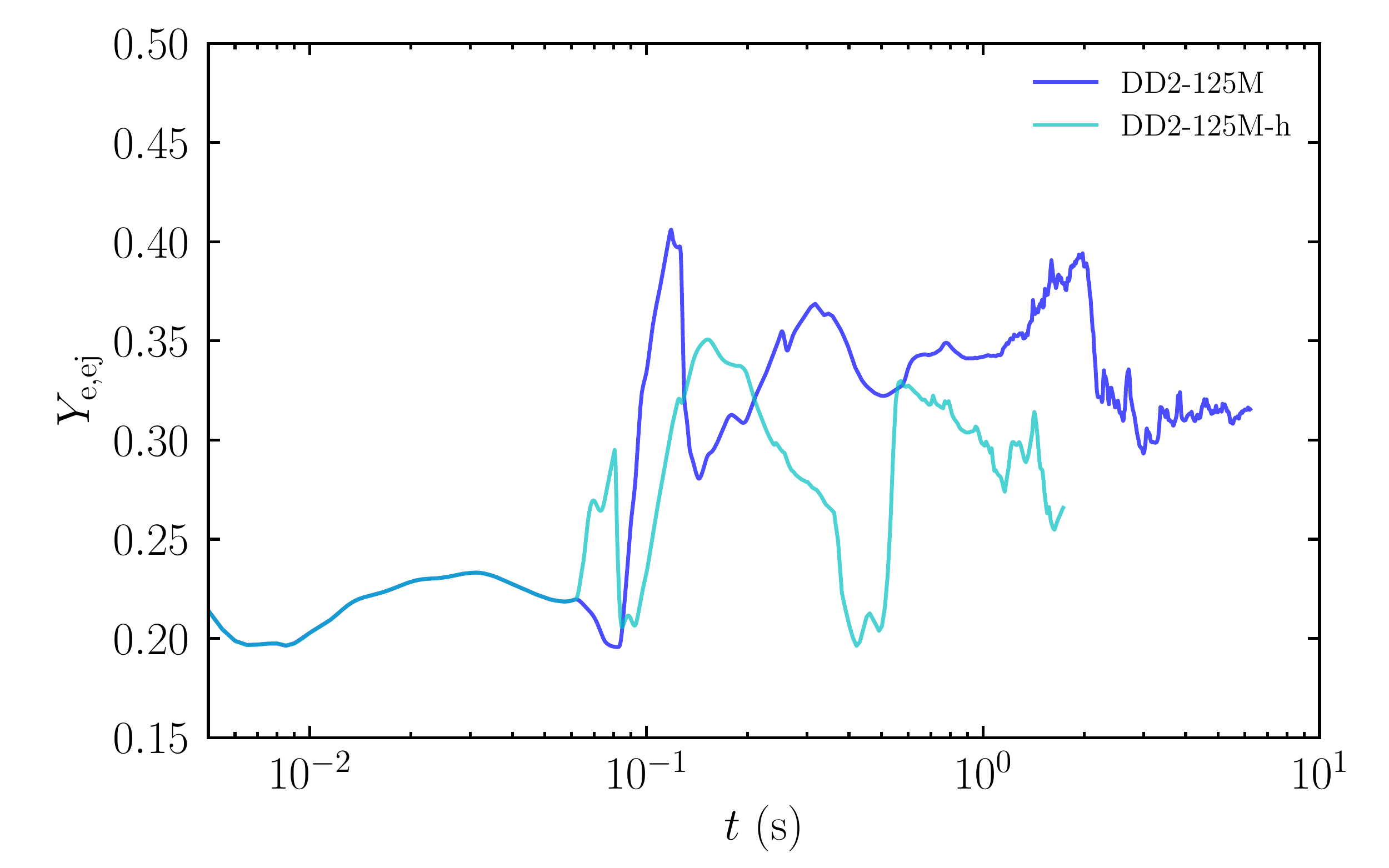}{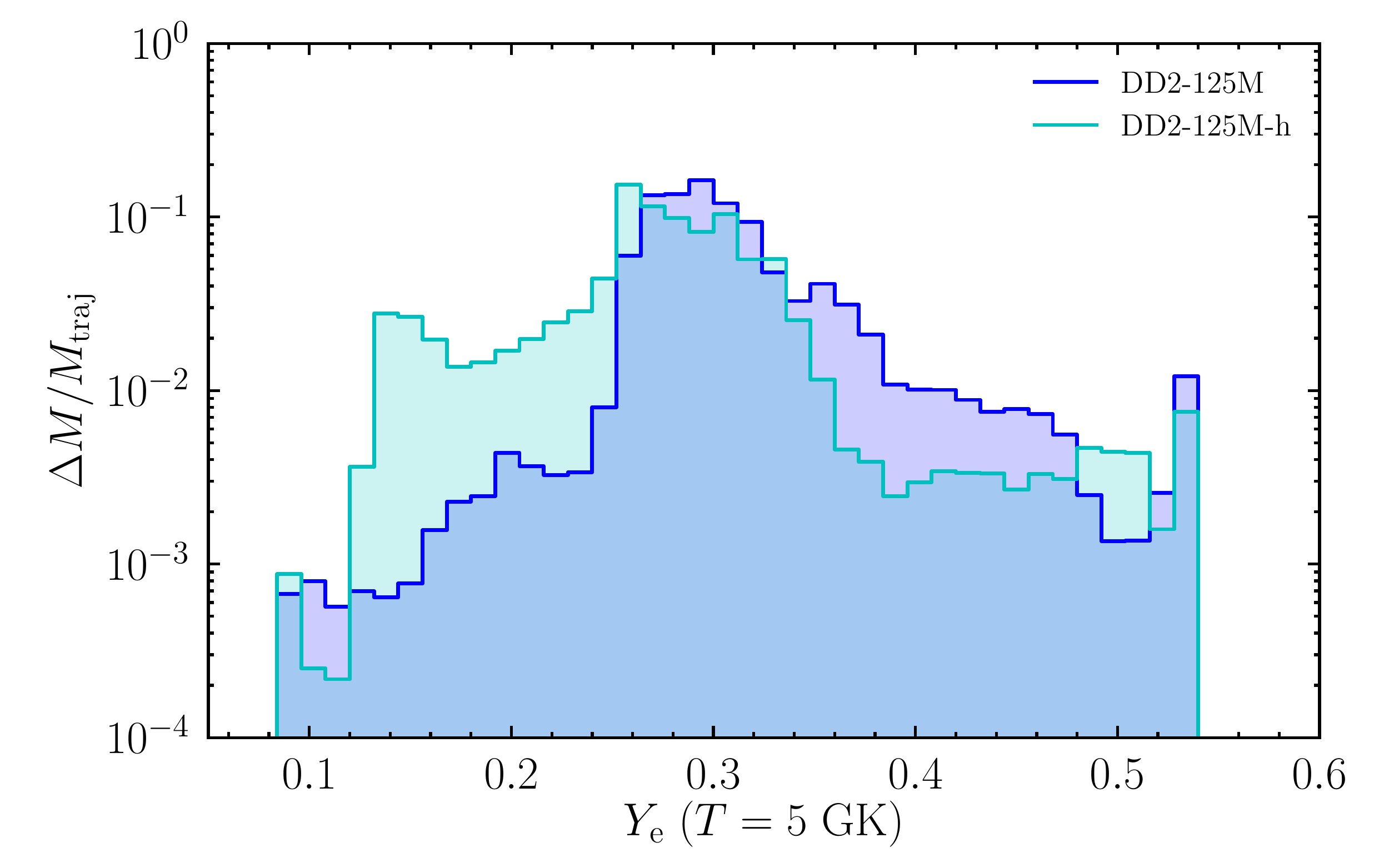}
\caption{The ratios of neutrino luminosities to viscous heating rates (top-left), ejecta masses (top-right), electron fractions (bottom-left), and histograms of the electron fraction (bottom-right) for models DD2-125M and DD2-125M-h.
}
\label{fig:highv}
\end{figure*}

Because the equilibrium value of the electron fraction changes steeply in the temperature region of $k_\mathrm{B}T \approx 1$--2\,MeV, at which the freeze-out of weak interaction occurs, the electron fraction of the ejecta should be affected significantly by the expansion timescale.
In the presence of strong viscous effect with $\alpha_\mathrm{vis}>0.04$ or the other short-timescale mass ejection processes (for example the Lorentz force by aligned magnetic fields; see \S \ref{subsec:mag}), the electron fraction of the ejecta would become lower (see \S \ref{subsec:highv}).

In the bottom panel of Fig.~\ref{fig:timescale}, the specific heating rates due to the viscous dissipation and neutrino absorption are shown together with the specific cooling rate due to the neutrino emission.
We note that the cooling rate shown in this figure is equal to the leakage rate in Eq.~\eqref{eq:rad}, in which the neutrino trapping is taken into account.
The heating rate is suppressed in the optically-thick region in our leakage-based neutrino radiation transfer method because a part of the neutrino absorption is assumed to balance with the neutrino emission (see \cite{fujibayashi2017a} for a detailed description of our method).
As a result, the heating rate for $k_\mathrm{B}T\gtrsim 2$\,MeV decreases with increasing temperature (i.e., high-density and high-optical depth region).
This figure illustrates that the viscous heating approximately balances the neutrino cooling for $k_\mathrm{B}T\gtrsim 1$ MeV (i.e., for the early phase of the evolution), while the viscous heating dominates over the neutrino cooling for $k_\mathrm{B}T\lesssim 1$ MeV (i.e., for the later phase after the neutrino cooling efficiency drops).
The late-time viscosity-driven mass ejection occurs in such conditions.

\subsection{DD2-125M-h: Higher Viscosity Model}\label{subsec:highv}

\begin{figure}
\epsscale{1.17}
\plotone{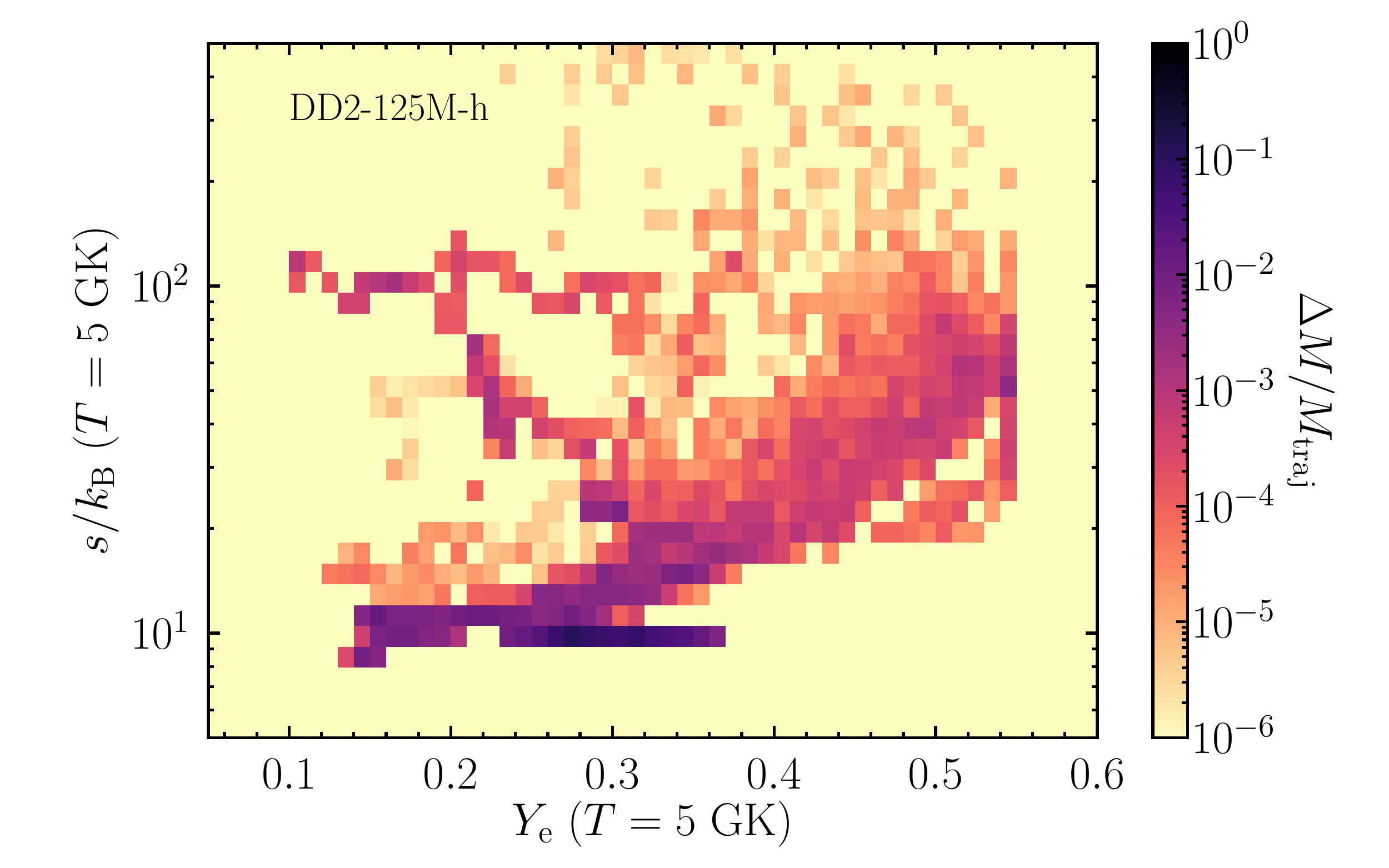}
\caption{Mass distribution of the ejecta in the $s$--$Y_\mathrm{e}$ plane for DD2-125M-h.
}
\label{fig:sye_highv}
\end{figure}

To explore the dependence of the results on the viscous coefficient (i.e., the timescale of the viscous processes), we perform a simulation of model DD2-125M-h, for which a higher viscous parameter ($\alpha_\mathrm{vis}$=0.1) is employed.
In the top-left panel of Fig.~\ref{fig:highv}, the ratio of neutrino luminosity to viscous heating rate for model DD2-125M-h is compared to that for model DD2-125M.
Due to the faster viscous expansion of the disk, the neutrino cooling efficiency drops and the mass ejection sets in earlier than for DD2-125M.
In addition, due to the larger viscous effect, the ejecta mass becomes larger: The total ejecta mass for this model is $M_\mathrm{ej,tot}\approx 0.2\,M_\odot$ at the end of the simulation, which is approximately twice as large as that for DD2-125M.
The reason for this enhancement is that the viscous timescale is shorter, and hence, a substantial fraction of material is accelerated outward by the neutrino and viscous heating before the material accretes onto the NS.

Due to the earlier mass ejection from the disk, the equilibrium value of the electron fraction at its freeze out ($t\approx 1$\,s for DD2-125M-h and $t\approx 2$\,s for DD2-125M) becomes lower \citep{Fujibayashi2020a}, and thus, the electron fraction of the ejecta becomes lower as well (see the bottom panels of Fig.~\ref{fig:highv}).
The bottom-right panel of Fig.~\ref{fig:highv} shows that the mass fraction with low electron fraction ($Y_\mathrm{e}<0.25$) for model DD2-125M-h is remarkably larger than for model DD2-125M.

Figure~\ref{fig:sye_highv} shows the mass distribution of the ejecta in the $s$--$Y_\mathrm{e}$ plane for model DD2-125M-h.
It is clearly found that the ejecta with such low-electron fraction is present in the low-entropy ($s/k_\mathrm{B}\sim 10$) region.
This illustrates that the electron fraction of the ejecta is sensitive to the onset time of the mass ejection.
As we discuss in \S~\ref{subsec:mag}, the mass ejection may proceed in a shorter timescale in the presence of strong magnetic fields.
The ejecta mass may be enhanced and the electron fraction of the ejecta may be decreased in the presence of such an efficient mass ejection process.

\subsection{Discussion: Possible Effect of the magnetic field in the post-merger evolution} \label{subsec:mag}

\begin{figure}
\epsscale{1.17}
\plotone{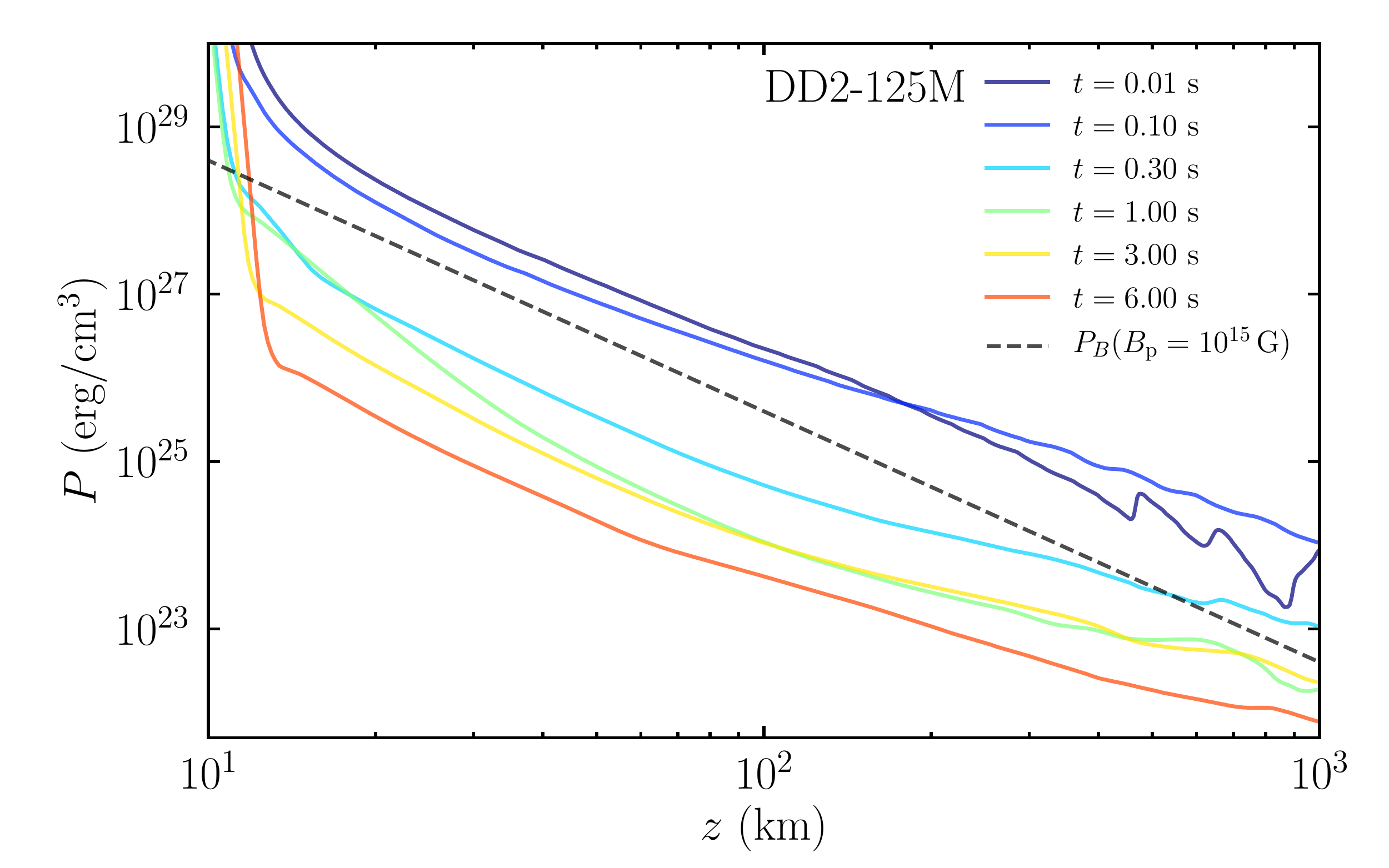}
\caption{Pressure profile along the pole of DD2-125M model at $t=0.01$, 0.1, 0.3, 1.0, 3.0, and 6.0 s.
The magnetic pressure assuming $B_\mathrm{p}=10^{15}$ G is also shown in a dashed line.
}
\label{fig:mag}
\end{figure}

The remnant NS and disk formed after the NS mergers would be strongly magnetized due to magnetohydrodynamical instabilities like the Kelvin-Helmholtz instability \citep{Price2006a} and the MRI \citep{Balbus1991a}, and thus the magnetic-field effect could play an important role for determining the ejecta dynamics and the activity of the remnant.
In this subsection, we discuss the possible effects of the magnetic field in its presence.

Figure~\ref{fig:dyn} shows that the density in the polar region decreases with time.
This indicates that the magnetic pressure by the strong magnetic field would modify the fluid dynamics in such a region.
Figure~\ref{fig:mag} shows the pressure profile for model DD2-125M along the polar direction for $t=0.01$--6 s together with the hypothetical magnetic pressure $P_B = B^2/8\pi$ assuming that the magnetic field has a dipole structure with the magnetic strength at the polar surface of the NS as $B_\mathrm{p}=10^{15}$\,G.
This figure suggests that the pressure along the polar direction substantially decreases after $\approx 0.3$\,s and that the magnetar-level magnetic field could determine the fluid dynamics along the pole: nearly force-free magnetosphere is likely to be produced.
Using the result of a force-free electromagnetic simulation by \cite{Spitkovsky2006a}, the electromagnetic luminosity is
\begin{align}
L &\approx \frac{1}{4} B_\mathrm{p}^2 R^6 \Omega^4 \notag \\
&= 6.6\times 10^{49}\,\mathrm{erg/s}\notag \\ &\times \biggl(\frac{B_\mathrm{p}}{10^{15}\,\mathrm{G}}\biggr)^2\biggl(\frac{R_\mathrm{NS}}{15\,\mathrm{km}}\biggr)^6\biggl(\frac{\Omega}{5000\,\mathrm{rad/s}}\biggr)^4.
\end{align}
Because the force-free condition would be satisfied only in the polar region, a part of this value would be released toward the polar direction.
If $\gtrsim 10$\% of $L$ is released, the long-lived and highly magnetized remnant NS, which should be rapidly rotating, could drive a high-Lorentz factor outflow and be an engine of gamma-ray bursts because the baryon loading problem is likely to be naturally avoided and a funnel structure would be helpful for confining the relativistic jet.

Because of the differential rotation of the remnant system, the toroidal magnetic field would be amplified by the winding effect.
It would result in the increase of magnetic pressure, and the development of tower-like outflow along the rotational axis \citep{Meier1999a}.
The electromagnetic luminosity due to this mechanism would be
\begin{align}
L &\sim 10^{50}\,\mathrm{erg/s}\notag \\ &\times \biggl(\frac{B}{10^{15}\,\mathrm{G}}\biggr)^2\biggl(\frac{R_\mathrm{NS}}{10\,\mathrm{km}}\biggr)^3\biggl(\frac{\Omega}{10^4\,\mathrm{rad/s}}\biggr)\biggl(\frac{\delta}{0.1}\biggr)^3,
\end{align}
where $(\delta R_\mathrm{NS})^3$ is an effective volume for which the magnetic-field amplification occurs.
At the same time the material at the polar surface of the NS would be stripped by the strong magnetic field, developing outflow toward the polar direction \citep{Shibata2011b}.

The effective viscosity may not be the only magnetohydrodynamical effect on the disk evolution.
If there is sufficiently strong aligned poloidal magnetic field in the disk, the Lorentz force would expel the disk material, in particular the neutron-rich material located in the region far from the central region, in a shorter timescale than the viscosity.
The material ejected by such an effect could be neutron-rich if its ejection timescale is short enough for the electron fraction of the material to freeze out earlier \citep{Fujibayashi2020a,Fernandez2019a,Siegel2018a}.

\section{Nucleosynthesis}
\label{sec:nucleosynthesis}

\subsection{Nuclear Reaction Network}
\label{subsec:network}

\begin{figure*}
\epsscale{1.17}
\plottwo{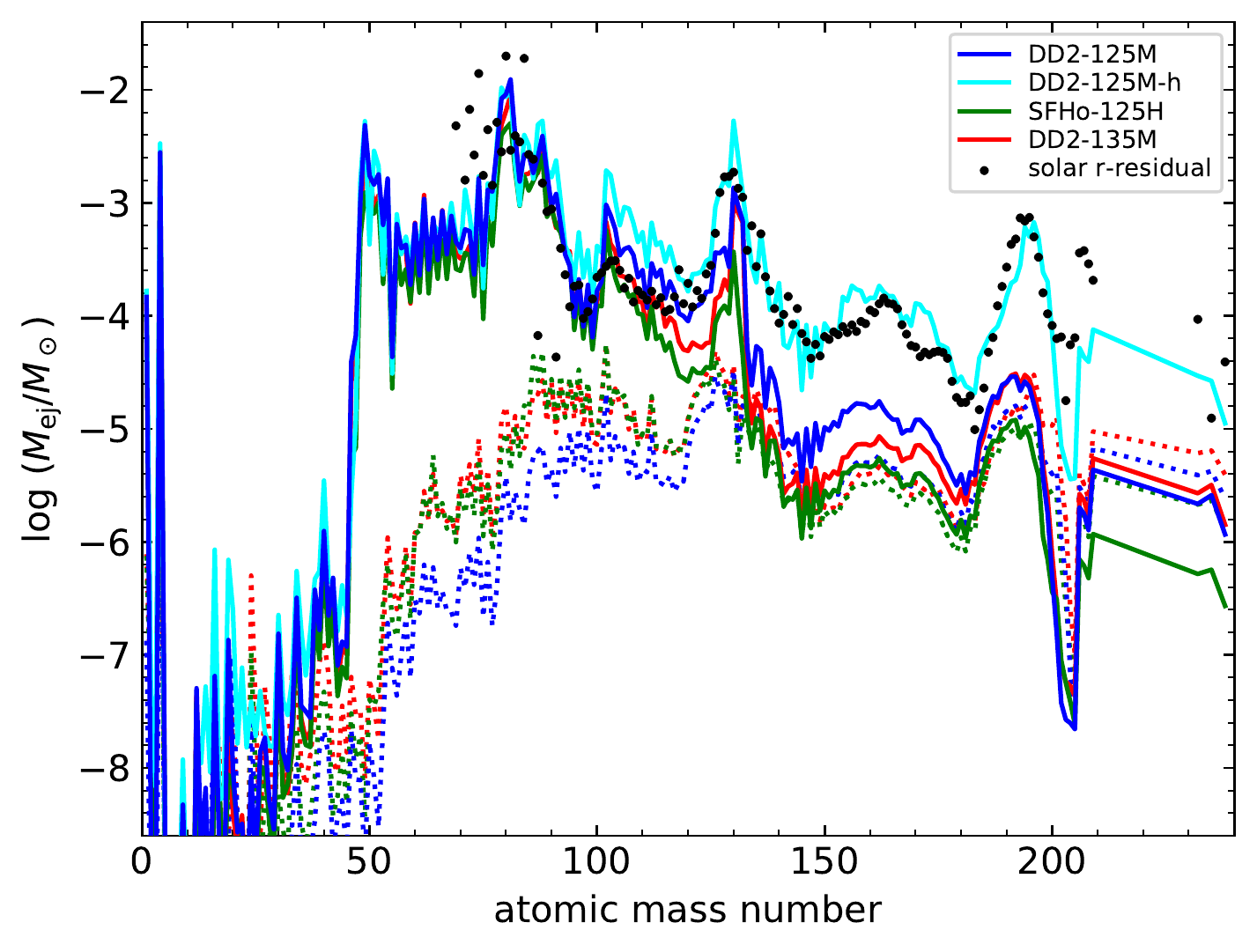}{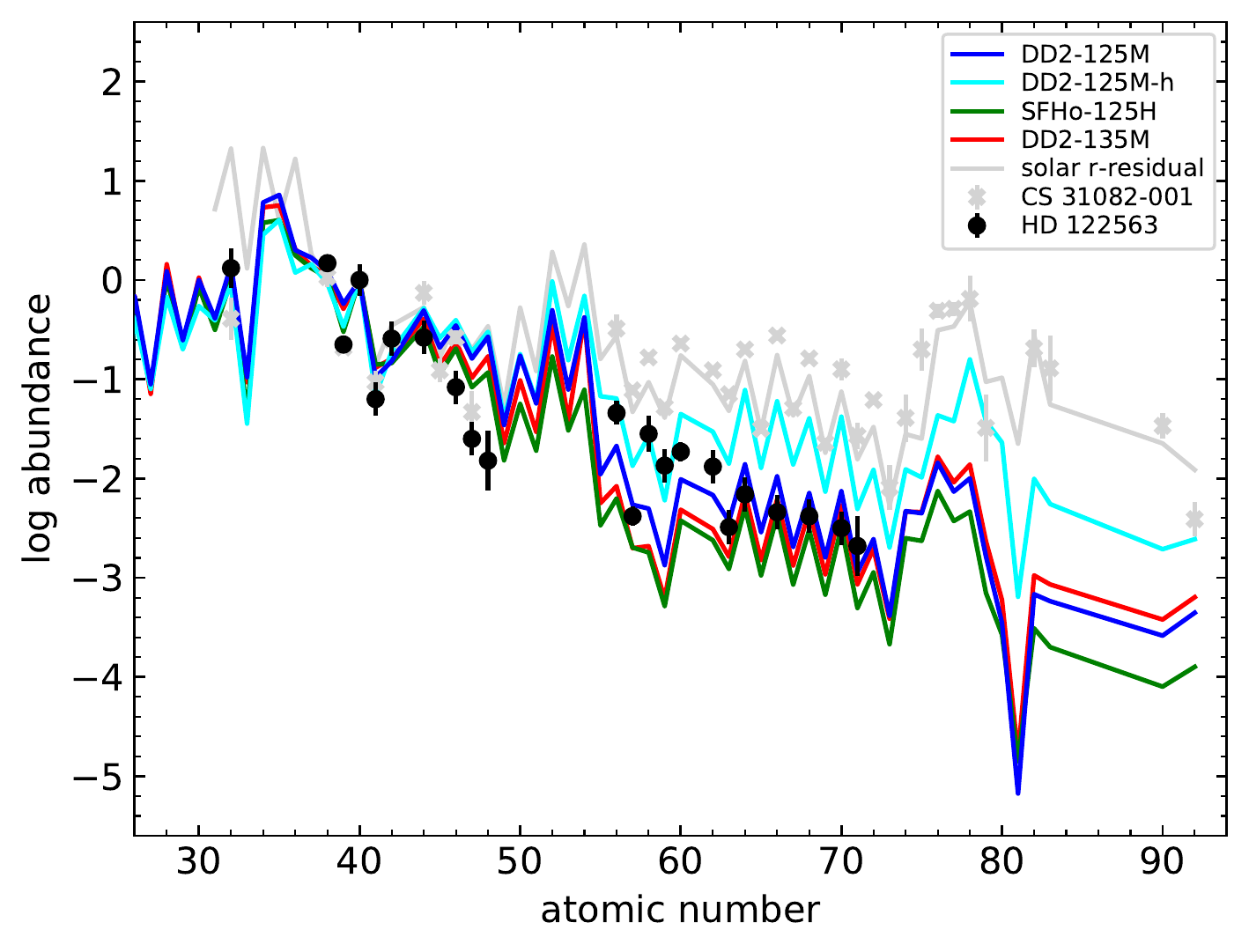}

\caption{Left: masses (solid lines; in units of $M_\odot$) of nucleosynthesis products at the end of simulation as functions of atomic mass number for models DD2-125M (blue), DD2-125M-h (cyan), SFHo-125H (green), and DD2-135M (red). The solar $r$-residual masses \citep{Prantzos2020a} also are plotted for comparison purposes, in which the values are shifted to match the mass of $^{82}$Se in DD2-135M. The dotted lines with different colors show the masses of dynamical ejecta for corresponding models computed in 3D. Right: comparison of the abundance patterns for explored models and the stellar abundances of CS~31082-001 \citep[crosses,][]{Siqueira2013a} and HD~122563 (circles, \citealt{Honda2006a}; Ge from \citealt{Cowan2005a}; Cd and Lu from \citealt{Roederer2012b}). Also plotted is the solar $r$-residual pattern \citep[gray line,][]{Prantzos2020a}. Each abundance distribution is normalized with respect to Zr ($Z = 40$).
}

\label{fig:aabun}
\end{figure*}

Nucleosynthetic yields for each tracer particle (\S~\ref{subsubsec:tracer}) are calculated in a post-processing step by using a nuclear reaction network code, \texttt{rNET}, as described in \citet{Wanajo2018a}. The network consists of 6300 isotopes ($Z = 1$--110) that are connected by experimentally evaluated rates when available (JINA REACLIB V2.0,\footnote{https://groups.nscl.msu.edu/jina/reaclib/db/index.php.} \citealt{Cyburt2010a}; Nuclear Wallet Cards\footnote{http://www.nndc.bnl.gov/wallet/}) and those from theoretical models otherwise. Theoretical rates for neutron, proton, and alpha capture \citep[TALYS,][]{Goriely2008a} and $\beta$-decay \citep[GT2,][]{Tachibana1990a} are based on a microscopic nuclear-mass prediction \citep[HFB-21,][]{Goriely2010a}.  Neutrino induced reactions are not included in the nucleosynthesis calculations, because they are expected to play only minor roles in our present models (except for setting the values of $Y_\mathrm{e}$ at $\gtrsim 10$~GK; see the middle panel of Fig.~\ref{fig:timescale}).

Each nucleosynthesis calculation starts when the temperature decreases to 10~GK with the initial composition of $1-Y_\mathrm{e}$ and $Y_\mathrm{e}$ for free neutrons and protons, respectively. At such high temperature, nuclear statistical equilibrium (NSE) immediately establishes. For a tracer particle in which the temperature does not reach 10~GK, the nucleosynthesis calculation starts from the beginning of each post-merger evolution, corresponding to $\approx 50$~ms after the merger. For these cases, therefore, the initial composition is adopted from the nucleosynthesis abundances at 50~ms in the dynamical ejecta of \citet{Wanajo2014a} with (almost) the same initial value of $Y_\mathrm{e}$, because an NSE condition is not assured (in particular for those with $< 5$~GK).

\subsection{Nuclesynthesis yields}
\label{subsec:yields}

\begin{figure}
\epsscale{1.17}
\plotone{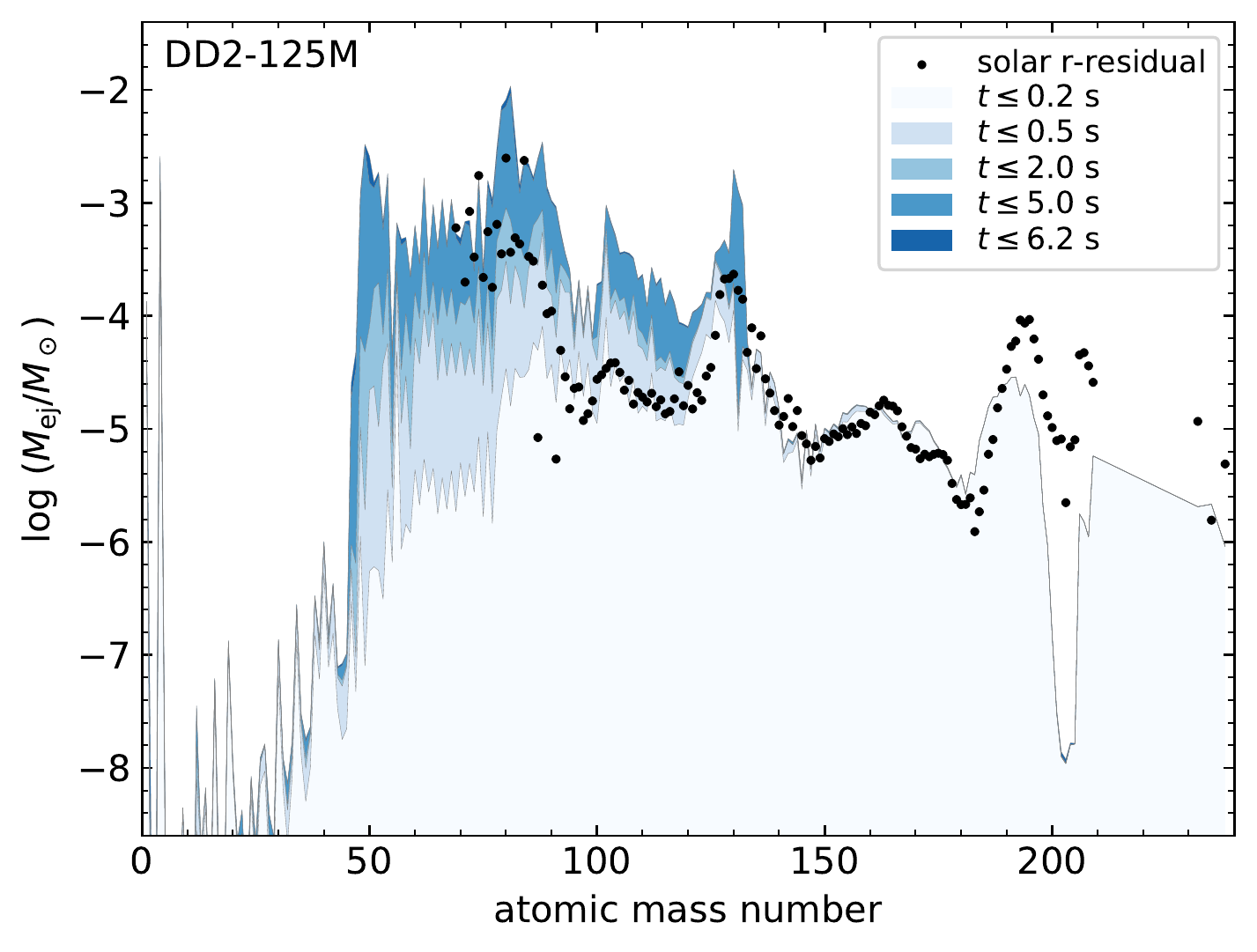}
\plotone{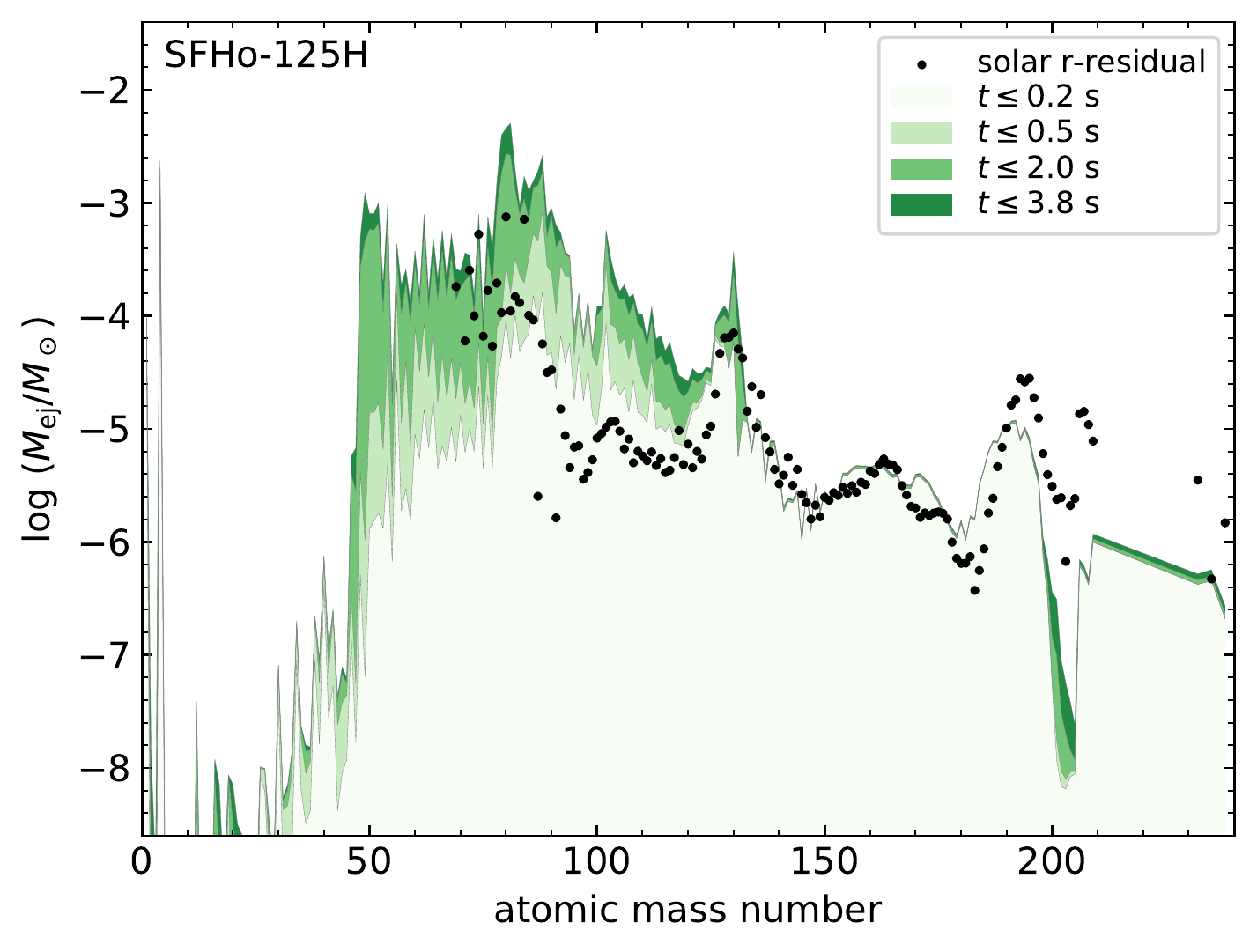}
\plotone{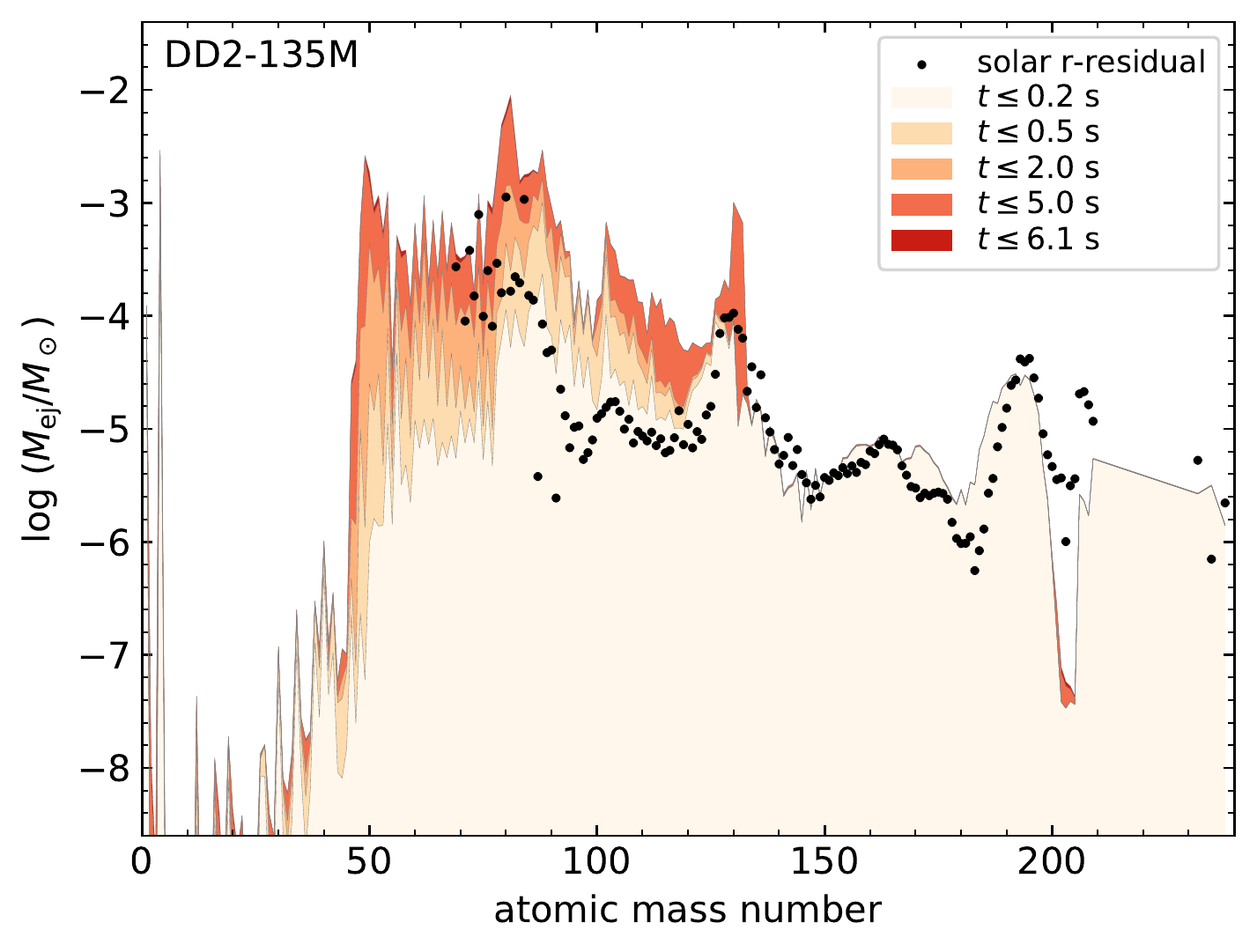}

\caption{Temporal variation of the ejecta masses (from the light to dark colors; in units of $M_\odot$) of nucleosynthesis products as functions of atomic mass number for models DD2-125M (top), SFHo-125H (middle), and DD2-135M (bottom). The solar $r$-residual masses \citep[][]{Prantzos2020a} also are plotted for comparison purposes, in which the values are shifted to match the mass of $^{151}$Eu at the end of simulation.  
}

\label{fig:time}
\end{figure}

The calculated ejecta masses of nucleosynthesis yields for models DD2-125M, DD2-125M-h, SFHo-125H, and DD2-135M are compared in the left panel of Fig.~\ref{fig:aabun} (solid lines). Here, the result for DD2-125M-h is added as a limiting case with a large value of $\alpha_\mathrm{vis} = 0.10$. The $r$-process residuals to the mass spectrum of the solar system abundances \citep[][hereafter the solar $r$-residuals]{Prantzos2020a} also are plotted, which are scaled to fit the ejecta mass of $^{82}$Se (one of the first peak abundances) in DD2-125M. Note that our 2D post-merger simulations contain the dynamical ejecta. In fact, the independent nucleosynthesis results (displayed by the dotted lines in the left panel of Fig.~\ref{fig:aabun}) by using the tracers from corresponding 3D simulations (Table~\ref{tab:3dmodel}) are in reasonable agreement with those from 2D simulations (solid lines, except for DD2-125M-h) for the heavy $r$-process nuclei ($A > 130$). For DD2-125M, the masses of these nuclei in 2D (blue solid line) are about a factor of a few larger than those in 3D (blue dotted line), indicating that the marginally bound material in the dynamical phase is pushed out by the subsequent early viscosity-driven outflows (\S~\ref{subsec:dyn}). It is interesting to note that the resultant abundance patterns (except for DD2-125M-h) are quite similar to those in the ejecta from BH accretion disks (expected to be formed after massive binary NS mergers) explored in \citet{Fujibayashi2020a}.

Overall, the abundance patterns are similar among the models with the same viscous parameter (i.e., except for DD2-125M-h), being independent of NS masses and EOSs adopted. While the abundances lighter than $A = 110$ are in good agreement with the solar $r$-residual pattern, the heavier nuclei are sizably underabundant (for $\alpha_\mathrm{vis} = 0.04$). This is a consequence of the fact that the relatively low entropies ($\sim 10$--20 $k_\mathrm{B}$; Fig.~\ref{fig:yesv}) and mild neutron-richness ($Y_\mathrm{e}\sim 0.25$--0.5, Fig.~\ref{fig:ye}) in the bulk of post-merger ejecta give the nucleosynthesis-relevant conditions for either of NSE, nuclear quasi-statistical equilibrium (QSE), or only a weak $r$-process \citep[e.g.,][]{Wanajo2011a, Wanajo2018a}.
Conversely, model DD2-125M-h results in the solar $r$-process-like abundance pattern because of its relatively neutron-rich ejecta ($Y_\mathrm{e}\sim 0.1$--0.35, bottom-right panel in Fig.~\ref{fig:highv}).

The outcomes for the models with our fiducial choice of $\alpha_\mathrm{vis} = 0.04$ (DD2-125M, SFHo-125H, and DD2-135M) conflict with the robustness of elemental abundance patterns over a wide range of atomic number (in particular for $Z \ge 56$ and lesser extent for $Z > 38$) among all the $r$-process-enhanced (or ``main" $r$-process) stars in the Galaxy \citep[e.g.,][]{Cowan2019a}. In the right panel of Fig.~\ref{fig:aabun}, the measured abundances of such a star, CS~31081-001 \citep[crosses,][]{Siqueira2013a}, are compared with those of our models, along with the solar $r$-residual pattern \citep[gray line,][]{Prantzos2020a}, which are normalized with respect to Zr ($Z = 40$). We find that the calculated abundance patterns except for DD2-125M-h are rather consistent with that of HD~122563 \citep[circles,][]{Honda2006a,Cowan2005a,Roederer2012b}, one of $r$-process-deficient metal-poor stars showing a descending abundance trend and referred to as a ``weak" $r$-process star \citep{Wanajo2006a}. Thus, low-mass NS binaries may be the sources of such weak $r$-process-like signatures found in metal-poor stars.\footnote{\label{foot:light} Note that the weak $r$-process-like stars observed to date exhibit no substantial enhancement of light neutron-capture elements \citep[e.g., having normal Sr/Fe values,][]{Aoki2017a}. It is not clear, therefore, if the abundances of such a star represent a single (or a few) nucleosynthesis event.} Alternatively, the viscosity in the disk should be effectively as large as $\alpha_\mathrm{vis} = 0.10$ that is adopted in DD2-125M-h, for reproducing the solar $r$-process abundance pattern.

As displayed in Fig.~\ref{fig:time}, each model with $\alpha_\mathrm{vis} = 0.04$ exhibits a solar $r$-process-like abundance pattern over a wide range of $A$ ($\sim 80$--200) only when the time elapsed for the post-merger phase is below 0.2~s. Even if a central remnant collapsed at this epoch (e.g., by a replacement of the binary mass with a larger one or the EOS with a softer one), the subsequent mass ejection from a BH accretion disk would add a substantial amount of material with $A < 130$ as found in \citet{Fujibayashi2020a}. We conclude, therefore, the binary NS systems explored in this study, that is, those ejecting small ($\sim 0.001\, M_\odot)$ and large ($> 0.05\, M_\odot)$ masses in the dynamical and post-merger phases, respectively, cannot be the predominant source of the Galactic $r$-process material, given that our choice of $\alpha_\mathrm{vis} = 0.04$ is representative. This implies that the frequency of the low-mass binary NS mergers, leading to long-lived massive NSs as remnants, would be rather low, given that the binary NS mergers were the predominant site of the $r$-process nucleosynthesis.
However, if the viscous effect is effectively large with $\alpha_\mathrm{vis} \sim 0.1$, or in other words, in the presence of an efficient mass ejection process in the early post-merger phase, this conclusion is significantly modified, as we already mentioned in the previous paragraph.

\subsection{Radioactive energies}
\label{subsec:radio}

\begin{figure*}
\epsscale{1.17}
\plottwo{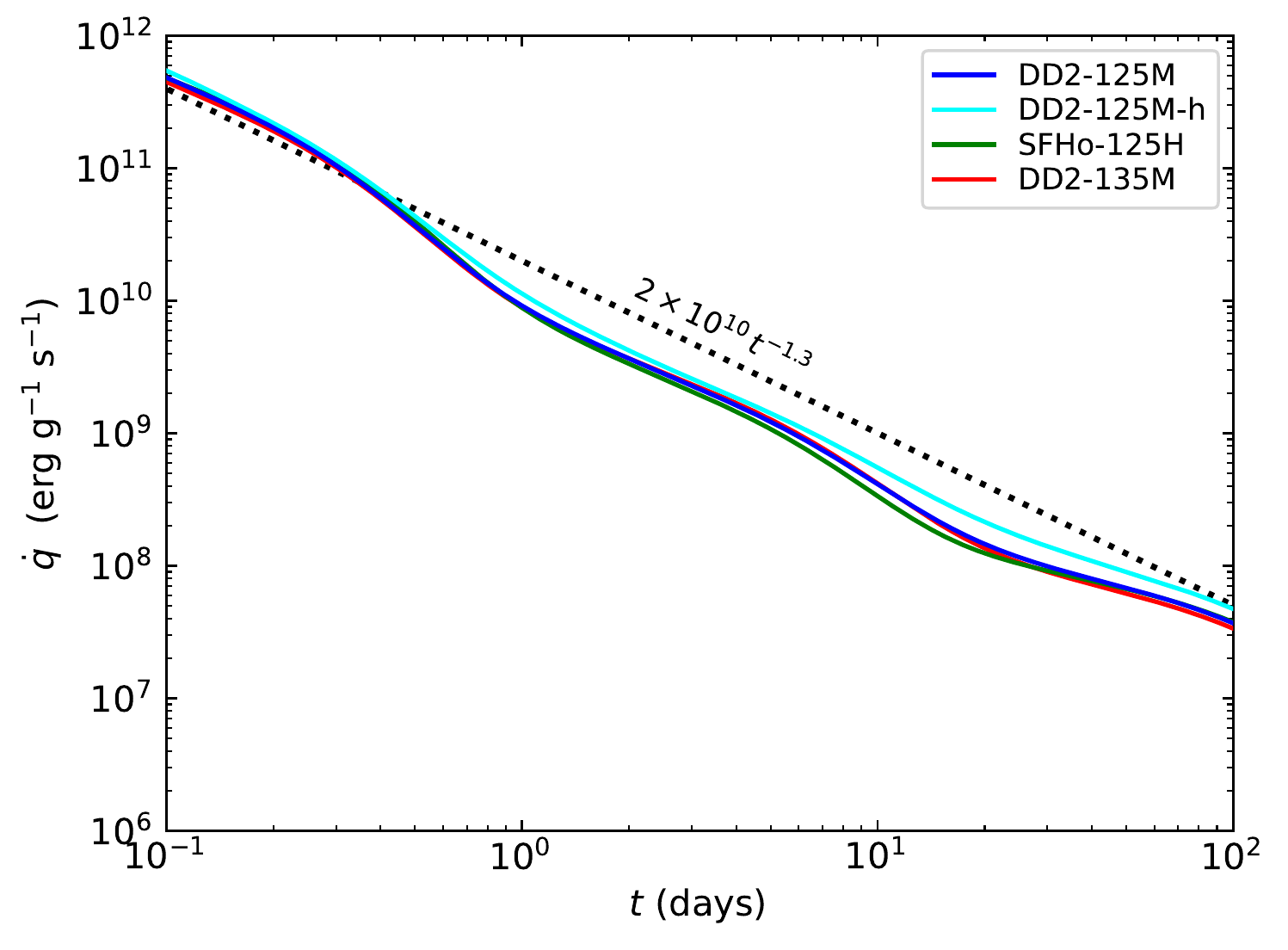}{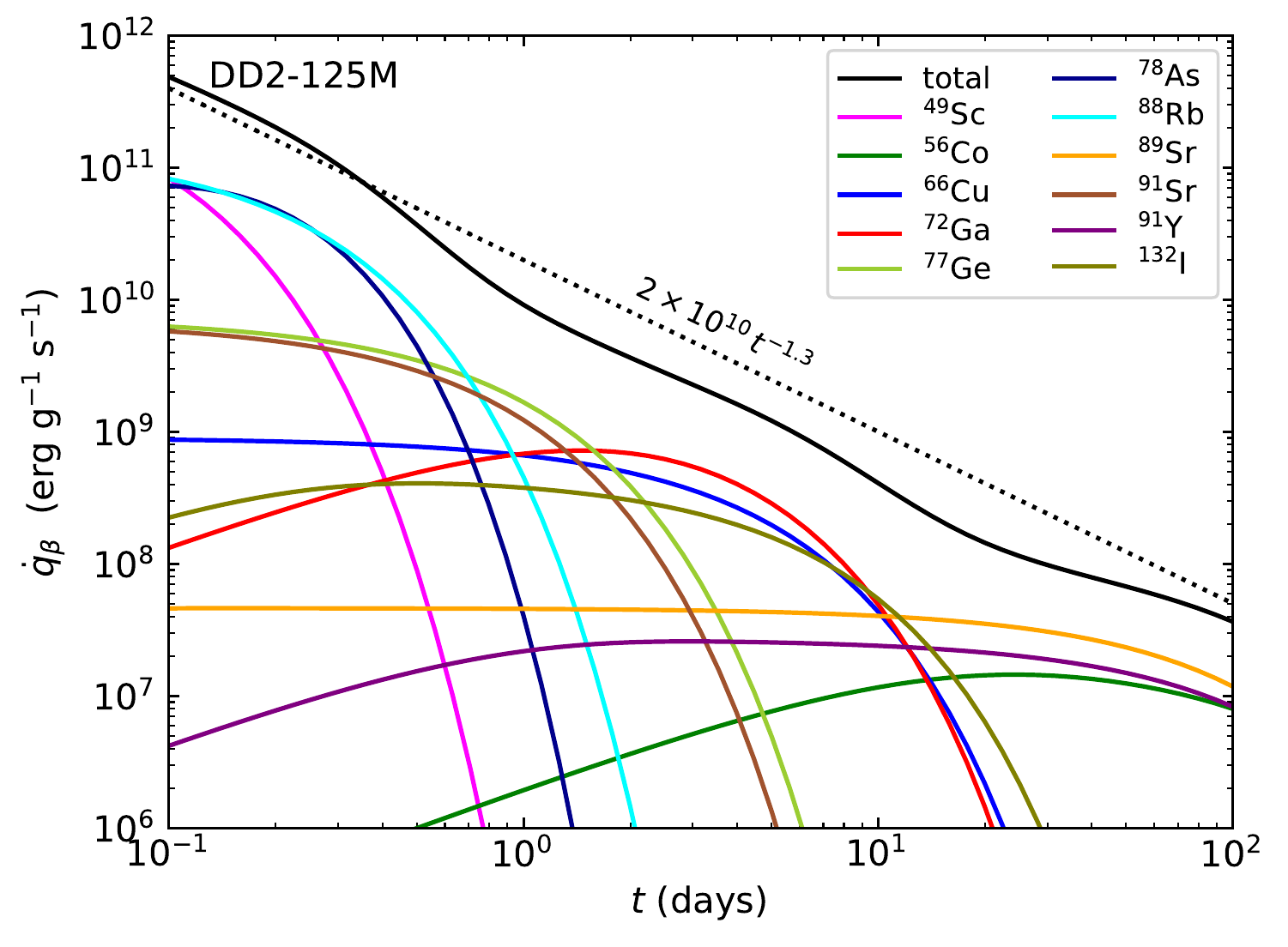}

\caption{Left: radioactive heating rates, $\dot{q}$, (in units of erg~g$^{-1}$~s$^{-1}$) in the ejecta for models DD2-125M (blue), DD2-125M-h (cyan), SFHo-125H (green), and DD2-135M (red). None of those rates follows the empirical (power-law type) heating rate indicated by the dotted line. Right: heating rates from the $\beta$-decays of individual isotopes, $\dot{q}_\beta$, (in units of erg~g$^{-1}$~s$^{-1}$) for DD2-125M. For readability, only the top 11 isotopes that have more than about 10\% contributions at the maxima are displayed in different colors. The black solid and dotted lines indicate the total and the empirical power-law type rates, respectively.
}

\label{fig:qdot}
\end{figure*}

We inspect the radioactive decays of the nuclei synthesized in the post-merger ejecta, which give rise to kilonova emission. For this purpose, the nuclear (specific) heating rates ($\dot{q}$; in units of erg~g$^{-1}$~s$^{-1}$) are computed from temporal evolution of nucleosynthesis for DD2-125M (blue), DD2-125M-h (cyan), SFHo-125H (green), and DD2-135M (red) as displayed in the left panel of Fig.~\ref{fig:qdot} (solid lines). The heating is chiefly due to $\beta$-decays; the contributions from fission and $\alpha$-decays are unimportant except for DD2-125M-h because of the small amount of trans-lead species synthesized in the ejecta (Fig.~\ref{fig:aabun}). The heating rates for all the models resemble each other, in particular for DD2-125M, SFHo-125H, and DD2-135M, because of their similar nucleosynthesis outcomes. However, these deviate from the empirical power-law like heating (with the index $-1.3$; dotted line) that has been found in previous studies as a result of the decaying second peak nuclei ($A \sim 130$) with various half-lives \citep{Metzger2010a, Wanajo2014a}. 

The reason for this deviation is evident from the right panel of Fig.~\ref{fig:qdot} that shows the heating rates from the $\beta$-decays of individual isotopes, $\dot{q}_\beta$, which have dominant contributions to the total heating rate, $\dot{q}$ (for DD2-125M as representative). The main contributors are those between the iron group and the first peak nuclei ($A \sim 50$--90) that dominate in the nucleosynthesis yields (Fig.~\ref{fig:aabun}). Only two second peak nuclei ($A = 131$ and 132) marginally contribute to the heating rates. Most of these isotopes (listed in Table~\ref{tab:decay}) have the half-lives (second column) either of a few hours, a few days, or several 10 days. As a result, these nuclei contribute to heating at different duration, that is, for $t < 1$~day, $t = 1$--10~days, and $t > 10$~days, making three bump-like structures in the curve of $\dot{q}$ (black solid line). For instance, the contributions of the two decay chains $^{66}$Ni$\rightarrow^{66}$Cu$\rightarrow^{66}$Zn and $^{72}$Zn$\rightarrow^{72}$Ga$\rightarrow^{72}$Ge with similar half-lives ($\sim 2$~days; Table~\ref{tab:decay}) stand out in the heating rate between 1 and 10~days \citep[as also found in][]{Wanajo2018b}. DD2-125M-h results in the larger heating rate than those in the other models because of the abundant 2nd peak elements in its ejecta coming into play.

\subsection{Comparison with the kilonova light curve of the NS merger GW170817}
\label{subsec:kilonova}

\begin{figure*}
\epsscale{1.17}
\plottwo{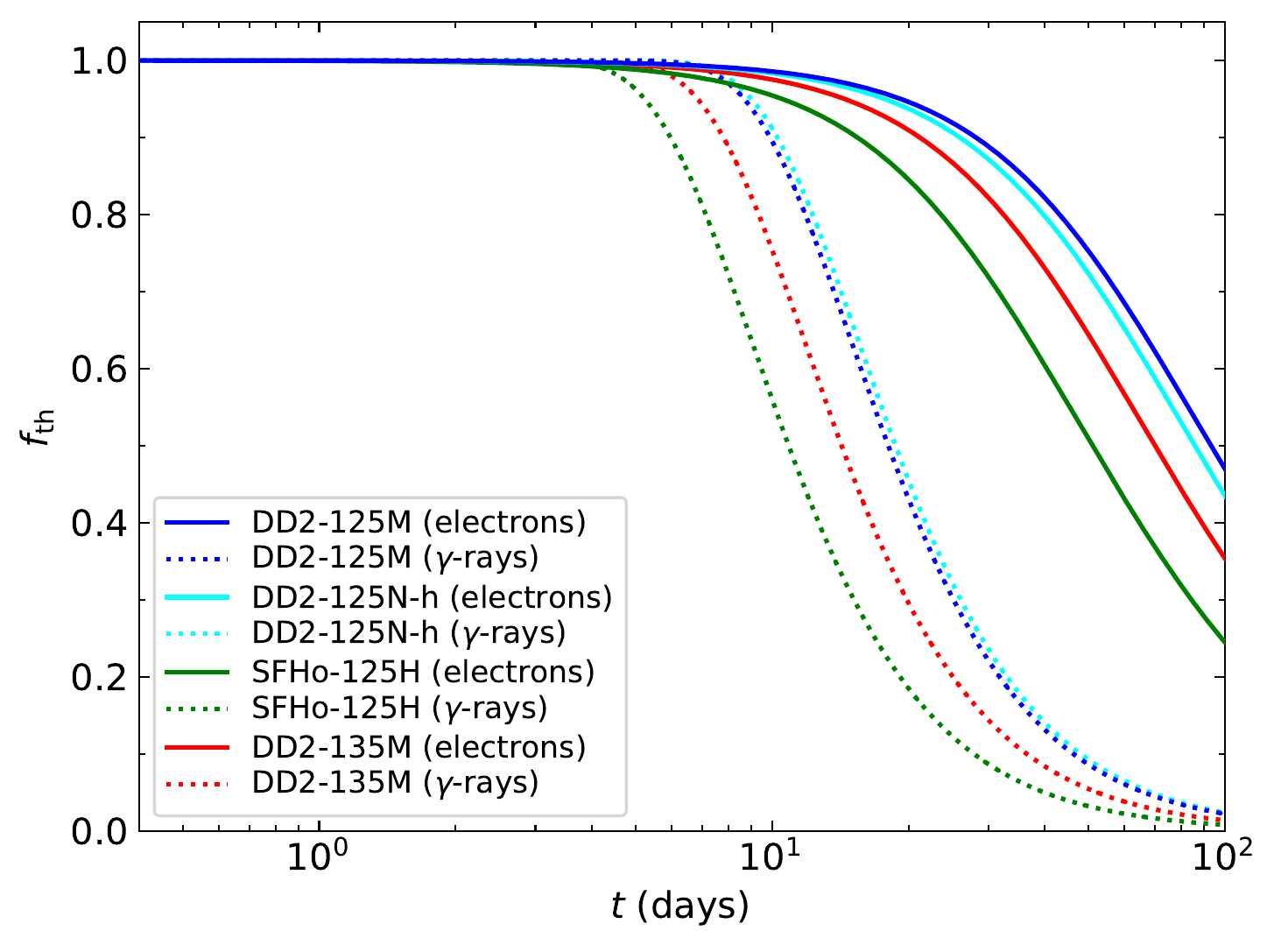}{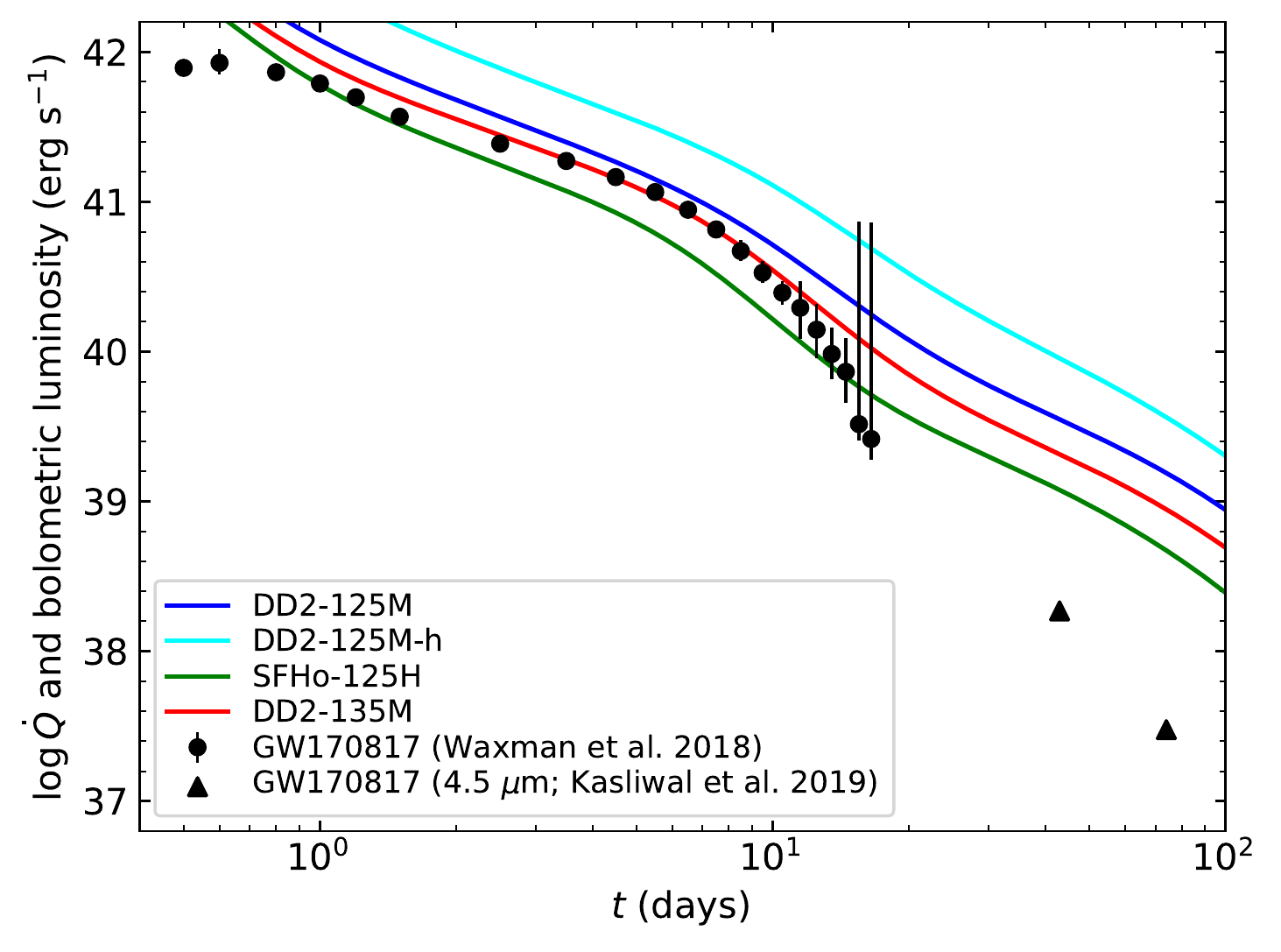}

\caption{Left: thermalization factors, $f_\mathrm{th}$, as functions of time for electrons (solid lines) and $\gamma$-rays (dotted lines) by using the analytical formula in \citet{Barnes2016a} with the ejecta masses ($M_\mathrm{ej,tot}$) and average velocities ($\langle V_\mathrm{ej} \rangle$) for models DD2-125M (blue), DD2-125M-h (cyan), SFHo-125H (green), and DD2-135M (red) listed in Table~\ref{tab:model}. Right: comparison of the total heating rates defined as $\dot{Q} \equiv M_\mathrm{ej,tot}\, f_\mathrm{th}\, \dot{q}$ (solid lines with different colors) in the ejecta (in units of erg s$^{-1}$) with the bolometric luminosities \citep{Waxman2018a} and those at 4.5~$\mu$m \citep[i.e., the lower limits,][]{Kasliwal2019a} of the kilonova associated with the NS merger GW170817. 
}

\label{fig:luminosity}
\end{figure*}

The total heating rates, $\dot{Q}$, in the ejecta for each model are calculated as the product of the specific heating rate, $\dot{q}$, the thermalization factor, $f_\mathrm{th}$, and the mass of ejecta, $M_\mathrm{ej,tot}$. The thermalization factors for the $\gamma$-rays, electrons, $\alpha$-particles, and fission fragments (the latter two are not shown) are obtained by using the analytic formula in \citet{Barnes2016a} with the ejecta mass and the average velocity estimated at the end of simulation for each model (10th and 11th columns in Table~\ref{tab:model}), as shown in the left panel of Fig.~\ref{fig:luminosity}. The factors ($f_\mathrm{th}$) for $\gamma$-rays rapidly decay after several days, while those for electrons slowly decrease at later times \citep[see also][]{Hotokezaka2020a}. In general, $f_\mathrm{th}$ is larger for more massive or slowly expanding ejecta. In our models, the average velocities ($\sim 0.1\, c$) as well as the heating rates are similar to each other and thus the ejecta mass is the main source of differences in $\dot{Q}$.

\begin{deluxetable*}{llllll}
\tabletypesize{\scriptsize}
\tablecaption{Dominant decay chains and energy partitions (MeV)\tablenotemark{a}}
\tablewidth{0pt}
\tablehead{
\colhead{Isotope} &
\colhead{} &
\colhead{Half-life} &
\colhead{Gamma-ray} &
\colhead{Electron} &
\colhead{Neutrino} 
}
\startdata
$^{66}$Ni  $\rightarrow$ $^{66}$Cu  &                          & 2.27 d & 0       & 0.0733 & 0.179\\
                                    & $\rightarrow$ $^{66}$Zn  & 5.12 m & 0.0978  & 1.07   & 1.48\\
$^{72}$Zn  $\rightarrow$ $^{72}$Ga  &                          & 1.94 d & 0.145   & 0.0805 & 0.194\\
                                    & $\rightarrow$ $^{72}$Ge  & 14.1 h & 2.77    & 0.471  & 0.759\\
$^{77}$Ge  $\rightarrow$ $^{77}$As  &                          & 11.3 h & 1.08    & 0.642  & 0.982\\
                                    & $\rightarrow$ $^{77}$Se  & 1.62 d & 0.00808 & 0.226  & 0.448\\
$^{78}$Ge  $\rightarrow$ $^{78}$As  &                          & 1.47 h & 0.278   & 0.227  & 0.450\\
                                    & $\rightarrow$ $^{78}$Se  & 1.51 h & 1.31    & 1.26   & 1.68\\
$^{88}$Kr  $\rightarrow$ $^{88}$Rb  &                          & 2.84 h & 1.95    & 0.364  & 0.606\\
                                    & $\rightarrow$ $^{88}$Sr  & 17.8 m & 0.677   & 2.05   & 2.58\\
$^{89}$Sr  $\rightarrow$ $^{89}$Y   &                          & 50.5 d & 0       & 0.585  & 0.912\\
$^{91}$Sr  $\rightarrow$ $^{91}$Y   &                          & 9.63 h & 0.707   & 0.641  & 0.992\\
                                    & $\rightarrow$ $^{91}$Zr  & 58.5 d & 0.00313 & 0.603  & 0.938\\
$^{131}$I  $\rightarrow$ $^{131}$Xe &                          & 8.03 d & 0.380   & 0.182  & 0.396\\
$^{132}$Te $\rightarrow$ $^{132}$I  &                          & 3.20 d & 0.212   & 0.0670 & 0.173\\
                                    & $\rightarrow$ $^{132}$Xe & 2.30 h & 2.26    & 0.486  & 0.837\\
\enddata
\tablenotetext{a}{Data are taken from the ENDF/B-VII.1 library \citep{Chadwick2011a}.}
\label{tab:decay}
\end{deluxetable*}

The computed curves of $\dot{Q}$ are compared with the bolometric luminosity of the kilonova \citep{Waxman2018a, Kasliwal2019a} associated with the NS merger GW170817. The right panel of Fig.~\ref{fig:luminosity} displays the results for models DD2-125M (blue), DD2-125M-h (cyan), SFHo-125H (green), and DD2-135M (red). According to the inferred lower-bound of the binary mass for GW170817  \citep[$2.73\, M_\odot$,][]{Abbot2017b}, the former three should be taken as the predictions for future low-mass binary events, while the latter may be taken as representative for this event. In fact, we find that the curve of $\dot{Q}$ for DD2-135M is in good agreement with the observed light curve (circles with error bars) between 1 and 10 days, including the bump-like signature at several days after the merger. During this epoch, the radioactive heating is predominantly due to the decay chains from $^{66}$Ni and $^{72}$Zn, which also is suggested in \citet{Wanajo2018a}.\footnote{\label{foot:steep} The steepening of the light curve can also be due to the fact that the photon diffusion wave crosses the bulk of the ejecta during the first several days \citep[e.g.,][]{Hotokezaka2020a}.} Despite the heating rate (Fig.~\ref{fig:qdot}) about a factor of two smaller than the empirical rate (dotted line) expected from a solar $r$-process-like abundance distribution \citep[e.g.,][]{Metzger2010a,Wanajo2014a}, the large $M_\mathrm{ej,tot}$ ($= 0.086\, M_\odot$) and the subsequent high $f_\mathrm{th}$ keep the $\dot{Q}$ sufficiently high.
Note that the slight overprediction of $\dot{Q}$ at $t \gtrsim 10$ days might be resolved by a photon diffusion effect \citep[][footnote~\ref{foot:steep}]{Hotokezaka2020a} or by a re-evaluation of the progressively increasing error bars at late times.
Conversely, the ejecta masses for the other models are larger or smaller to match the light curve of the merger GW170817. The behaviors of $\dot{Q}$ are, however, similar to each other.

For early times ($t < 1$~days), the radioactive energy has been lost by the adiabatic expansion of the ejecta and thus the heating rate is larger than the bolometric luminosity of the kilonova. During this epoch, the two $\beta$-decay chains near the first peak ($A \sim 80$) at the isobars $A = 78$ and 88 (Table~\ref{tab:decay} and the right panel of Fig.~\ref{fig:qdot}) play dominant roles for brightening the kilonova emission. It is interesting to note that the identification of the Sr absorption line in the ejecta of the merger GW170817 \citep{Watson2019a} may also be indicative of the contribution of the $\beta$-decay chain at $A = 88$ to the luminosity of the kilonova near its peak.

For late times ($t > 10$~days), the main contributors to the kilonova emission are also the isotopes near the first peak (except for DD2-125M-h), as a result of the successive $\beta$-decay at the isobars $A = 89$ and 91 (Table~\ref{tab:decay} and the right panel of Fig.~\ref{fig:qdot}). The relatively long half-lives of $^{89}$Sr (50.5~days) and $^{91}$Y (58.5~days) can give rise to the long-lasting kilonova emission. At this epoch, the Spitzer 4.5~$\mu$m detection (the right panel of Fig.~\ref{fig:luminosity}; triangles) places the tight lower bound, although the bolometric luminosity itself is highly uncertain. The half-lives of $^{89}$Sr and $^{91}$Y are similar to that of the spontaneous fission of $^{254}$Cf (60.5~days), which is suggested to be a dominant heating source of kilonovae at late times \citep{Wanajo2014a,Hotokezaka2016a,Zhu2018a,Wu2019a}. This would make it difficult to confirm the presence of the heaviest $r$-process nuclei, $^{254}$Cf, from the observed kilonova light curves at late times as suggested by \citet{Zhu2018a}. Note that the decay chain (successive electron capture)  of $^{56}$Ni$\rightarrow^{56}$Co$\rightarrow^{56}$Fe (with the half-lives of 6.08~days and 77.2~days) plays a subdominant role to the late-time heating because of their energy deposition exclusively in the form of $\gamma$-rays with the diminishing $f_\mathrm{th}$ for $t > 10$~days (dotted lines in the left panel of Fig.~\ref{fig:luminosity}).

In our explored models, the dynamical ejecta have negligible contributions to the radioactive heating that powers the kilonova emission. Therefore, solely the comparison of $\dot{Q}$ with the bolometric luminosity of the kilonova does not provide any constraint on the production of heavy $r$-process elements. However, the mass fraction of lanthanides in each model, $X_\mathrm{la}$, serves as another constraint. The values for the models with $\alpha_\mathrm{vis} = 0.04$, $X_\mathrm{la} \sim 0.002$--0.004 (last column in Table~\ref{tab:model}), fall within the observational value, $\sim 0.001$--0.01, in the kilonova ejecta of the merger GW170817 \citep[e.g.,][]{Waxman2018a}. This indicates that the heavy $r$-process elements synthesized in the dynamical ejecta become the dominant opacity source of the kilonova, although the amount is insufficient to reproduce the solar $r$-process-like pattern.

The similar behaviors of $\dot{Q}$ evolution among our explored models (Fig.~\ref{fig:luminosity}) imply that the low-mass binary events (to be measured in the future) likely exhibit GW170817-like kilonova light curves. The brightness depends, however, on the binary mass as well as the true EOS and viscous parameter.

\section{Summary and Conclusions}
\label{sec:summary}

We explored the post-merger evolution of binary mergers with nearly the lowest mass of double NSs, including the early dynamical mass-ejection phase for self-consistency. Equal-mass binaries with each mass of $1.25\, M_\odot$ (and $1.35\, M_\odot$ for comparison purposes) were explored, in which two different (referred to as ``stiff" and ``soft", respectively) nuclear EOSs, DD2 \citep{banik2014a} and SFHo \citep{steiner2013a}, were adopted. For both the dynamical and post-merger phases, we performed general-relativistic numerical simulations with approximate neutrino transport taken into account \citep{sekiguchi2010a,fujibayashi2017a}. 

The early dynamical phase was computed in 3D \citep[Kiuchi et al., in preparation, see also][]{Sekiguchi2015a,Sekiguchi2016a}. The spatial distributions of physical quantities at $\approx 50$~ms were mapped into the axisymmetric 2D space as the initial condition for the subsequent post-merger evolution using the viscous-hydrodynamics code described in \citet{fujibayashi2017a,fujibayashi2018a}. The viscous parameter of $\alpha_\mathrm{vis} = 0.04$ was adopted as a fiducial value, and one model of $\alpha_\mathrm{vis} = 0.10$ was added as a limiting high-viscosity case (DD2-125M-h). The massive NSs survived until the simulations had been terminated ($\sim 4$--6~s after the merger) for our representative models (DD2-125M, SFHo-125H, and DD2-135M). The mass of ejecta (unbound material at the end of simulation) was $M_\mathrm{ej,tot} = 0.06$--$0.11\, M_\odot$, being about 30\% of the initial disk masses as also found in previous studies \citep{Metzger2014a,Lippuner2017a,fujibayashi2018a}. The amount of ejecta was larger for a smaller-mass ($1.25\, M_\odot$) binary as well as for a stiffer EOS (DD2) among our models as a result of the larger (initial) disk mass. The overall ejecta properties were found to be similar to each other as characterized by $\langle V_\mathrm{ej} \rangle/c \sim 0.09$--0.11, $\langle s_\mathrm{ej} \rangle/k_\mathrm{B} \sim$ 15--19, and $\langle Y_\mathrm{e, ej} \rangle \sim$ 0.32--0.34.
For the high-viscosity model (DD2-125M-h), the ejecta was more massive ($M_\mathrm{ej,tot} = 0.2\, M_\odot$) as well as more neutron-rich ($\langle Y_\mathrm{e, ej} \rangle \sim 0.28$).
This illustrates that the onset time of the mass ejection would be one of the key quantities that characterizes the property of the ejecta.  

The nucleosynthesis yields were obtained in a post-processing step by using $\approx 10000$--40000 tracer particles deduced from each post-merger model. The resulting abundance trends were similar among all the representative models (DD2-125M, SFHo-125H, and DD2-135M) as anticipated from their ejecta properties. While the early dynamical ejecta were dominated by the heavy $r$-process nuclei with $A > 130$, the subsequent post-merger ejecta added the lighter nuclei with an order of magnitude more massive amount. As a result, the heavier $r$-process components were sizably underabundant, although the abundance distributions between $A = 80$ and 110 were in reasonable agreement with the solar $r$-process pattern. Accordingly, the mass fraction of lanthanides, $X_\mathrm{la} \sim 0.002$--0.004, is substantially smaller than those in the corresponding early dynamical component ($\sim 0.07$--0.2). By contrast, the high-viscosity model (DD2-125M-h) resulted in a solar $r$-process-like abundance pattern with a larger amount of $X_\mathrm{la}$ ($\sim 0.03$) because of the more neutron-rich ejecta.

The resultant radioactive heating, an energy source of kilonova emission, was predominantly due to the species between the iron group and first $r$-process peak elements ($A \sim 50$--90), e.g., $^{66}$Ni, $^{72}$Zn, $^{78}$Ge, $^{88}$Kr, and $^{89}$Sr, rather than those near the second peak ($A \sim 130$) as found in earlier work \citep[e.g.,][]{Metzger2010a,Wanajo2014a}. The radioactive heating rates did not exhibit power-law behavior and were a factor of a few smaller than the empirical rate of $\approx 2 \times 10^{10}\, t^{-1.3}$~erg~g$^{-1}$~s$^{-1}$ after $\sim$ 1 day. Nevertheless, the total heating rates for 1--10~days were in good agreement (DD2-135M) or even higher (DD2-125M) than the observed luminosity of the kilonova associated with GW170817 because of their large ejecta mass ($0.09\, M_\odot$ and $0.11\, M_\odot$, respectively). Except for the absolute values, the time variations of total heating rates were similar among our representative models due to a resemblance of their nucleosynthetic outcomes.

Our results of the present study as summarized above lead us to several important conclusions. First of all, the long-lived massive NSs with their lifetime longer than seconds are the common outcomes of low-mass binary NS mergers. This is confirmed by using the stiff (DD2) and soft (SFHo) EOSs, given these EOSs bracketing the properties of the true EOS. Here, we intend ``low-mass" binaries to have the system mass about $2.5\, M_\odot$. This value resides near the lower bound for the observed Galactic NS binaries \citep[e.g.,][]{Tauris2017a} such as PSR J1946+2052 \citep[$2.50 \pm 0.04\, M_\odot$,][]{Stovall2018a} as well as that predicted from the theoretical studies of core-collapse supernovae \citep[CCSNe; e.g.,][]{Mueller2016a,Suwa2018a}. As the lowest mass of a single NS obtained from such calculations is $\sim 1.2\, M_\odot$, our assumption of taking equal-mass NSs may be justified.

It is also concluded that the mergers of such low-mass NSs are very rare in reality, given our choice of $\alpha_\mathrm{vis} = 0.04$ (or less) being realistic, because the nucleosynthetic products in our representative models are dominated only by the light $r$-process nuclei with $A < 100$--130. To date, all the $r$-process-enhanced stars in the Galactic halo \citep[e.g.,][]{Cowan2019a} as well as in the ultra-faint dwarf galaxies Reticulum~II \citep{Ji2016a}, Tucana~III \citep{Hansen2017a}, and Grus~II \citep{Hansen2020a} exhibit the abundance distributions that closely follow the solar $r$-process pattern. Provided that such abundance signatures recorded the nucleosynthetic histories of single NS events \citep[e.g.,][]{Ishimaru2015a,Ojima2018a}, the majority of NS mergers would also have the solar $r$-process-like abundance distributions in their ejecta. This may exclude a possibility that the binaries resulting from the CCSNe near the low-mass end of progenitors are the main channel for binary NS mergers \citep[e.g.,][]{Podsiadlowski2004a}.

Although they are expected to be rare, such low-mass NS binaries do exist in the Galaxy as evidenced by the discovery of the double NS system, J1946+2052 \citep[][]{Stovall2018a}.
Therefore, the mergers of low-mass NSs may be detected in the future as well. It is expected that the kilonovae of such events share similar properties with those of GW170817, that is, the bright blue emission followed by the fading red component as well as the steepening of the light curve at several days after the merger \citep{Waxman2018a}. The presence of such Galactic low-mass NS binaries may also imply future discovery of stars enhanced with only light neutron-capture elements (i.e., with a high Sr/Fe; see footnote~\ref{foot:light}) by the spectroscopic surveys of metal-poor stars. If the mass ejection is as efficient as that in the model of $\alpha_\mathrm{vis} = 0.10$, however, the merger may eject a large amount of neutron-rich material resulting in a solar $r$-process-like distribution.
If this is the case, we will observe a bright red emission in the kilonovae.
A detailed lightcurve prediction for this case will be presented in the forthcoming paper (Kawaguchi et al., in preparation).

It should be noted that future GW170817-like kilonovae cannot necessarily be an indication of low-mass NS merger events; obviously GW170817 with the total mass of 2.73--$2.78\, M_\odot$ \citep[][]{Abbot2017a} was not the case. Moreover, our results do not exclude GW170817 being a typical NS merger, although our model DD2-135M (with the total mass of $2.7\, M_\odot$ that is consistent with the lower bound for GW170817) results in similar outcomes to those in DD2-125M and SFHo-125H, which show non-solar $r$-process abundance patterns. The reasons are that the fate of a merger remnant depends on the nuclear EOS; model SFHo-135 leaves a hyper-massive NS immediately collapsing into a BH. We also focus on the equal-mass NS binaries only, which may not be the case for GW170817. The agreement of the total heating rate ($\dot{Q}$) for DD2-135M with the light curve of the kilonova of GW170817 (between 1 and 10 days) does not necessarily indicate the non solar $r$-process abundances in the ejecta, either. In fact, the solar $r$-process-like ejecta abundances including the first peak (and co-produced $^{66}$Ni) can explain the light curve of the kilonova associated with GW170817 \citep{Wanajo2018b,Kawaguchi2018a,Wu2019a}.

Finally, we stress that our conclusions presented here are based on the numerical simulations adopting the prescription of the parameterized viscous heating ($\alpha_\mathrm{vis} = 0.04$ as representative) that drives mass ejection from the accretion disk. As the magnetic turbulence is the predominant source of viscosity in the disk, our results in this study should be testified by a long-term magnetohydrodynamical simulation with a sufficient grid resolution.
In addition, the effect of the neutrino irradiation on the electron fraction of the ejecta is still uncertain because of our approximate method of the neutrino transport.
An elaborate treatment of the neutrino transport is needed for a more quantitative study.
The impact of these effects will be investigated in our future work.

\acknowledgements

This work was in part supported by Grant-in-Aid for Scientific Research (Grant Nos. 16K17706, 16H02183, 18H01213, 19K14720, and 20H00158) of Japanese MEXT/JSPS. We also thank all the participants of a workshop entitled “Nucleosynthesis and electromagnetic counterparts of neutron-star mergers” at Yukawa Institute for Theoretical Physics, Kyoto University (No. YITP-T-18-06) for many useful discussions. Numerical computations were performed at Oakforest-PACS at Information Technology Center of the University of Tokyo, XC50 at National Astronomical Observatory of Japan, XC40 at Yukawa Institute for Theoretical Physics, and Sakura and Cobra of the Max Planck Computing and Data Facility.

\appendix
\section{Angular Momentum Loss due to Beamed Neutrino Emission} \label{app:angloss}
Here, we estimate the amount of the angular momentum loss due to the beamed neutrino emission from the neutrino sphere of the remnant massive NSs of rapid rotation.
This effect was already discussed in \cite{baumgarte1998a}, but we estimate the loss rate in the general relativistic momentum formalism \citep{shibata2011a}.
The source term for the momentum density of the fluid due to the neutrino interactions is found in Eq.~(7.7) in \cite{shibata2011a}, and thus, the evolution of the momentum density of the fluid due to radiation reaction is described by the terms
\begin{eqnarray}
\del_t(\rho_* h u_i)\big |_\mathrm{rad} = - \sqrt{\gamma} \alpha S^\mu \gamma_{\mu i} \label{eq:rr},
\end{eqnarray}
where $S^\mu$ is the energy-momentum dissipation rate per volume due to the neutrino emission.
This is written as
\begin{eqnarray}
S^\mu =  \int \frac{d^3k}{(2\pi)^3} k^\mu B(k) = \int d\omega \frac{\omega^3}{(2\pi)^3} \int d\bar{\Omega} (u^\mu + \ell^\mu) B(\omega, \bar{\Omega}), \label{eq:source-moment}
\end{eqnarray}
where $B$ is the so-called collision integral that appears in the Boltzmann equation due to the interaction between radiation and fluid.
Note that $B$ includes the emission, absorption, pair processes, and it is evaluated in the fluid rest frame, so that the momentum of the radiation particle is decomposed as $k^\mu = \omega (u^\mu + \ell^\mu)$ with the energy of the particle in the fluid rest frame $\omega = -k^\mu u_\mu$ and a unit vector $\ell^\mu$, which satisfies $u^\mu \ell_\mu = 0$.
$\ell^\mu$ represents the spatial direction of the momentum of each neutrino in the fluid rest frame and is parametrized by the solid angle $\bar{\Omega}$.

If the angle dependence of the expression in Eq.~\eqref{eq:source-moment} is small, that is, the source term is nearly isotropic, we obtain
\begin{eqnarray}
S^\mu = u^\mu \int d\omega \frac{\omega^3}{(2\pi)^3} \int d\bar{\Omega} B(\omega,\bar{\Omega}) = u^\mu \dot{Q}_\mathrm{rad}, \label{eq:source-iso}
\end{eqnarray}
where $\dot{Q}_\mathrm{rad}$ is the net energy emission rate density.
This assumption is valid for the case that the source term is dominated by neutrino cooling due to the electron/positron capture, or thermal production, which has no preferred direction of the neutrino emission.

We obtain the angular momentum loss rate in a volume $V$ due to the neutrino emission by integrating the angular component of Eq.~\eqref{eq:rr} as
\begin{eqnarray}
\frac{dJ}{dt}\bigg |_\mathrm{rad} =  - \int_V d^3x \sqrt{\gamma} \alpha u_\phi \dot{Q}_\mathrm{rad}, \label{eq:djdt}
\end{eqnarray}
where we defined the angular momentum in a volume $V$ as
\begin{eqnarray}
J = \int_V d^3x \rho_* h u_\phi.  \label{eq:angj}
\end{eqnarray}

By taking the Newtonian limit of this expression as $\gamma = \alpha = h = 1$ and $u_\phi = R^2 \Omega$, where $\Omega$ is the angular velocity, we obtain
\begin{eqnarray}
\frac{dJ}{dt} &=& - \int_V d^3x R^2 \Omega \dot{Q}_\mathrm{rad} \approx - R_\mathrm{NS}{}^2 \Omega L = - \frac{JL}{\kappa M}\nonumber\\
&\approx& 6\times 10^{47}\,\mathrm{erg\,s/s} \,\biggl(\frac{\kappa}{0.4}\biggr) \biggl(\frac{J}{\mathrm{10^{49}\,erg\,s}}\biggr) \biggl(\frac{L}{\mathrm{10^{53}\,erg/s}}\biggr)  \biggl(\frac{M}{\mathrm{2.6\,M_\odot}}\biggr)^{-1}, \label{eq:oom}
\end{eqnarray}
where $R_\mathrm{NS}$ is the NS radius and $L$ is the neutrino luminosity, provided that the neutrino sphere is equal to the NS radius.
We assumed that the NS would be rigidly rotating and $J=\kappa M R_\mathrm{NS}{}^2\Omega$, where $\kappa M R_\mathrm{NS}{}^2$ is the moment of inertia of the NS.
This assumption is expected to be valid for $\gtrsim 1$\,s after the merger due to its viscous evolution.
It is found that the order of the magnitude of the angular momentum loss rate agrees with the numerical result presented in \S \ref{sec:hydro}. 

\section{Model DD2-135L and DD2-135M-v14}\label{app:mod}

\begin{figure*}
\epsscale{1.17}
\plottwo{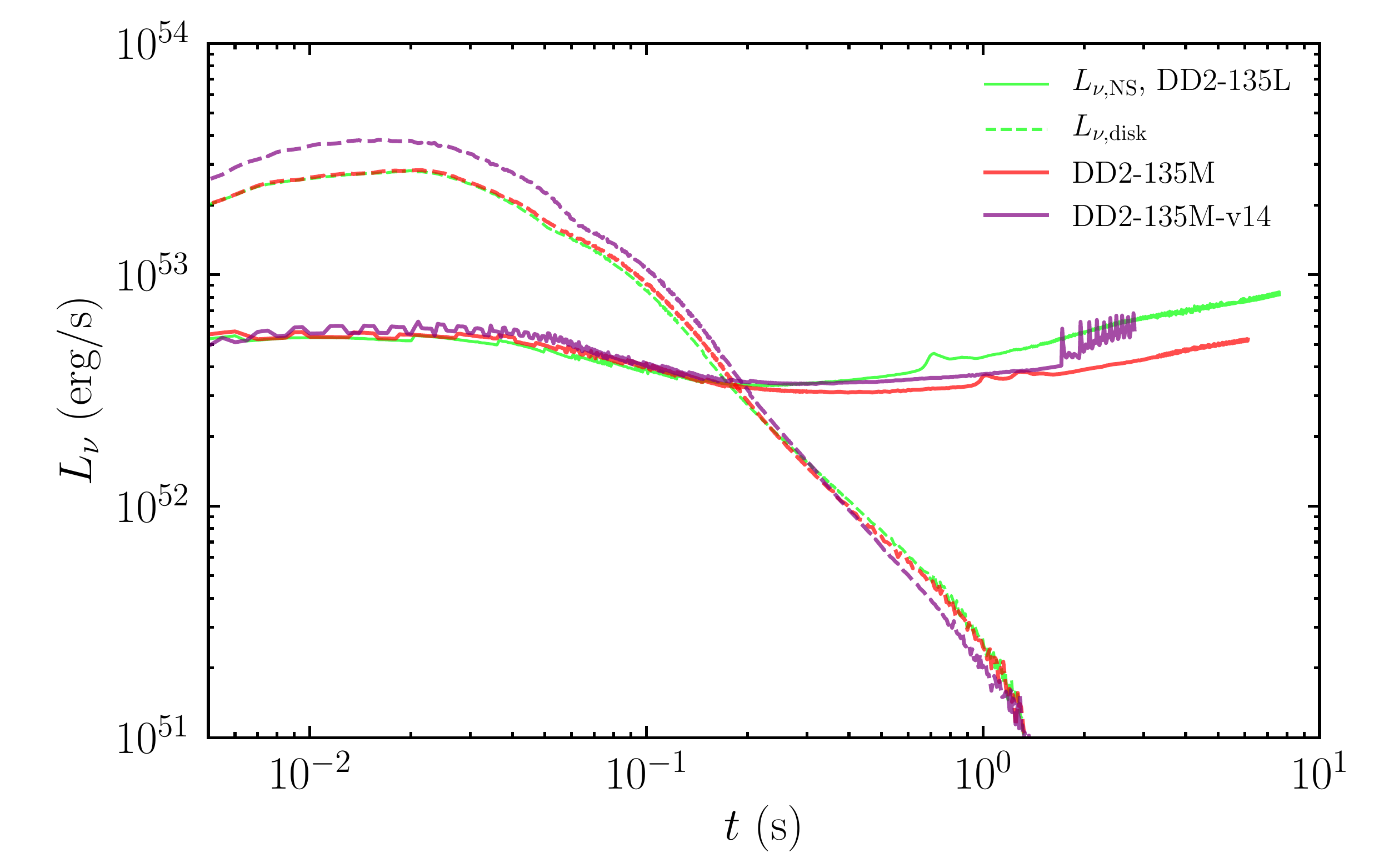}{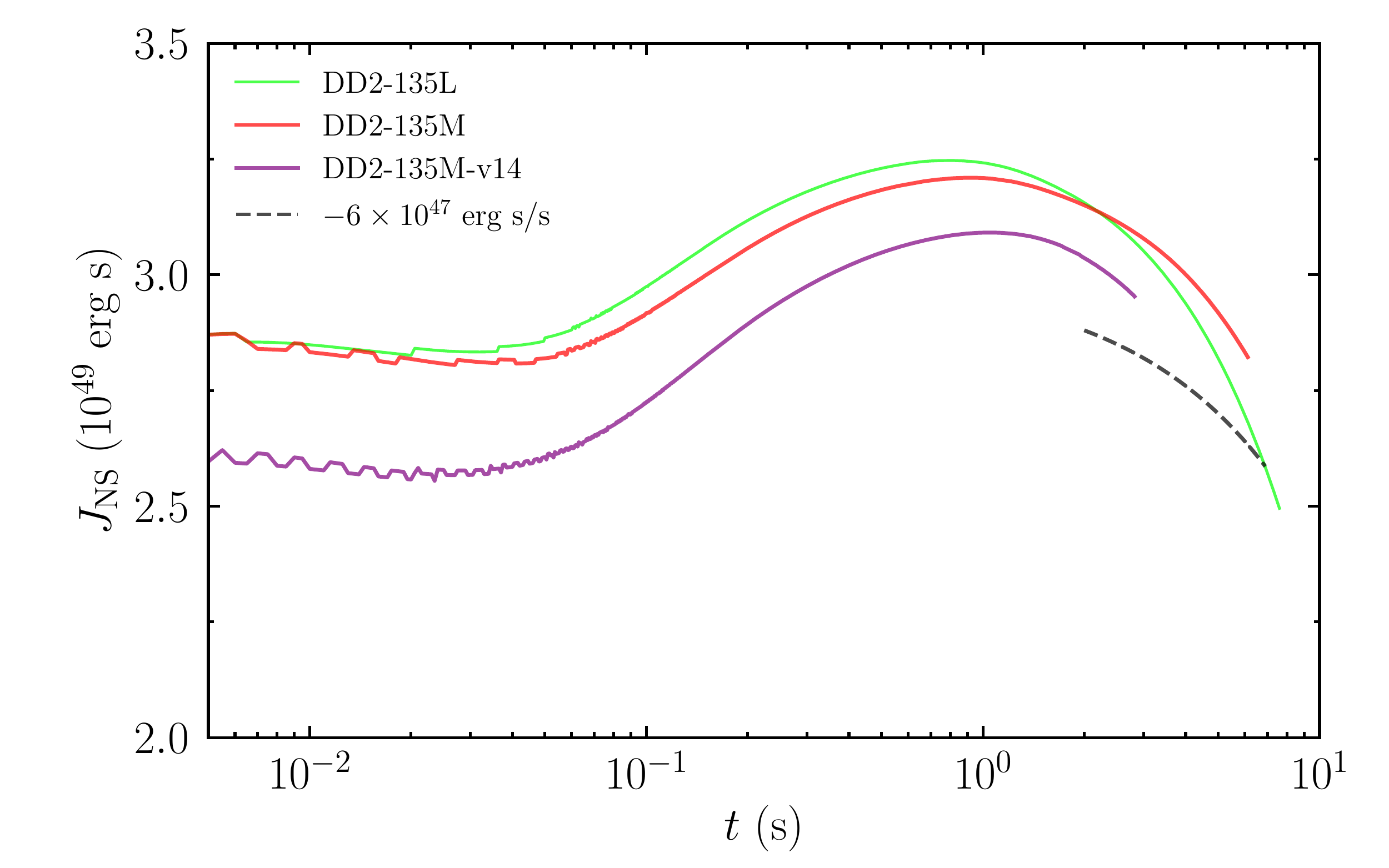}
\plottwo{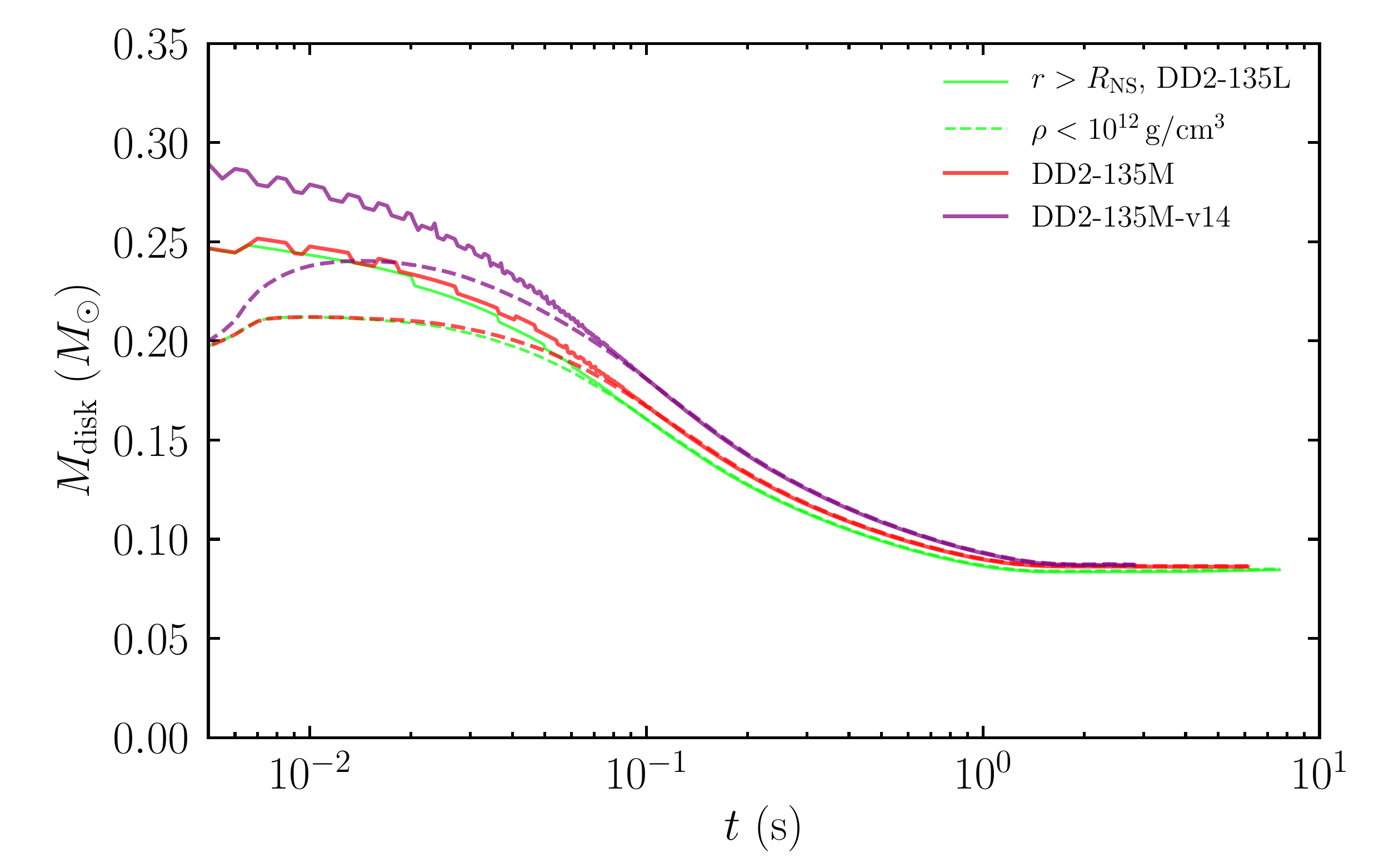}{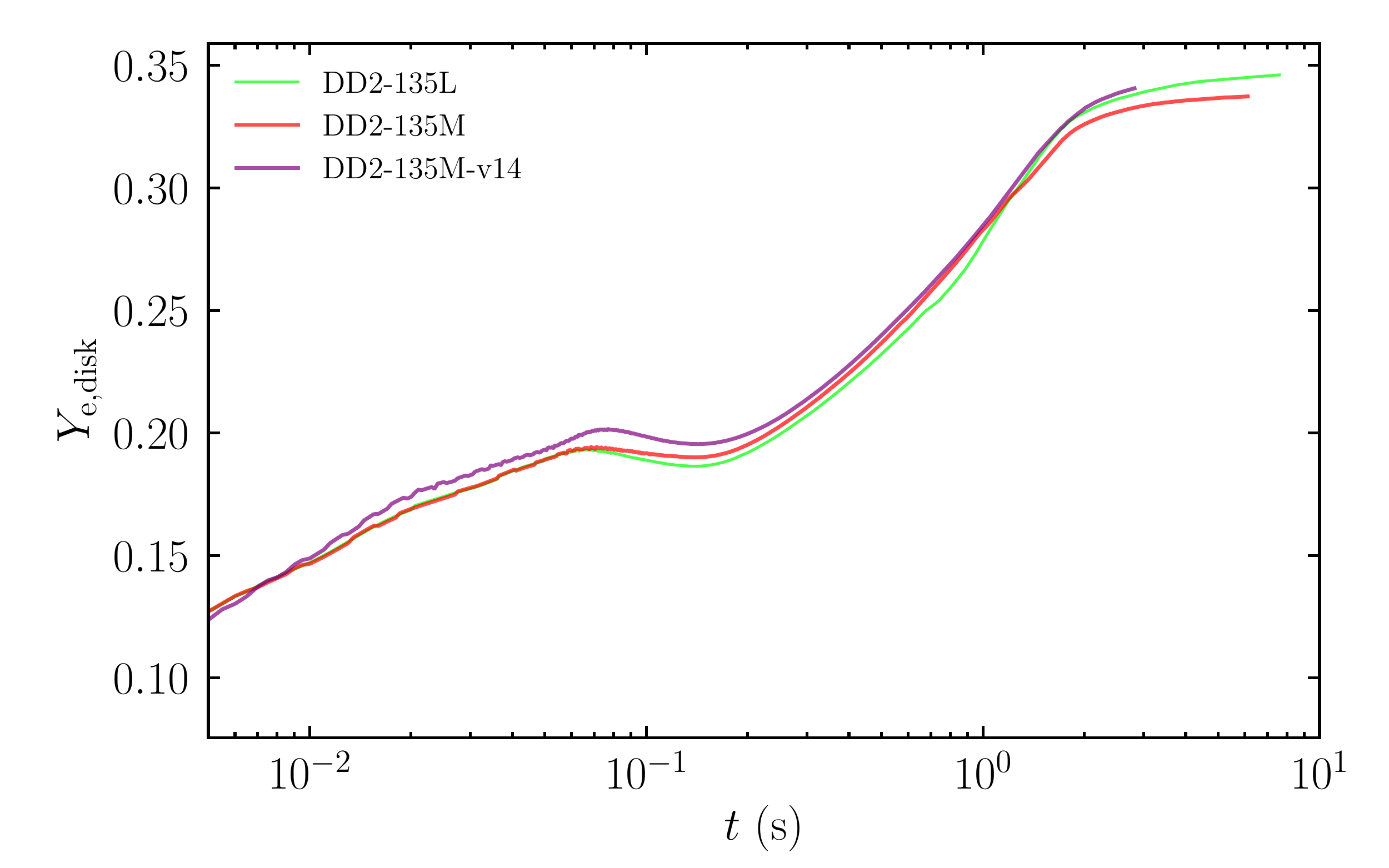}
\plottwo{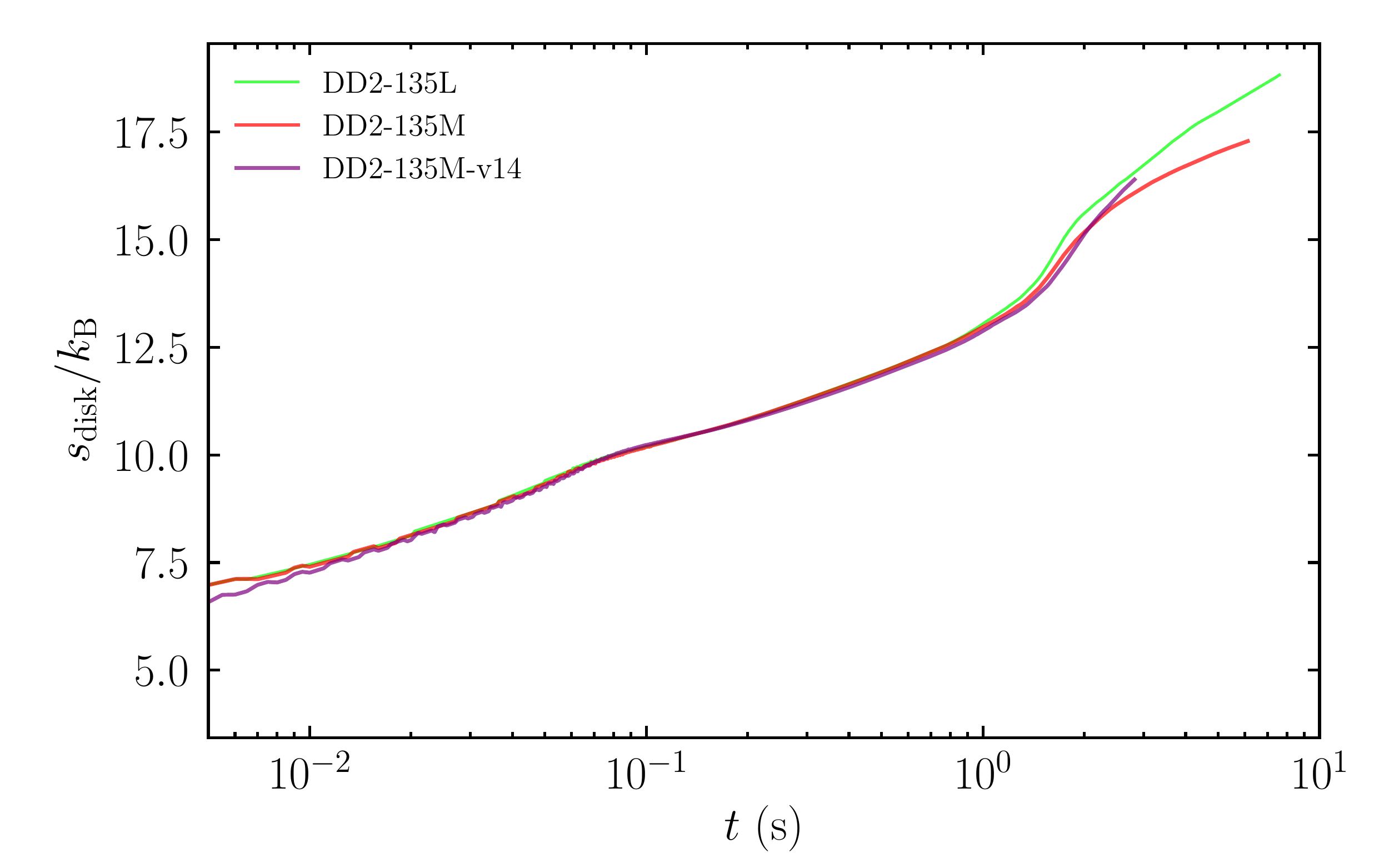}{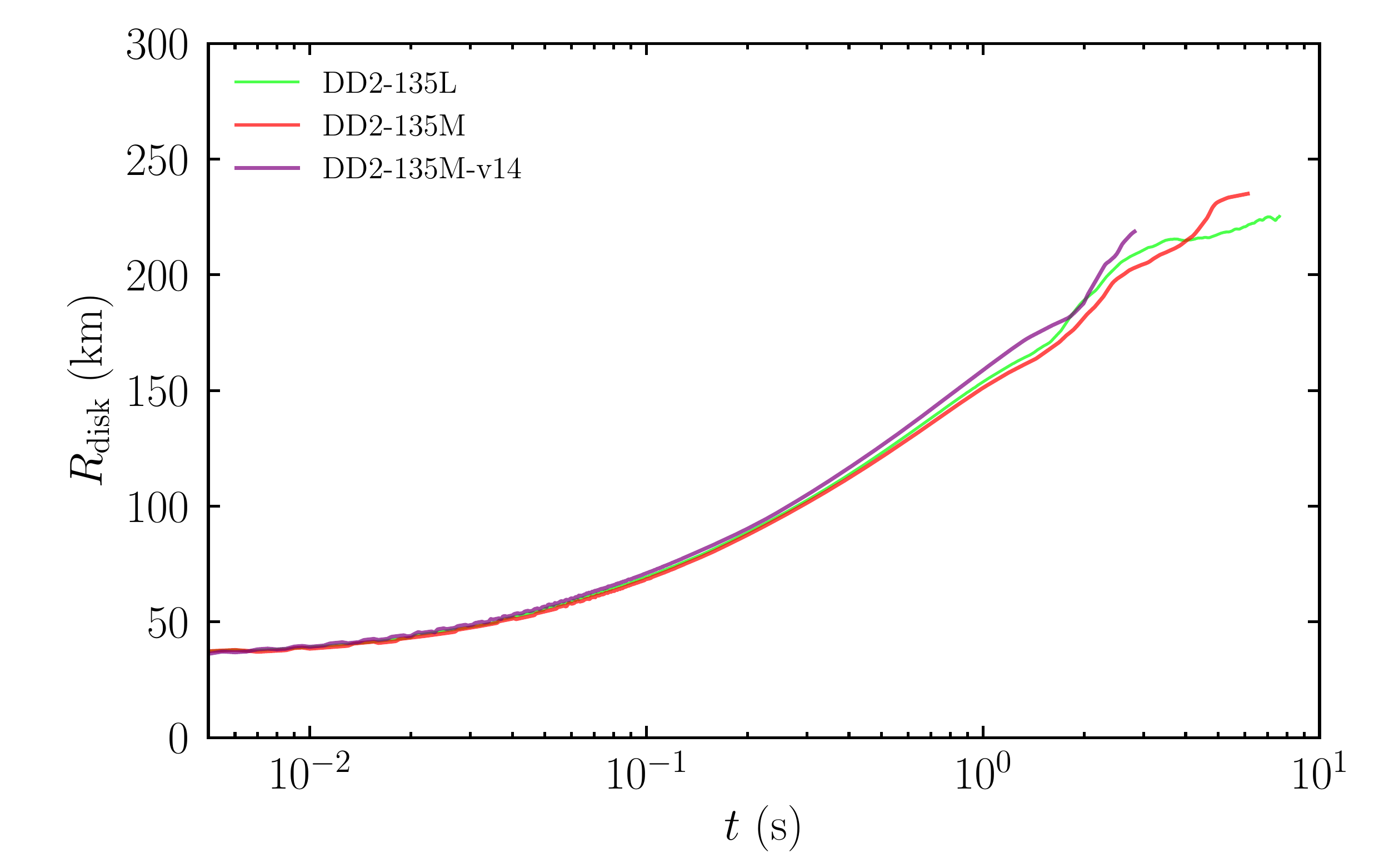}

\caption{Quantities of the disk and the NS for models DD2-135L, DD2-135M, and DD2-135-v14.
}
\label{fig:rem-dd2135}
\end{figure*}

\begin{figure*}
\epsscale{1.17}

\plottwo{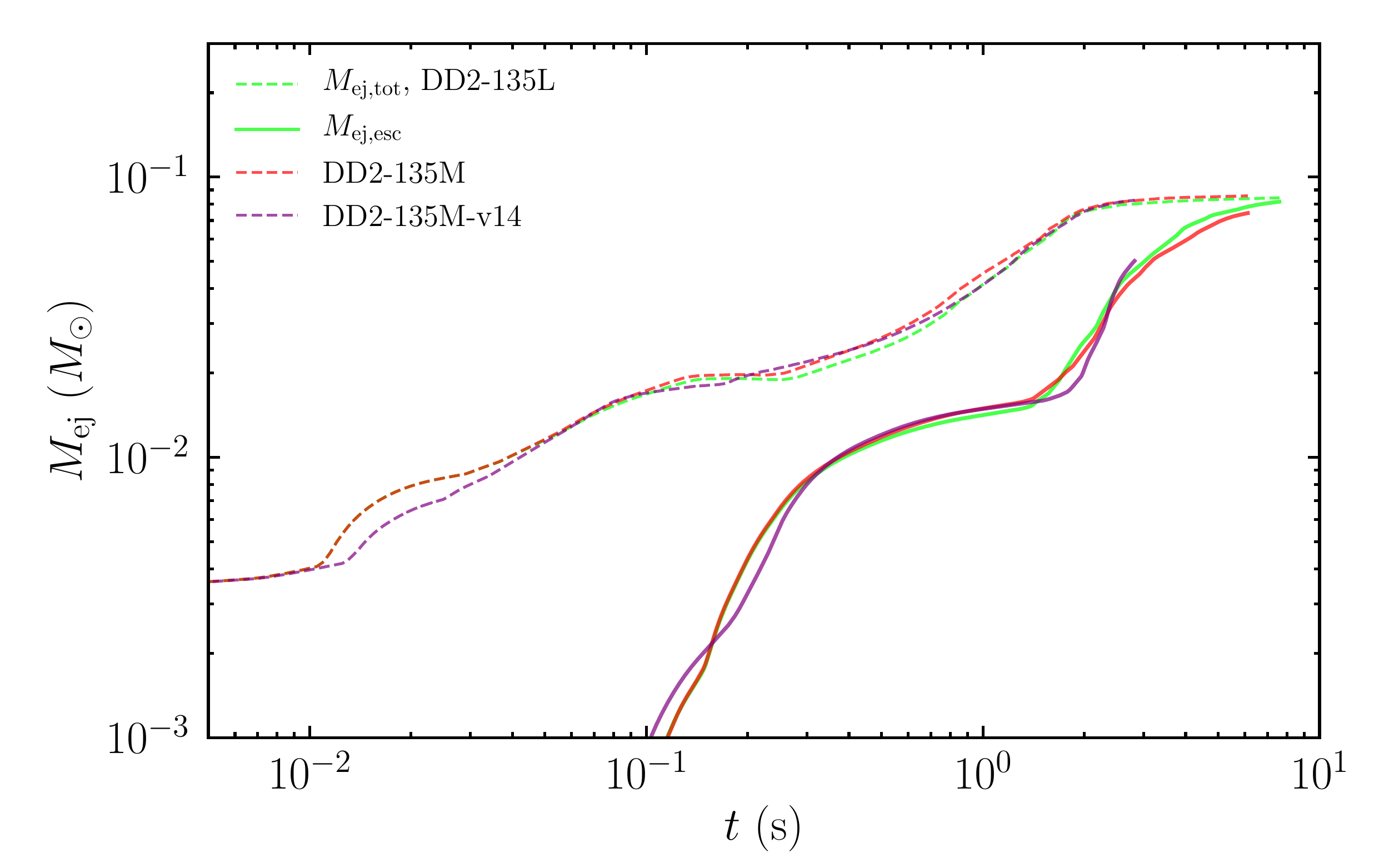}{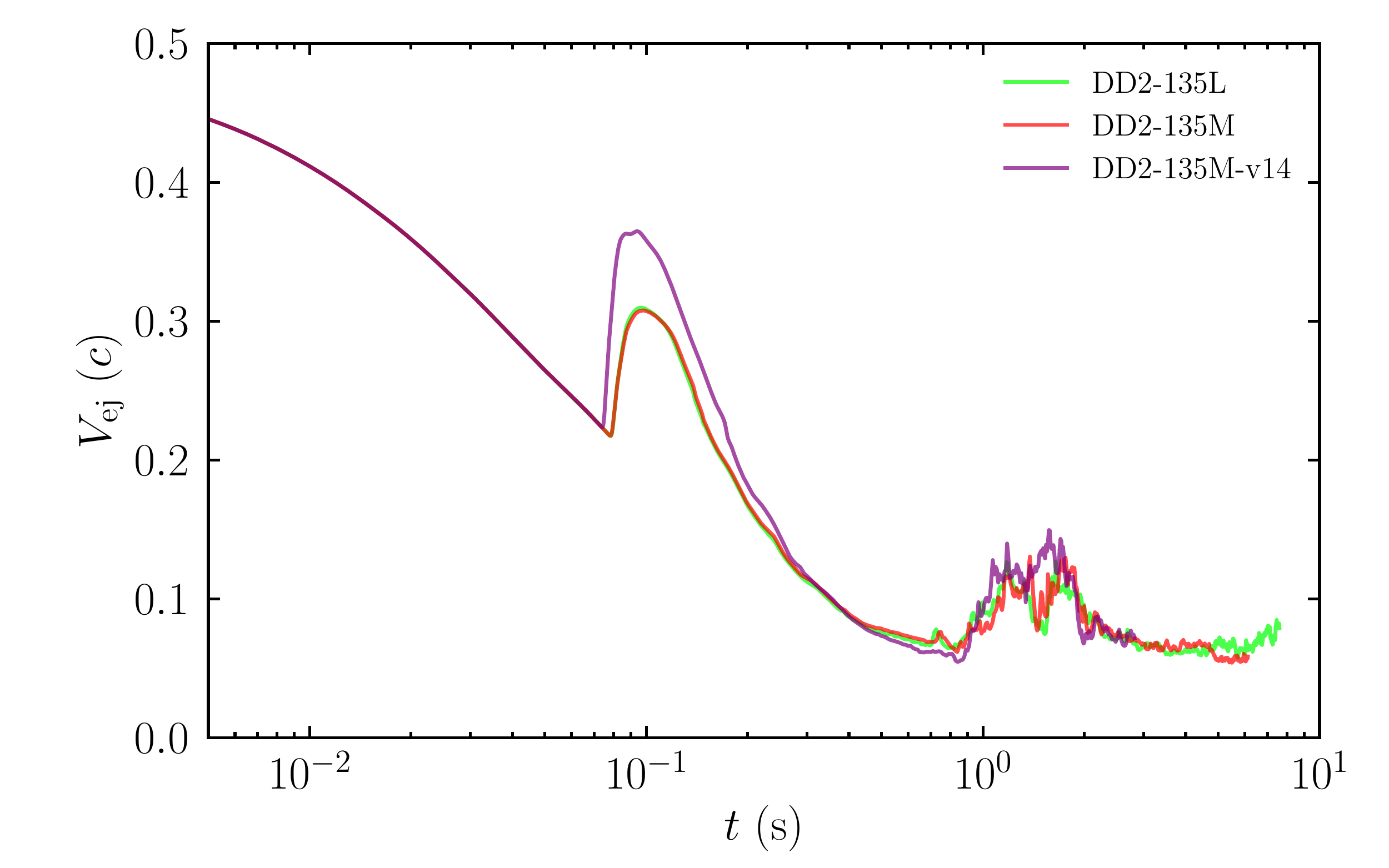}
\plottwo{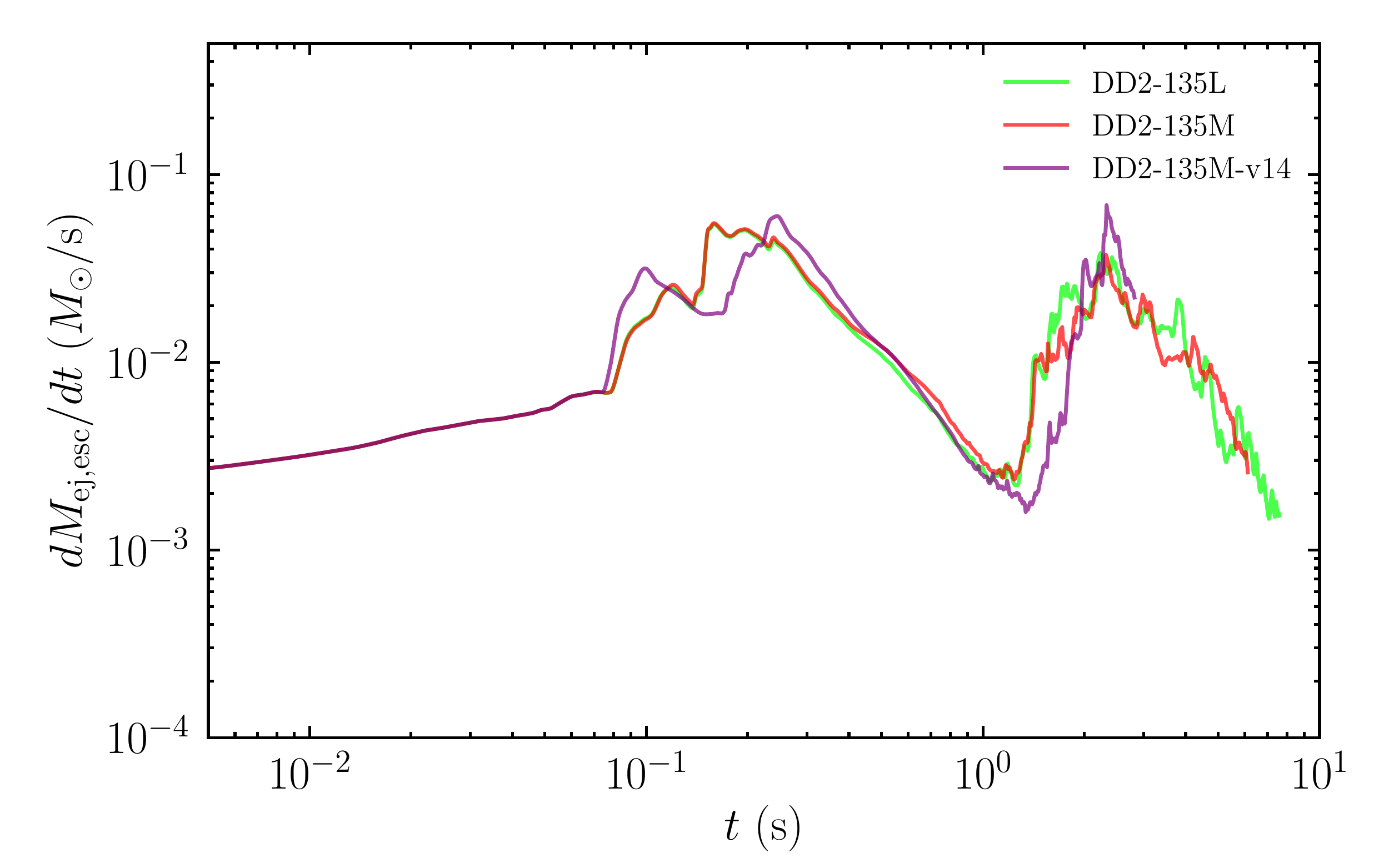}{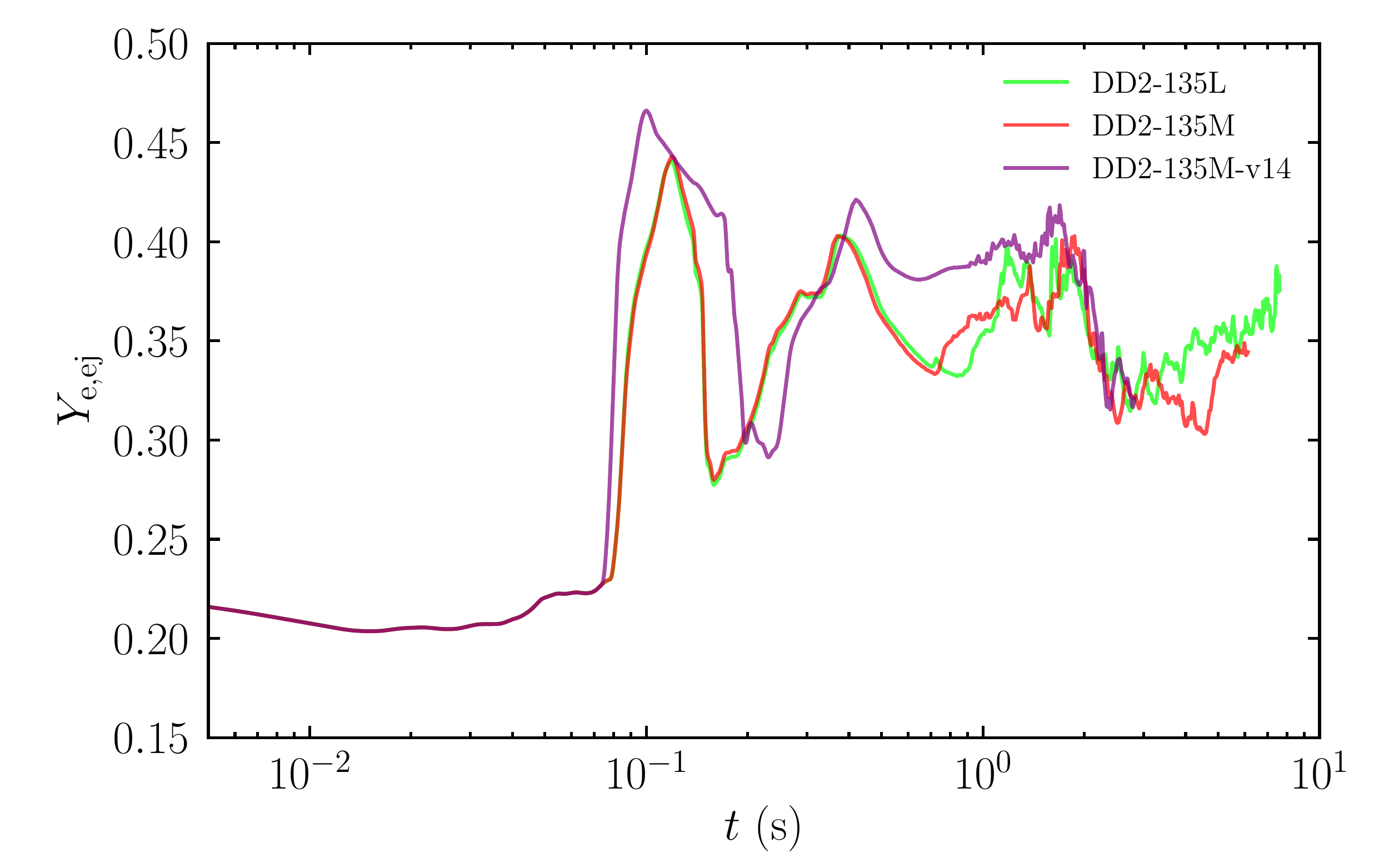}

\caption{Quantities of the ejecta for models DD2-135L, DD2-135M, and DD2-135-v14.
}
\label{fig:ejecta-dd2135}
\end{figure*}

In this appendix, we compare the results of DD2-135L, DD2-135M, and DD2-135M-v14 to investigate the dependence of the grid resolution and the viscous effect inside the NS.
First, we focus on the dependence of the grid resolution on the results by comparing the models DD2-135M and DD2-135L.
The top-left panel of Fig.~\ref{fig:rem-dd2135} shows that the neutrino luminosity of the NS in DD2-135L starts to deviate from that of DD2-135M at $t\approx 0.3$\,s.
Even for DD2-135M, the luminosity of NS starts to increase at $t\approx 1$\,s.
This feature was already found in our earlier work \citep{fujibayashi2017a}, in which the higher resolution model has lower neutrino luminosity at late time of massive NS evolution.
Thus, we would always slightly overestimate the effects of neutrinos in this work.
However, as seen in Fig.~\ref{fig:rem-dd2135} and Fig.~\ref{fig:ejecta-dd2135}, the average electron fraction and entropy of the disk as well as the quantities related to the ejecta do not change much by increasing the grid resolution.
Therefore, we conclude that this spurious increase of the neutrino luminosity does not induce a serious error in evaluating the ejecta properties.

In the top-right panel of Fig.~\ref{fig:rem-dd2135}, we also show the evolution of the angular momentum of the NS.
Due to the lower neutrino luminosity for higher resolution models, the decrease of the angular momentum of the NS is slower.
Thus, the dissipation of the angular momentum of the NS would be overestimated in this work.

Second, we compare models DD2-135M and DD2-135M-v14, for which the viscosity is switched off in the high-density ($\rho>10^{14}\,\mathrm{g/cm^3}$) region.
The mass and the other properties of the ejecta agree well with each other for these two models.
This indicates that the viscosity inside the NS does not significantly affect the ejecta properties.

We find that the slow increase of the angular velocity inside the NS found even for model DD2-135M-v14 (see the bottom panel of Fig.~\ref{fig:dodx}) is due to the redistribution of the angular momentum by the fluid motion.
To clarify this, we define the increase rate of the average specific angular momentum inside a spherical shell with enclosed mass $m$, $j_m$, due to the convective and viscous angular momentum transport by
\begin{align}
\frac{dj_m}{dt} = - \frac{1}{m}\int_{r=r_m} ds \biggl[ \biggl(v^r - \frac{dr_m}{dt}\biggr) \rho_* h u_\phi + \nu h \rho_* \tau^0{}_{\phi k} \gamma^{k r} \biggr],
\end{align}
where $r_m$ is the spherical radius of the mass shell, $ds$ is the area element of the sphere with the radius $r_m$, and $dr_m/dt$ is the radial velocity of the mass shell defined by
\begin{align}
\frac{dr_m}{dt} \int _{r=r_m} ds \rho_* = \int_{r=r_m} ds \rho_* v^r.
\end{align}

\begin{figure*}
\epsscale{1.17}
\plottwo{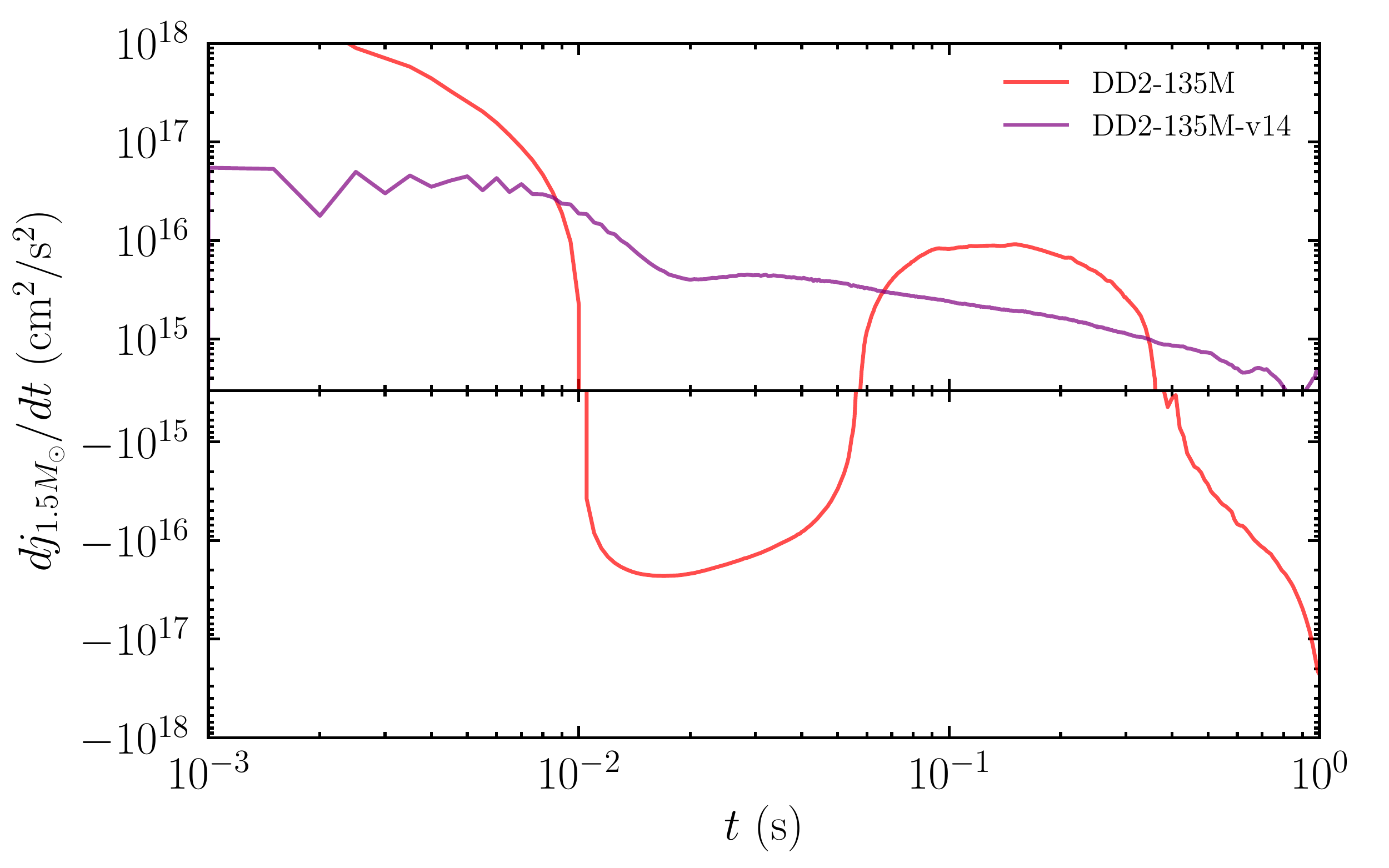}{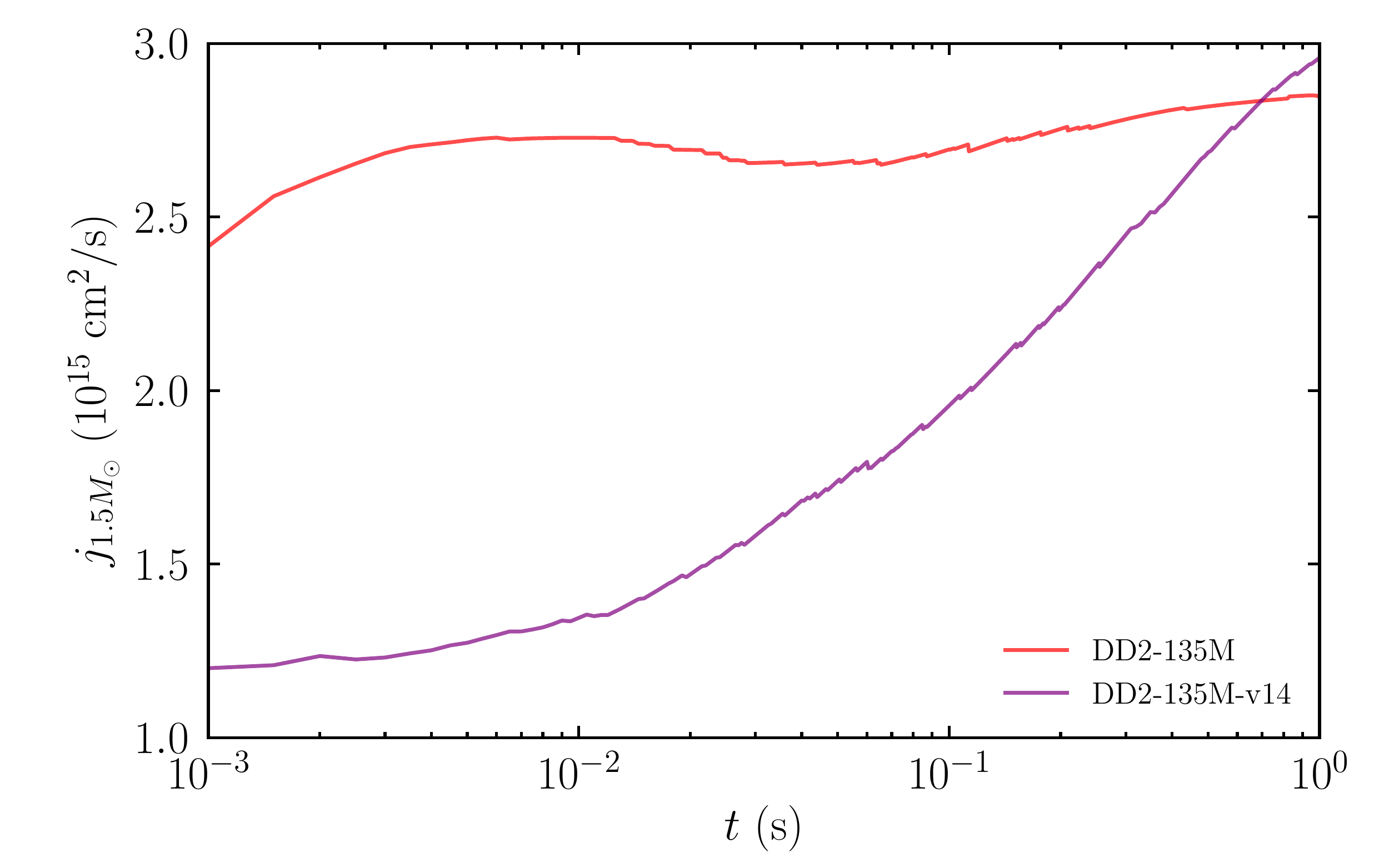}
\caption{
Time evolution of the increase rate of the average specific angular momentum of the material inside the mass shell with $m=1.5M_\odot$ (left) and the average specific angular momentum in the region for models DD2-135M and DD2-135M-v14.
}
\label{fig:j-dd2135}
\end{figure*}

In Fig.~\ref{fig:j-dd2135}, $dj_m/dt$ and $j_m$ with the mass shell $m=1.5M_\odot$ as functions of time are shown for models DD2-135M and DD2-135M-v14.
The mass shell is located at the radius of $\approx 7$\,km with the density of $\approx 4\times 10^{14}$\,g/cm$^3$, for which the slope of the angular velocity along the equatorial plane is steepest at $t=0$ (see the top panel of Fig.~\ref{fig:dodx}).
For model DD2-135M, $dj_m/dt$ has a positive value with the timescale of $j_m(dj_m/dt)^{-1}\sim 10^{-3}$\,s for $t\lesssim 0.01$\,s due to the viscous angular momentum transport.
This timescale is consistent with the viscous timescale inside the NS (see Eq.~\eqref{eq:vis}).
Then it becomes negative for 0.01\,s $\lesssim t \lesssim$ 0.06\,s because of the outward viscous angular momentum transport.
The positive value of $dj_m/dt$ for 0.06\,s $\lesssim t \lesssim$ 0.3\,s is reflected from the spin-up of the NS due to the mass accretion from the disk.
After the mass accretion becomes weak, $dj_m/dt$ becomes negative again due to the viscous effect.
For model DD2-135M-v14, the viscous coefficient $\nu$ is zero for the region with $\rho \ge 10^{14}$\,g/cm$^3$, and thus, the viscous angular momentum transport does not occur at this radius, and thus, the positive value of $dj_m/dt$ is due to the convective angular momentum transport inside the NS.
For this case, the timescale is $j_m(dj_m/dt)^{-1}\sim 1$\,s at $t\sim 0.1$\,s.

Compared to the inviscid model DD2-135M-v0, the increase of the angular velocity for $x\lesssim 10$\,km is significant for DD2-135M-v14.
Therefore, the convective motion inside the NS would be enhanced by the modification of the stellar structure caused by the viscous effects in the outer part of the NS.
Specifically, we speculate that the decrease of the angular velocity and enhanced neutrino cooling at $x\sim 10$\,km due to the viscous effect play roles to enhance the convection inside the NS by weakening the centrifugal force and enhancing the negative entropy gradient in that region.

\bibliography{apj-jour,reference}

\begin{thebibliography}{}
\expandafter\ifx\csname natexlab\endcsname\relax\def\natexlab#1{#1}\fi
\providecommand{\url}[1]{\href{#1}{#1}}
\providecommand{\dodoi}[1]{doi:~\href{http://doi.org/#1}{\nolinkurl{#1}}}
\providecommand{\doeprint}[1]{\href{http://ascl.net/#1}{\nolinkurl{http://ascl.net/#1}}}
\providecommand{\doarXiv}[1]{\href{https://arxiv.org/abs/#1}{\nolinkurl{https://arxiv.org/abs/#1}}}

\bibitem[{{Abadie} {et~al.}(2010){Abadie}, {Abbott}, {Abbott}, {Abernathy},
  {Adams}, {Adhikari}, {Ajith}, {Allen}, {Allen}, {Amador Ceron}, \&
  et~al.}]{Abadie2010a}
{Abadie}, J., {Abbott}, B.~P., {Abbott}, R., {et~al.} 2010, Nuclear Instruments
  and Methods in Physics Research A, 624, 223,
  \dodoi{10.1016/j.nima.2010.07.089}

\bibitem[{{Abbott} {et~al.}(2017{\natexlab{a}}){Abbott}, {Abbott}, {Abbott},
  {Acernese}, {Ackley}, {Adams}, {Adams}, {Addesso}, {Adhikari}, {Adya}, \&
  et~al.}]{Abbot2017a}
{Abbott}, B.~P., {Abbott}, R., {Abbott}, T.~D., {et~al.} 2017{\natexlab{a}},
  Physical Review Letters, 119, 161101, \dodoi{10.1103/PhysRevLett.119.161101}

\bibitem[{{Abbott} {et~al.}(2017{\natexlab{b}}){Abbott}, {Abbott}, {Abbott},
  {Acernese}, {Ackley}, {Adams}, {Adams}, {Addesso}, {Adhikari}, {Adya}, \&
  et~al.}]{Abbot2017b}
---. 2017{\natexlab{b}}, \apjl, 848, L13, \dodoi{10.3847/2041-8213/aa920c}

\bibitem[{{Abbott} {et~al.}(2020){Abbott}, {Abbott}, {Abbott}, {Abraham},
  {Acernese}, {Ackley}, {Adams}, {Adhikari}, {Adya}, {Affeldt}, \&
  et~al.}]{Abbott2020a}
---. 2020, \apjl, 892, L3, \dodoi{10.3847/2041-8213/ab75f5}

\bibitem[{{Accadia} {et~al.}(2011){Accadia}, {Acernese}, {Antonucci}, {Astone},
  {Ballardin}, {Barone}, {Barsuglia}, {Basti}, {Bauer}, {Beker}, {Belletoile},
  {Birindelli}, {Bitossi}, {Bizouard}, {Blom}, {Bondu}, {Bonelli}, {Bonnand},
  {Boschi}, {Bosi}, {Bouhou}, {Braccini}, {Bradaschia}, {Brillet}, {Brisson},
  {Budzy{\'n}ski}, {Bulik}, {Bulten}, {Buskulic}, {Buy}, {Cagnoli}, {Calloni},
  {Campagna}, {Canuel}, {Carbognani}, {Cavalier}, {Cavalieri}, {Cella},
  {Cesarini}, {Chaibi}, {Chassande Mottin}, {Chincarini}, {Cleva}, {Coccia},
  {Colacino}, {Colas}, {Colla}, {Colombini}, {Corsi}, {Coulon}, {Cuoco},
  {D'Antonio}, {Dattilo}, {Davier}, {Day}, {De Rosa}, {Debreczeni}, {del
  Prete}, {Di Fiore}, {Di Lieto}, {Emilio}, {Di Virgilio}, {Dietz}, {Drago},
  {Fafone}, {Ferrante}, {Fidecaro}, {Fiori}, {Flaminio}, {Forte}, {Fournier},
  {Franc}, {Frasca}, {Frasconi}, {Freise}, {Galimberti}, {Gammaitoni},
  {Garufi}, {G{\'a}sp{\'a}r}, {Gemme}, {Genin}, {Gennai}, {Giazotto}, {Gouaty},
  {Granata}, {Greverie}, {Guidi}, {Hayau}, {Heitmann}, {Hello}, {Hild}, {Huet},
  {Jaranowski}, {Kowalska}, {Kr{\'o}lak}, {Leroy}, {Letendre}, {Li}, {Liguori},
  {Lorenzini}, {Loriette}, {Losurdo}, {Majorana}, {Maksimovic}, {Man},
  {Mantovani}, {Marchesoni}, {Marion}, {Marque}, {Martelli}, {Masserot},
  {Michel}, {Milano}, {Minenkov}, {Mohan}, {Morgado}, {Morgia}, {Mosca},
  {Moscatelli}, {Mours}, {Neri}, {Nocera}, {Pagliaroli}, {Palladino},
  {Palomba}, {Paoletti}, {Pardi}, {Parisi}, {Pasqualetti}, {Passaquieti},
  {Passuello}, {Persichetti}, {Pichot}, {Piergiovanni}, {Pietka}, {Pinard},
  {Poggiani}, {Prato}, {Prodi}, {Punturo}, {Puppo}, {Rabeling}, {R{\'a}cz},
  {Rapagnani}, {Re}, {Regimbau}, {Ricci}, {Robinet}, {Rocchi}, {Rolland},
  {Romano}, {Rosi{\'n}ska}, {Ruggi}, {Sassolas}, {Sentenac}, {Sperandio},
  {Sturani}, {Swinkels}, {Tacca}, {Toncelli}, {Tonelli}, {Torre}, {Tournefier},
  {Travasso}, {Vajente}, {van den Brand}, {van der Putten}, {Vasuth},
  {Vavoulidis}, {Vedovato}, {Verkindt}, {Vetrano}, {Vicer{\'e}}, {Vinet},
  {Vocca}, {Ward}, {Was}, \& {Yvert}}]{Accadia2011a}
{Accadia}, T., {Acernese}, F., {Antonucci}, F., {et~al.} 2011, Classical and
  Quantum Gravity, 28, 025005, \dodoi{10.1088/0264-9381/28/2/025005}

\bibitem[{{Akutsu} {et~al.}(2018){Akutsu}, {Ando}, {Araki}, {Araya}, {Arima},
  {Aritomi}, {Asada}, {Aso}, {Atsuta}, {Awai}, \& et~al.}]{Akutsu2018a}
{Akutsu}, T., {Ando}, M., {Araki}, S., {et~al.} 2018, Progress of Theoretical
  and Experimental Physics, 2018, 013F01, \dodoi{10.1093/ptep/ptx180}

\bibitem[{{Alcubierre} {et~al.}(2001){Alcubierre}, {Br{\"u}gmann}, {Holz},
  {Takahashi}, {Brandt}, {Seidel}, {Thornburg}, \&
  {Ashtekar}}]{Alcubierre2001a}
{Alcubierre}, M., {Br{\"u}gmann}, B., {Holz}, D., {et~al.} 2001, International
  Journal of Modern Physics D, 10, 273, \dodoi{10.1142/S0218271801000834}

\bibitem[{{Alexander} {et~al.}(2017){Alexander}, {Berger}, {Fong}, {Williams},
  {Guidorzi}, {Margutti}, {Metzger}, {Annis}, {Blanchard}, {Brout}, {Brown},
  {Chen}, {Chornock}, {Cowperthwaite}, {Drout}, {Eftekhari}, {Frieman}, {Holz},
  {Nicholl}, {Rest}, {Sako}, {Soares-Santos}, \& {Villar}}]{Alexander2017a}
{Alexander}, K.~D., {Berger}, E., {Fong}, W., {et~al.} 2017, \apjl, 848, L21,
  \dodoi{10.3847/2041-8213/aa905d}

\bibitem[{{Aoki} {et~al.}(2017){Aoki}, {Ishimaru}, {Aoki}, \&
  {Wanajo}}]{Aoki2017a}
{Aoki}, M., {Ishimaru}, Y., {Aoki}, W., \& {Wanajo}, S. 2017, \apj, 837, 8,
  \dodoi{10.3847/1538-4357/aa5d08}

\bibitem[{{Arnowitt} {et~al.}(1960){Arnowitt}, {Deser}, \&
  {Misner}}]{arnowitt1960a}
{Arnowitt}, R., {Deser}, S., \& {Misner}, C.~W. 1960, Physical Review, 118,
  1100, \dodoi{10.1103/PhysRev.118.1100}

\bibitem[{{Balbus} \& {Hawley}(1991)}]{Balbus1991a}
{Balbus}, S.~A., \& {Hawley}, J.~F. 1991, \apj, 376, 214,
  \dodoi{10.1086/170270}

\bibitem[{{Banik} {et~al.}(2014){Banik}, {Hempel}, \&
  {Bandyopadhyay}}]{banik2014a}
{Banik}, S., {Hempel}, M., \& {Bandyopadhyay}, D. 2014, \apjs, 214, 22,
  \dodoi{10.1088/0067-0049/214/2/22}

\bibitem[{{Barnes} {et~al.}(2016){Barnes}, {Kasen}, {Wu}, \&
  {Mart{\'{\i}}nez-Pinedo}}]{Barnes2016a}
{Barnes}, J., {Kasen}, D., {Wu}, M.-R., \& {Mart{\'{\i}}nez-Pinedo}, G. 2016,
  \apj, 829, 110, \dodoi{10.3847/0004-637X/829/2/110}

\bibitem[{{Baumgarte} \& {Shapiro}(1998)}]{baumgarte1998a}
{Baumgarte}, T.~W., \& {Shapiro}, S.~L. 1998, \apj, 504, 431,
  \dodoi{10.1086/306067}

\bibitem[{{Baumgarte} \& {Shapiro}(1999)}]{baumgarte1999}
---. 1999, \prd, 59, 024007, \dodoi{10.1103/PhysRevD.59.024007}

\bibitem[{{Bauswein} {et~al.}(2013){Bauswein}, {Goriely}, \&
  {Janka}}]{Bauswein2013a}
{Bauswein}, A., {Goriely}, S., \& {Janka}, H.~T. 2013, \apj, 773, 78,
  \dodoi{10.1088/0004-637X/773/1/78}

\bibitem[{{Bovard} {et~al.}(2017){Bovard}, {Martin}, {Guercilena}, {Arcones},
  {Rezzolla}, \& {Korobkin}}]{Bovard2017a}
{Bovard}, L., {Martin}, D., {Guercilena}, F., {et~al.} 2017, \prd, 96, 124005,
  \dodoi{10.1103/PhysRevD.96.124005}

\bibitem[{{Chadwick} {et~al.}(2011){Chadwick}, {Herman},
  {Oblo{\v{z}}insk{\'y}}, {Dunn}, {Danon}, {Kahler}, {Smith}, {Pritychenko},
  {Arbanas}, \& {Arcilla}}]{Chadwick2011a}
{Chadwick}, M.~B., {Herman}, M., {Oblo{\v{z}}insk{\'y}}, P., {et~al.} 2011,
  Nuclear Data Sheets, 112, 2887, \dodoi{10.1016/j.nds.2011.11.002}

\bibitem[{{Chen} \& {Beloborodov}(2007)}]{Chen2007a}
{Chen}, W.-X., \& {Beloborodov}, A.~M. 2007, \apj, 657, 383,
  \dodoi{10.1086/508923}

\bibitem[{{Cowan} {et~al.}(2019){Cowan}, {Sneden}, {Lawler}, {Aprahamian},
  {Wiescher}, {Langanke}, {Mart{\'\i}nez-Pinedo}, \& {Thielemann}}]{Cowan2019a}
{Cowan}, J.~J., {Sneden}, C., {Lawler}, J.~E., {et~al.} 2019, arXiv e-prints,
  arXiv:1901.01410.
\newblock \doarXiv{1901.01410}

\bibitem[{{Cowan} {et~al.}(2005){Cowan}, {Sneden}, {Beers}, {Lawler},
  {Simmerer}, {Truran}, {Primas}, {Collier}, \& {Burles}}]{Cowan2005a}
{Cowan}, J.~J., {Sneden}, C., {Beers}, T.~C., {et~al.} 2005, \apj, 627, 238,
  \dodoi{10.1086/429952}

\bibitem[{{Cyburt} {et~al.}(2010){Cyburt}, {Amthor}, {Ferguson}, {Meisel},
  {Smith}, {Warren}, {Heger}, {Hoffman}, {Rauscher}, {Sakharuk}, {Schatz},
  {Thielemann}, \& {Wiescher}}]{Cyburt2010a}
{Cyburt}, R.~H., {Amthor}, A.~M., {Ferguson}, R., {et~al.} 2010, \apjs, 189,
  240, \dodoi{10.1088/0067-0049/189/1/240}

\bibitem[{{Eichler} {et~al.}(1989){Eichler}, {Livio}, {Piran}, \&
  {Schramm}}]{Eichler1989a}
{Eichler}, D., {Livio}, M., {Piran}, T., \& {Schramm}, D.~N. 1989, \nat, 340,
  126, \dodoi{10.1038/340126a0}

\bibitem[{{Fern{\'a}ndez} {et~al.}(2020){Fern{\'a}ndez}, {Foucart}, \&
  {Lippuner}}]{Fernandez2020a}
{Fern{\'a}ndez}, R., {Foucart}, F., \& {Lippuner}, J. 2020, arXiv e-prints,
  arXiv:2005.14208.
\newblock \doarXiv{2005.14208}

\bibitem[{{Fern{\'a}ndez} \& {Metzger}(2013)}]{Fernandez2013a}
{Fern{\'a}ndez}, R., \& {Metzger}, B.~D. 2013, \mnras, 435, 502,
  \dodoi{10.1093/mnras/stt1312}

\bibitem[{{Fern{\'a}ndez} {et~al.}(2019){Fern{\'a}ndez}, {Tchekhovskoy},
  {Quataert}, {Foucart}, \& {Kasen}}]{Fernandez2019a}
{Fern{\'a}ndez}, R., {Tchekhovskoy}, A., {Quataert}, E., {Foucart}, F., \&
  {Kasen}, D. 2019, \mnras, 482, 3373, \dodoi{10.1093/mnras/sty2932}

\bibitem[{{Foucart} {et~al.}(2016){Foucart}, {Haas}, {Duez}, {O'Connor}, {Ott},
  {Roberts}, {Kidder}, {Lippuner}, {Pfeiffer}, \& {Scheel}}]{Foucart2016a}
{Foucart}, F., {Haas}, R., {Duez}, M.~D., {et~al.} 2016, \prd, 93, 044019,
  \dodoi{10.1103/PhysRevD.93.044019}

\bibitem[{{Freiburghaus} {et~al.}(1999){Freiburghaus}, {Rosswog}, \&
  {Thielemann}}]{Freiburghaus1999a}
{Freiburghaus}, C., {Rosswog}, S., \& {Thielemann}, F.~K. 1999, \apjl, 525,
  L121, \dodoi{10.1086/312343}

\bibitem[{{Fujibayashi} {et~al.}(2018){Fujibayashi}, {Kiuchi}, {Nishimura},
  {Sekiguchi}, \& {Shibata}}]{fujibayashi2018a}
{Fujibayashi}, S., {Kiuchi}, K., {Nishimura}, N., {Sekiguchi}, Y., \&
  {Shibata}, M. 2018, \apj, 860, 64, \dodoi{10.3847/1538-4357/aabafd}

\bibitem[{{Fujibayashi} {et~al.}(2017){Fujibayashi}, {Sekiguchi}, {Kiuchi}, \&
  {Shibata}}]{fujibayashi2017a}
{Fujibayashi}, S., {Sekiguchi}, Y., {Kiuchi}, K., \& {Shibata}, M. 2017, \apj,
  846, 114, \dodoi{10.3847/1538-4357/aa8039}

\bibitem[{{Fujibayashi} {et~al.}(2020){Fujibayashi}, {Shibata}, {Wanajo},
  {Kiuchi}, {Kyutoku}, \& {Sekiguchi}}]{Fujibayashi2020a}
{Fujibayashi}, S., {Shibata}, M., {Wanajo}, S., {et~al.} 2020, arXiv e-prints,
  arXiv:2001.04467.
\newblock \doarXiv{2001.04467}

\bibitem[{{Gonz{\'a}lez} {et~al.}(2007){Gonz{\'a}lez}, {Audit}, \&
  {Huynh}}]{gonzalez2007a}
{Gonz{\'a}lez}, M., {Audit}, E., \& {Huynh}, P. 2007, \aap, 464, 429,
  \dodoi{10.1051/0004-6361:20065486}

\bibitem[{{Goriely} {et~al.}(2015){Goriely}, {Bauswein}, {Just}, {Pllumbi}, \&
  {Janka}}]{Goriely2015a}
{Goriely}, S., {Bauswein}, A., {Just}, O., {Pllumbi}, E., \& {Janka}, H.~T.
  2015, \mnras, 452, 3894, \dodoi{10.1093/mnras/stv1526}

\bibitem[{{Goriely} {et~al.}(2010){Goriely}, {Chamel}, \&
  {Pearson}}]{Goriely2010a}
{Goriely}, S., {Chamel}, N., \& {Pearson}, J.~M. 2010, \prc, 82, 035804,
  \dodoi{10.1103/PhysRevC.82.035804}

\bibitem[{{Goriely} {et~al.}(2008){Goriely}, {Hilaire}, \&
  {Koning}}]{Goriely2008a}
{Goriely}, S., {Hilaire}, S., \& {Koning}, A.~J. 2008, \aap, 487, 767,
  \dodoi{10.1051/0004-6361:20078825}

\bibitem[{{Guilet} {et~al.}(2017){Guilet}, {Bauswein}, {Just}, \&
  {Janka}}]{Guilet2017a}
{Guilet}, J., {Bauswein}, A., {Just}, O., \& {Janka}, H.-T. 2017, \mnras, 471,
  1879, \dodoi{10.1093/mnras/stx1739}

\bibitem[{{Hansen} {et~al.}(2017){Hansen}, {Simon}, {Marshall}, {Li},
  {Carollo}, {DePoy}, {Nagasawa}, {Bernstein}, {Drlica-Wagner}, {Abdalla},
  {Allam}, {Annis}, {Bechtol}, {Benoit-L{\'e}vy}, {Brooks}, {Buckley-Geer},
  {Carnero Rosell}, {Carrasco Kind}, {Carretero}, {Cunha}, {da Costa}, {Desai},
  {Eifler}, {Fausti Neto}, {Flaugher}, {Frieman}, {Garc{\'\i}a-Bellido},
  {Gaztanaga}, {Gerdes}, {Gruen}, {Gruendl}, {Gschwend}, {Gutierrez}, {James},
  {Krause}, {Kuehn}, {Kuropatkin}, {Lahav}, {Miquel}, {Plazas}, {Romer},
  {Sanchez}, {Santiago}, {Scarpine}, {Smith}, {Soares-Santos}, {Sobreira},
  {Suchyta}, {Swanson}, {Tarle}, {Walker}, \& {DES
  Collaboration}}]{Hansen2017a}
{Hansen}, T.~T., {Simon}, J.~D., {Marshall}, J.~L., {et~al.} 2017, \apj, 838,
  44, \dodoi{10.3847/1538-4357/aa634a}

\bibitem[{{Hansen} {et~al.}(2020){Hansen}, {Marshall}, {Simon}, {Li},
  {Bernstein}, {Pace}, {Ferguson}, {Nagasawa}, {Kuehn}, {Carollo}, {Geha},
  {James}, {Walker}, {Diehl}, {Aguena}, {Allam}, {Avila}, {Bertin}, {Brooks},
  {Buckley-Geer}, {Burke}, {Carnero Rosell}, {Carrasco Kind}, {Carretero},
  {Costanzi}, {da Costa}, {Desai}, {De Vicente}, {Doel}, {Eckert}, {Eifler},
  {Everett}, {Ferrero}, {Frieman}, {Garc{\'\i}a-Bellido}, {Gaztanaga},
  {Gerdes}, {Gruen}, {Gruendl}, {Gschwend}, {Gutierrez}, {Hinton}, {Hollowood},
  {Honscheid}, {Kuropatkin}, {Maia}, {March}, {Miquel}, {Palmese},
  {Paz-Chinch{\'o}n}, {Plazas}, {Sanchez}, {Santiago}, {Scarpine}, {Serrano},
  {Smith}, {Soares-Santos}, {Suchyta}, {Swanson}, {Tarle}, {Varga}, \&
  {Wilkinson}}]{Hansen2020a}
{Hansen}, T.~T., {Marshall}, J.~L., {Simon}, J.~D., {et~al.} 2020, arXiv
  e-prints, arXiv:2005.10767.
\newblock \doarXiv{2005.10767}

\bibitem[{{Hilditch} {et~al.}(2013){Hilditch}, {Bernuzzi}, {Thierfelder},
  {Cao}, {Tichy}, \& {Br{\"u}gmann}}]{Hilditch2013a}
{Hilditch}, D., {Bernuzzi}, S., {Thierfelder}, M., {et~al.} 2013, \prd, 88,
  084057, \dodoi{10.1103/PhysRevD.88.084057}

\bibitem[{{Honda} {et~al.}(2006){Honda}, {Aoki}, {Ishimaru}, {Wanajo}, \&
  {Ryan}}]{Honda2006a}
{Honda}, S., {Aoki}, W., {Ishimaru}, Y., {Wanajo}, S., \& {Ryan}, S.~G. 2006,
  \apj, 643, 1180, \dodoi{10.1086/503195}

\bibitem[{{Hotokezaka} {et~al.}(2013){Hotokezaka}, {Kiuchi}, {Kyutoku},
  {Muranushi}, {Sekiguchi}, {Shibata}, \& {Taniguchi}}]{Hotokezaka2013a}
{Hotokezaka}, K., {Kiuchi}, K., {Kyutoku}, K., {et~al.} 2013, \prd, 88, 044026,
  \dodoi{10.1103/PhysRevD.88.044026}

\bibitem[{{Hotokezaka} \& {Nakar}(2020)}]{Hotokezaka2020a}
{Hotokezaka}, K., \& {Nakar}, E. 2020, \apj, 891, 152,
  \dodoi{10.3847/1538-4357/ab6a98}

\bibitem[{{Hotokezaka} \& {Piran}(2015)}]{Hotokezaka2015a}
{Hotokezaka}, K., \& {Piran}, T. 2015, \mnras, 450, 1430,
  \dodoi{10.1093/mnras/stv620}

\bibitem[{{Hotokezaka} {et~al.}(2016){Hotokezaka}, {Wanajo}, {Tanaka}, {Bamba},
  {Terada}, \& {Piran}}]{Hotokezaka2016a}
{Hotokezaka}, K., {Wanajo}, S., {Tanaka}, M., {et~al.} 2016, \mnras, 459, 35,
  \dodoi{10.1093/mnras/stw404}

\bibitem[{{Ishimaru} {et~al.}(2015){Ishimaru}, {Wanajo}, \&
  {Prantzos}}]{Ishimaru2015a}
{Ishimaru}, Y., {Wanajo}, S., \& {Prantzos}, N. 2015, \apjl, 804, L35,
  \dodoi{10.1088/2041-8205/804/2/L35}

\bibitem[{{Israel} \& {Stewart}(1979)}]{Israel1979a}
{Israel}, W., \& {Stewart}, J.~M. 1979, Annals of Physics, 118, 341,
  \dodoi{10.1016/0003-4916(79)90130-1}

\bibitem[{{Janka} {et~al.}(2012){Janka}, {Hanke}, {H{\"u}depohl}, {Marek},
  {M{\"u}ller}, \& {Obergaulinger}}]{Janka2012b}
{Janka}, H.-T., {Hanke}, F., {H{\"u}depohl}, L., {et~al.} 2012, Progress of
  Theoretical and Experimental Physics, 2012, 01A309,
  \dodoi{10.1093/ptep/pts067}

\bibitem[{{Ji} {et~al.}(2016){Ji}, {Frebel}, {Chiti}, \& {Simon}}]{Ji2016a}
{Ji}, A.~P., {Frebel}, A., {Chiti}, A., \& {Simon}, J.~D. 2016, \nat, 531, 610,
  \dodoi{10.1038/nature17425}

\bibitem[{{Just} {et~al.}(2015){Just}, {Bauswein}, {Ardevol Pulpillo},
  {Goriely}, \& {Janka}}]{Just2015a}
{Just}, O., {Bauswein}, A., {Ardevol Pulpillo}, R., {Goriely}, S., \& {Janka},
  H.~T. 2015, \mnras, 448, 541, \dodoi{10.1093/mnras/stv009}

\bibitem[{{Kasliwal} {et~al.}(2019){Kasliwal}, {Kasen}, {Lau}, {Perley},
  {Rosswog}, {Ofek}, {Hotokezaka}, {Chary}, {Sollerman}, \&
  {Goobar}}]{Kasliwal2019a}
{Kasliwal}, M.~M., {Kasen}, D., {Lau}, R.~M., {et~al.} 2019, \mnras, L14,
  \dodoi{10.1093/mnrasl/slz007}

\bibitem[{{Kawaguchi} {et~al.}(2018){Kawaguchi}, {Shibata}, \&
  {Tanaka}}]{Kawaguchi2018a}
{Kawaguchi}, K., {Shibata}, M., \& {Tanaka}, M. 2018, \apjl, 865, L21,
  \dodoi{10.3847/2041-8213/aade02}

\bibitem[{{Kawanaka} \& {Mineshige}(2007)}]{Kawanaka2007a}
{Kawanaka}, N., \& {Mineshige}, S. 2007, \apj, 662, 1156,
  \dodoi{10.1086/517985}

\bibitem[{{Kiuchi} {et~al.}(2018){Kiuchi}, {Kyutoku}, {Sekiguchi}, \&
  {Shibata}}]{kiuchi2018a}
{Kiuchi}, K., {Kyutoku}, K., {Sekiguchi}, Y., \& {Shibata}, M. 2018, \prd, 97,
  124039, \dodoi{10.1103/PhysRevD.97.124039}

\bibitem[{{Kulkarni}(2005)}]{Kulkarni2005a}
{Kulkarni}, S.~R. 2005, arXiv Astrophysics e-prints

\bibitem[{{Lattimer} \& {Schramm}(1974)}]{Lattimer1974a}
{Lattimer}, J.~M., \& {Schramm}, D.~N. 1974, \apjl, 192, L145,
  \dodoi{10.1086/181612}

\bibitem[{{Levermore}(1984)}]{levermore1984a}
{Levermore}, C.~D. 1984, \jqsrt, 31, 149, \dodoi{10.1016/0022-4073(84)90112-2}

\bibitem[{{Li} \& {Paczy{\'n}ski}(1998)}]{Li1998a}
{Li}, L.-X., \& {Paczy{\'n}ski}, B. 1998, \apjl, 507, L59,
  \dodoi{10.1086/311680}

\bibitem[{{Liebend{\"o}rfer} {et~al.}(2003){Liebend{\"o}rfer}, {Mezzacappa},
  {Messer}, {Martinez-Pinedo}, {Hix}, \& {Thielemann}}]{Liebendoerfer2003a}
{Liebend{\"o}rfer}, M., {Mezzacappa}, A., {Messer}, O.~E.~B., {et~al.} 2003,
  \nphysa, 719, C144, \dodoi{10.1016/S0375-9474(03)00984-9}

\bibitem[{{Lippuner} {et~al.}(2017){Lippuner}, {Fern{\'a}ndez}, {Roberts},
  {Foucart}, {Kasen}, {Metzger}, \& {Ott}}]{Lippuner2017a}
{Lippuner}, J., {Fern{\'a}ndez}, R., {Roberts}, L.~F., {et~al.} 2017, \mnras,
  472, 904, \dodoi{10.1093/mnras/stx1987}

\bibitem[{{Lyman} {et~al.}(2018){Lyman}, {Lamb}, {Levan}, {Mandel}, {Tanvir},
  {Kobayashi}, {Gompertz}, {Hjorth}, {Fruchter}, {Kangas}, {Steeghs}, {Steele},
  {Cano}, {Copperwheat}, {Evans}, {Fynbo}, {Gall}, {Im}, {Izzo}, {Jakobsson},
  {Milvang-Jensen}, {O'Brien}, {Osborne}, {Palazzi}, {Perley}, {Pian},
  {Rosswog}, {Rowlinson}, {Schulze}, {Stanway}, {Sutton}, {Th{\"o}ne}, {de
  Ugarte Postigo}, {Watson}, {Wiersema}, \& {Wijers}}]{Lyman2018a}
{Lyman}, J.~D., {Lamb}, G.~P., {Levan}, A.~J., {et~al.} 2018, Nature Astronomy,
  2, 751, \dodoi{10.1038/s41550-018-0511-3}

\bibitem[{{Marronetti} {et~al.}(2008){Marronetti}, {Tichy}, {Br{\"u}gmann},
  {Gonz{\'a}lez}, \& {Sperhake}}]{marronetti2008}
{Marronetti}, P., {Tichy}, W., {Br{\"u}gmann}, B., {Gonz{\'a}lez}, J., \&
  {Sperhake}, U. 2008, \prd, 77, 064010, \dodoi{10.1103/PhysRevD.77.064010}

\bibitem[{{Meier}(1999)}]{Meier1999a}
{Meier}, D.~L. 1999, \apj, 522, 753, \dodoi{10.1086/307671}

\bibitem[{{Metzger} \& {Berger}(2012)}]{Metzger2012a}
{Metzger}, B.~D., \& {Berger}, E. 2012, \apj, 746, 48,
  \dodoi{10.1088/0004-637X/746/1/48}

\bibitem[{{Metzger} \& {Fern{\'a}ndez}(2014)}]{Metzger2014a}
{Metzger}, B.~D., \& {Fern{\'a}ndez}, R. 2014, \mnras, 441, 3444,
  \dodoi{10.1093/mnras/stu802}

\bibitem[{{Metzger} {et~al.}(2010){Metzger}, {Mart{\'{\i}}nez-Pinedo},
  {Darbha}, {Quataert}, {Arcones}, {Kasen}, {Thomas}, {Nugent}, {Panov}, \&
  {Zinner}}]{Metzger2010a}
{Metzger}, B.~D., {Mart{\'{\i}}nez-Pinedo}, G., {Darbha}, S., {et~al.} 2010,
  \mnras, 406, 2650, \dodoi{10.1111/j.1365-2966.2010.16864.x}

\bibitem[{{Meyer}(1989)}]{Meyer1989a}
{Meyer}, B.~S. 1989, \apj, 343, 254, \dodoi{10.1086/167702}

\bibitem[{{M{\"u}ller}(2016)}]{Mueller2016a}
{M{\"u}ller}, B. 2016, \pasa, 33, e048, \dodoi{10.1017/pasa.2016.40}

\bibitem[{{Nishimura} {et~al.}(2015){Nishimura}, {Takiwaki}, \&
  {Thielemann}}]{nishimura2015a}
{Nishimura}, N., {Takiwaki}, T., \& {Thielemann}, F.-K. 2015, \apj, 810, 109,
  \dodoi{10.1088/0004-637X/810/2/109}

\bibitem[{{Ojima} {et~al.}(2018){Ojima}, {Ishimaru}, {Wanajo}, {Prantzos}, \&
  {Fran{\c{c}}ois}}]{Ojima2018a}
{Ojima}, T., {Ishimaru}, Y., {Wanajo}, S., {Prantzos}, N., \& {Fran{\c{c}}ois},
  P. 2018, \apj, 865, 87, \dodoi{10.3847/1538-4357/aada11}

\bibitem[{{Palenzuela} {et~al.}(2015){Palenzuela}, {Liebling}, {Neilsen},
  {Lehner}, {Caballero}, {O'Connor}, \& {Anderson}}]{Palenzuela2015a}
{Palenzuela}, C., {Liebling}, S.~L., {Neilsen}, D., {et~al.} 2015, \prd, 92,
  044045, \dodoi{10.1103/PhysRevD.92.044045}

\bibitem[{{Perego} {et~al.}(2014){Perego}, {Rosswog}, {Cabez{\'o}n},
  {Korobkin}, {K{\"a}ppeli}, {Arcones}, \& {Liebend{\"o}rfer}}]{Perego2014a}
{Perego}, A., {Rosswog}, S., {Cabez{\'o}n}, R.~M., {et~al.} 2014, \mnras, 443,
  3134, \dodoi{10.1093/mnras/stu1352}

\bibitem[{{Podsiadlowski} {et~al.}(2004){Podsiadlowski}, {Langer},
  {Poelarends}, {Rappaport}, {Heger}, \& {Pfahl}}]{Podsiadlowski2004a}
{Podsiadlowski}, P., {Langer}, N., {Poelarends}, A.~J.~T., {et~al.} 2004, \apj,
  612, 1044, \dodoi{10.1086/421713}

\bibitem[{{Prantzos} {et~al.}(2020){Prantzos}, {Abia}, {Cristallo}, {Limongi},
  \& {Chieffi}}]{Prantzos2020a}
{Prantzos}, N., {Abia}, C., {Cristallo}, S., {Limongi}, M., \& {Chieffi}, A.
  2020, \mnras, 491, 1832, \dodoi{10.1093/mnras/stz3154}

\bibitem[{{Price} \& {Rosswog}(2006)}]{Price2006a}
{Price}, D.~J., \& {Rosswog}, S. 2006, Science, 312, 719,
  \dodoi{10.1126/science.1125201}

\bibitem[{{Radice} {et~al.}(2018){Radice}, {Perego}, {Hotokezaka}, {Fromm},
  {Bernuzzi}, \& {Roberts}}]{Radice2018a}
{Radice}, D., {Perego}, A., {Hotokezaka}, K., {et~al.} 2018, \apj, 869, 130,
  \dodoi{10.3847/1538-4357/aaf054}

\bibitem[{{Roederer} {et~al.}(2012{\natexlab{a}}){Roederer}, {Lawler}, {Cowan},
  {Beers}, {Frebel}, {Ivans}, {Schatz}, {Sobeck}, \& {Sneden}}]{Roederer2012a}
{Roederer}, I.~U., {Lawler}, J.~E., {Cowan}, J.~J., {et~al.}
  2012{\natexlab{a}}, \apjl, 747, L8, \dodoi{10.1088/2041-8205/747/1/L8}

\bibitem[{{Roederer} {et~al.}(2012{\natexlab{b}}){Roederer}, {Lawler},
  {Sobeck}, {Beers}, {Cowan}, {Frebel}, {Ivans}, {Schatz}, {Sneden}, \&
  {Thompson}}]{Roederer2012b}
{Roederer}, I.~U., {Lawler}, J.~E., {Sobeck}, J.~S., {et~al.}
  2012{\natexlab{b}}, \apjs, 203, 27, \dodoi{10.1088/0067-0049/203/2/27}

\bibitem[{{Sekiguchi}(2010)}]{sekiguchi2010a}
{Sekiguchi}, Y. 2010, Progress of Theoretical Physics, 124, 331.
\newblock \doarXiv{1009.3320}

\bibitem[{{Sekiguchi} {et~al.}(2015){Sekiguchi}, {Kiuchi}, {Kyutoku}, \&
  {Shibata}}]{Sekiguchi2015a}
{Sekiguchi}, Y., {Kiuchi}, K., {Kyutoku}, K., \& {Shibata}, M. 2015, \prd, 91,
  064059, \dodoi{10.1103/PhysRevD.91.064059}

\bibitem[{{Sekiguchi} {et~al.}(2016){Sekiguchi}, {Kiuchi}, {Kyutoku},
  {Shibata}, \& {Taniguchi}}]{Sekiguchi2016a}
{Sekiguchi}, Y., {Kiuchi}, K., {Kyutoku}, K., {Shibata}, M., \& {Taniguchi}, K.
  2016, \prd, 93, 124046, \dodoi{10.1103/PhysRevD.93.124046}

\bibitem[{{Shakura} \& {Sunyaev}(1973)}]{Shakura1973a}
{Shakura}, N.~I., \& {Sunyaev}, R.~A. 1973, \aap, 500, 33

\bibitem[{{Shibata}(2000)}]{Shibata2000a}
{Shibata}, M. 2000, Progress of Theoretical Physics, 104, 325,
  \dodoi{10.1143/PTP.104.325}

\bibitem[{{Shibata}(2016)}]{Shibata2016a}
---. 2016, {Numerical Relativity}, \dodoi{10.1142/9692}

\bibitem[{{Shibata} {et~al.}(2017{\natexlab{a}}){Shibata}, {Fujibayashi},
  {Hotokezaka}, {Kiuchi}, {Kyutoku}, {Sekiguchi}, \& {Tanaka}}]{Shibata2017a}
{Shibata}, M., {Fujibayashi}, S., {Hotokezaka}, K., {et~al.}
  2017{\natexlab{a}}, \prd, 96, 123012, \dodoi{10.1103/PhysRevD.96.123012}

\bibitem[{{Shibata} \& {Hotokezaka}(2019)}]{Shibata2019a}
{Shibata}, M., \& {Hotokezaka}, K. 2019, Annual Review of Nuclear and Particle
  Science, 69, 41, \dodoi{10.1146/annurev-nucl-101918-023625}

\bibitem[{{Shibata} {et~al.}(2011{\natexlab{a}}){Shibata}, {Kiuchi},
  {Sekiguchi}, \& {Suwa}}]{shibata2011a}
{Shibata}, M., {Kiuchi}, K., {Sekiguchi}, Y., \& {Suwa}, Y. 2011{\natexlab{a}},
  Progress of Theoretical Physics, 125, 1255, \dodoi{10.1143/PTP.125.1255}

\bibitem[{{Shibata} {et~al.}(2017{\natexlab{b}}){Shibata}, {Kiuchi}, \&
  {Sekiguchi}}]{shibata2017b}
{Shibata}, M., {Kiuchi}, K., \& {Sekiguchi}, Y.-i. 2017{\natexlab{b}}, \prd,
  95, 083005, \dodoi{10.1103/PhysRevD.95.083005}

\bibitem[{{Shibata} \& {Nakamura}(1995)}]{shibata1995a}
{Shibata}, M., \& {Nakamura}, T. 1995, \prd, 52, 5428,
  \dodoi{10.1103/PhysRevD.52.5428}

\bibitem[{{Shibata} {et~al.}(2011{\natexlab{b}}){Shibata}, {Suwa}, {Kiuchi}, \&
  {Ioka}}]{Shibata2011b}
{Shibata}, M., {Suwa}, Y., {Kiuchi}, K., \& {Ioka}, K. 2011{\natexlab{b}},
  \apjl, 734, L36, \dodoi{10.1088/2041-8205/734/2/L36}

\bibitem[{{Shibata} \& {Taniguchi}(2006)}]{Shibata2006a}
{Shibata}, M., \& {Taniguchi}, K. 2006, \prd, 73, 064027,
  \dodoi{10.1103/PhysRevD.73.064027}

\bibitem[{{Shibata} {et~al.}(2005){Shibata}, {Taniguchi}, \&
  {Ury{\={u}}}}]{Shibata2005a}
{Shibata}, M., {Taniguchi}, K., \& {Ury{\={u}}}, K. 2005, \prd, 71, 084021,
  \dodoi{10.1103/PhysRevD.71.084021}

\bibitem[{{Siegel} \& {Metzger}(2018)}]{Siegel2018a}
{Siegel}, D.~M., \& {Metzger}, B.~D. 2018, \apj, 858, 52,
  \dodoi{10.3847/1538-4357/aabaec}

\bibitem[{{Siqueira Mello} {et~al.}(2013){Siqueira Mello}, {Spite}, {Barbuy},
  {Spite}, {Caffau}, {Hill}, {Wanajo}, {Primas}, {Plez}, {Cayrel}, {Andersen},
  {Nordstr{\"o}m}, {Sneden}, {Beers}, {Bonifacio}, {Fran{\c{c}}ois}, \&
  {Molaro}}]{Siqueira2013a}
{Siqueira Mello}, C., {Spite}, M., {Barbuy}, B., {et~al.} 2013, \aap, 550,
  A122, \dodoi{10.1051/0004-6361/201219949}

\bibitem[{{Spitkovsky}(2006)}]{Spitkovsky2006a}
{Spitkovsky}, A. 2006, \apjl, 648, L51, \dodoi{10.1086/507518}

\bibitem[{{Steiner} {et~al.}(2013){Steiner}, {Hempel}, \&
  {Fischer}}]{steiner2013a}
{Steiner}, A.~W., {Hempel}, M., \& {Fischer}, T. 2013, \apj, 774, 17,
  \dodoi{10.1088/0004-637X/774/1/17}

\bibitem[{{Stovall} {et~al.}(2018){Stovall}, {Freire}, {Chatterjee},
  {Demorest}, {Lorimer}, {McLaughlin}, {Pol}, {van Leeuwen}, {Wharton},
  {Allen}, {Boyce}, {Brazier}, {Caballero}, {Camilo}, {Camuccio}, {Cordes},
  {Crawford}, {Deneva}, {Ferdman}, {Hessels}, {Jenet}, {Kaspi}, {Knispel},
  {Lazarus}, {Lynch}, {Parent}, {Patel}, {Pleunis}, {Ransom}, {Scholz},
  {Seymour}, {Siemens}, {Stairs}, {Swiggum}, \& {Zhu}}]{Stovall2018a}
{Stovall}, K., {Freire}, P.~C.~C., {Chatterjee}, S., {et~al.} 2018, \apjl, 854,
  L22, \dodoi{10.3847/2041-8213/aaad06}

\bibitem[{{Suwa} {et~al.}(2018){Suwa}, {Yoshida}, {Shibata}, {Umeda}, \&
  {Takahashi}}]{Suwa2018a}
{Suwa}, Y., {Yoshida}, T., {Shibata}, M., {Umeda}, H., \& {Takahashi}, K. 2018,
  \mnras, 481, 3305, \dodoi{10.1093/mnras/sty2460}

\bibitem[{{Symbalisty} \& {Schramm}(1982)}]{Symbalisty1982a}
{Symbalisty}, E., \& {Schramm}, D.~N. 1982, \aplett, 22, 143

\bibitem[{{Tachibana} {et~al.}(1990){Tachibana}, {Yamada}, \&
  {Yoshida}}]{Tachibana1990a}
{Tachibana}, T., {Yamada}, M., \& {Yoshida}, Y. 1990, Progress of Theoretical
  Physics, 84, 641, \dodoi{10.1143/ptp/84.4.641}

\bibitem[{{Tauris} {et~al.}(2017){Tauris}, {Kramer}, {Freire}, {Wex}, {Janka},
  {Langer}, {Podsiadlowski}, {Bozzo}, {Chaty}, {Kruckow}, {van den Heuvel},
  {Antoniadis}, {Breton}, \& {Champion}}]{Tauris2017a}
{Tauris}, T.~M., {Kramer}, M., {Freire}, P.~C.~C., {et~al.} 2017, \apj, 846,
  170, \dodoi{10.3847/1538-4357/aa7e89}

\bibitem[{{Thorne}(1981)}]{Thorne1981a}
{Thorne}, K.~S. 1981, \mnras, 194, 439, \dodoi{10.1093/mnras/194.2.439}

\bibitem[{{Timmes} \& {Swesty}(2000)}]{timmes2000a}
{Timmes}, F.~X., \& {Swesty}, F.~D. 2000, \apjs, 126, 501,
  \dodoi{10.1086/313304}

\bibitem[{{Villar} {et~al.}(2017){Villar}, {Guillochon}, {Berger}, {Metzger},
  {Cowperthwaite}, {Nicholl}, {Alexander}, {Blanchard}, {Chornock},
  {Eftekhari}, {Fong}, {Margutti}, \& {Williams}}]{Villar2017a}
{Villar}, V.~A., {Guillochon}, J., {Berger}, E., {et~al.} 2017, \apjl, 851,
  L21, \dodoi{10.3847/2041-8213/aa9c84}

\bibitem[{{Vincent} {et~al.}(2020){Vincent}, {Foucart}, {Duez}, {Haas},
  {Kidder}, {Pfeiffer}, \& {Scheel}}]{Vincent2020a}
{Vincent}, T., {Foucart}, F., {Duez}, M.~D., {et~al.} 2020, \prd, 101, 044053,
  \dodoi{10.1103/PhysRevD.101.044053}

\bibitem[{{Wanajo}(2018)}]{Wanajo2018b}
{Wanajo}, S. 2018, \apj, 868, 65, \dodoi{10.3847/1538-4357/aae0f2}

\bibitem[{{Wanajo} \& {Ishimaru}(2006)}]{Wanajo2006a}
{Wanajo}, S., \& {Ishimaru}, Y. 2006, \nphysa, 777, 676,
  \dodoi{10.1016/j.nuclphysa.2005.10.012}

\bibitem[{{Wanajo} {et~al.}(2011){Wanajo}, {Janka}, \&
  {M{\"u}ller}}]{Wanajo2011a}
{Wanajo}, S., {Janka}, H.-T., \& {M{\"u}ller}, B. 2011, \apj, 726, L15,
  \dodoi{10.1088/2041-8205/726/2/L15}

\bibitem[{{Wanajo} {et~al.}(2018){Wanajo}, {M{\"u}ller}, {Janka}, \&
  {Heger}}]{Wanajo2018a}
{Wanajo}, S., {M{\"u}ller}, B., {Janka}, H.-T., \& {Heger}, A. 2018, \apj, 852,
  40, \dodoi{10.3847/1538-4357/aa9d97}

\bibitem[{{Wanajo} {et~al.}(2014){Wanajo}, {Sekiguchi}, {Nishimura}, {Kiuchi},
  {Kyutoku}, \& {Shibata}}]{Wanajo2014a}
{Wanajo}, S., {Sekiguchi}, Y., {Nishimura}, N., {et~al.} 2014, \apj, 789, L39,
  \dodoi{10.1088/2041-8205/789/2/L39}

\bibitem[{{Watson} {et~al.}(2019){Watson}, {Hansen}, {Selsing}, {Koch},
  {Malesani}, {Andersen}, {Fynbo}, {Arcones}, {Bauswein}, {Covino}, {Grado},
  {Heintz}, {Hunt}, {Kouveliotou}, {Leloudas}, {Levan}, {Mazzali}, \&
  {Pian}}]{Watson2019a}
{Watson}, D., {Hansen}, C.~J., {Selsing}, J., {et~al.} 2019, \nat, 574, 497,
  \dodoi{10.1038/s41586-019-1676-3}

\bibitem[{{Waxman} {et~al.}(2018){Waxman}, {Ofek}, {Kushnir}, \&
  {Gal-Yam}}]{Waxman2018a}
{Waxman}, E., {Ofek}, E.~O., {Kushnir}, D., \& {Gal-Yam}, A. 2018, \mnras, 481,
  3423, \dodoi{10.1093/mnras/sty2441}

\bibitem[{{Woosley} {et~al.}(2002){Woosley}, {Heger}, \&
  {Weaver}}]{Woosley2002a}
{Woosley}, S.~E., {Heger}, A., \& {Weaver}, T.~A. 2002, Reviews of Modern
  Physics, 74, 1015, \dodoi{10.1103/RevModPhys.74.1015}

\bibitem[{{Wu} {et~al.}(2019){Wu}, {Barnes}, {Mart{\'\i}nez-Pinedo}, \&
  {Metzger}}]{Wu2019a}
{Wu}, M.-R., {Barnes}, J., {Mart{\'\i}nez-Pinedo}, G., \& {Metzger}, B.~D.
  2019, \prl, 122, 062701, \dodoi{10.1103/PhysRevLett.122.062701}

\bibitem[{{Zhu} {et~al.}(2018){Zhu}, {Wollaeger}, {Vassh}, {Surman}, {Sprouse},
  {Mumpower}, {M{\"o}ller}, {McLaughlin}, {Korobkin}, {Kawano}, {Jaffke},
  {Holmbeck}, {Fryer}, {Even}, {Couture}, \& {Barnes}}]{Zhu2018a}
{Zhu}, Y., {Wollaeger}, R.~T., {Vassh}, N., {et~al.} 2018, \apjl, 863, L23,
  \dodoi{10.3847/2041-8213/aad5de}

\end{thebibliography}

\end{document}